# On the weak coupling limit of the periodic quantum Lorentz gas[*]

MASSIMILIANO GUBINELLI, University of Oxford

VISHNU SANJAY[A], Gran Sasso Science Institute

We report partial progress on the weak coupling limit behavior of observables for the periodic quantum Lorentz gas. Our results indicate that for certain observables, the limit behavior is trivial and can be described via a transport equation, while for other observables, the existence of the limit hinges on the regularity properties at resonant momenta of a certain Bloch-Wigner transform. We are currently unable to prove or disprove this regularity property, and so the weak coupling limit for these observables remains an open question. A novelty of this work is the use of the sewing lemma in the derivation of the kinetic scaling limit for almost every mometum.

## TABLE OF CONTENTS

---

A. vishnu.sanjay@gssi.it

# 1  Introduction and prior work

Consider the linear Schrödinger equation with a time-independent potential $V$: for $x \in \mathbb{R}^d, t \in \mathbb{R}$

$$i\partial_t \varphi(t,x) = -\Delta_x \varphi(t,x) + \lambda V(x)\varphi(t,x), \qquad \varphi(0,x) = \varphi_0(x). \tag{1.1}$$

Here $\lambda \in \mathbb{R}_{\geqslant 0}$ is a small coupling constant. The potential $V$ is assumed to be $\mathbb{Z}^d$-periodic, i.e., $V(x+n) = V(x), \forall x \in \mathbb{R}^d, \forall n \in \mathbb{Z}^d$. We refer to this model as the periodic quantum Lorentz gas. As a matter of convenience, we assume throughout this document that

$$V \in \cap_{m \in \mathbb{N}_0} H^m([0,1]^d). \tag{1.2}$$

In practice, we only need that $V \in \cap_{m \leqslant M} H^m([0,1]^d)$ for some sufficiently large $M \in \mathbb{N}_0$.

Instead of working with the wavefunction directly, we study the problem in phase space. Recall the definition of the Wigner transform, for $\varphi \in L^2(\mathbb{R}^d; \mathbb{C})$,

$$W_\varphi(x,k) := \int_{\mathbb{R}^d} \mathrm{d}y\, e^{2\pi i k \cdot y} \varphi\left(x - \frac{y}{2}\right) \varphi^*\left(x + \frac{y}{2}\right), \quad x, k \in \mathbb{R}^d. \tag{1.3}$$

The Wigner function has many properties that a phase space density would have. Formally, integrating in momenta,

$$\int_{\mathbb{R}^d} \mathrm{d}k\, W_\varphi(x,k) = |\varphi(x)|^2.$$

This quantity corresponds to the probability of finding the particle at position $x$. Thus, similar to a phase space density, one recovers information on the position of the particle by integrating over the momenta. However, the Wigner transform is not necessarily positive. We refer the reader to [21], [9] and [24] for detailed studies of the Wigner transform and its properties.

In the weak coupling regime, one first sets $\lambda = \varepsilon^{1/2}$ for some $0 < \varepsilon \ll 1$. Since the coupling constant is small, one needs to wait for a long time to see the effects of the potential. This is commonly done by working with a rescaled Wigner function:

$$W^\varepsilon(t,x,k) := W_\varphi\left(\frac{t}{\varepsilon}, \frac{x}{\varepsilon}, k\right) \tag{1.4}$$

with initial conditions $W_0^\varepsilon(x,k) = W_{\varphi_{0,\varepsilon}}(x,k)$, where $\varphi$ is the solution to (1.1) with initial data $\varphi_{0,\varepsilon}$.

*Remark* 1.1. Note that when rescaling the Wigner function as in equation (1.4), one needs to choose initial data $\varphi_{0,\varepsilon}$ for equation (1.1) depending on $\varepsilon$ such that their $L^2$-norms grow, in order to move from the "microscopic scale" ($\varepsilon = 1$) to the "macroscopic scale" ($\varepsilon = 0$). For instance, the WKB family of initial data used in [14] corresponds to

$$\varphi_{0,\varepsilon}(x) = h(\varepsilon x)e^{2\pi i S(\varepsilon x)/\varepsilon}$$

for $h, S \in \mathscr{S}(\mathbb{R}^d)$. Then one can compute that as $\varepsilon \to 0$,

$$W^{\varepsilon}_{\varphi_{0,\varepsilon}}(x,k) \rightharpoonup |h|^2(x)\delta(k - \nabla S(x))$$

weakly as a tempered distribution. In particular, the $L^\infty$-norm of this family of rescaled Wigner initial data blows up. See [24] for other possible families of initial data.

Physically measurable quantities such as $|\varphi(x)|^2$ will be referred to as observables. For our purposes, we will restrict to studying observables of the form

$$\int_{\mathbb{R}^d} \mathrm{d}k \int_{\mathbb{R}^d} \mathrm{d}x\, W^\varepsilon(t,x,k) F(x,k), \tag{1.5}$$

for some $F \in \mathscr{S}(\mathbb{R}^{2d})$. We are interested in the time evolution of such expressions, in the limit $\varepsilon \to 0$.

We are currently unable to answer this question for every $F \in \mathscr{S}(\mathbb{R}^{2d})$, but present some partial progress towards this goal. In an accompanying work (in preparation) we will demonstrate that for observables supported away from a certain zero measure set of resonant momenta, the limiting behavior can be described by the solution of a transport equation.

The classical Lorentz gas is a model in kinetic theory, describing a moving particle in an environment of fixed obstacles. The particle obeys Newton's laws of motion and undergoes elastic collisions with the obstacles it encounters in its trajectory. Typically in the literature, one assumes that the configuration of scatterers is randomly distributed according to a certain law, or that it is periodic. Two important scaling limits have been identified for this model, where interesting mesoscopic effects are seen in the limit. The first is the weak coupling limit, where the obstacles have a weak effect on the particle, but on a larger space and time scale, the number of collisions is large enough to have a cumulative effect. This is similar to what we have described above. The second is the low-density limit, where the collisions occur much less frequently, but where each collision has a strong effect. See [16], [30], [23], [11] and [4] for the pioneering results in the study of kinetic limits for the classical random Lorentz gas. The low-density limit of the periodic Lorentz gas was derived in [25].

Here, we are studying the weak coupling limit of the quantum analogue of the classical Lorentz gas. The wavefunction models the particle, and the obstacles are modeled via the potential term in (1.1). In the case when $V$ is a Gaussian random field with smooth covariance, in the weak-coupling scaling, for WKB initial data, it is known that for all $F \in \mathscr{S}(\mathbb{R}^d), t \in \mathbb{R}$

$$\mathbb{E}\left[\int_{\mathbb{R}^d} \mathrm{d}k \int_{\mathbb{R}^d} \mathrm{d}x\, W^\varepsilon(t,x,k) F(x,k)\right] \xrightarrow{\varepsilon \to 0} \int_{\mathbb{R}^d} \mathrm{d}k \int_{\mathbb{R}^d} \mathrm{d}x\, f(t,x,k) F(x,k)$$

where $f$ satisfies a linear Boltzmann equation. This was shown in [14], extending the result of [29], where a local in time version of the above result was obtained. See [2] for a heuristic justification of this result. In [12] the authors derive the low-density limit of the quantum Lorentz gas with a random potential. There are also works investigating the problem with time-dependent potentials, see for instance [3] and [18]. In [7], the average wave function was studied instead of the Wigner transform.

For the periodic quantum Lorentz gas, the low-density limit was investigated in [20], but the limiting behavior has not been identified yet. The problem has also been studied in the torus rather than the full space, see [6] and [5]. To the best of our knowledge, the weak coupling limit in the full space for the observables as in expression (1.5) has not yet been established for the periodic quantum Lorentz gas.

We also mention that the periodic Schrödinger equation has been investigated in the so called semiclassical limit. See [26] and [17].

Finally, regarding techniques, we mention that, unlike in the previously cited works, we use the sewing lemma and tools from the theory of unbounded rough drivers. We refer the reader to [1] and [10], for the theory of unbounded rough drivers.

In Section 2, we introduce and state our main result, Theorem 2.8, taking the limit $\varepsilon \to 0$ in a certain topology for an object related to the rescaled Wigner transform. Section 3 contains the proof of Theorem 2.8. Section 4 is a study of the observables, showing that in the topology relevant to the study of the observables, there is an obstruction to proceeding as in the proof of Theorem 2.8. We characterize this obstruction in terms of regularity properties of a certain phase space object. Appendices B.1 and B.2 are devoted to the proofs of minor lemmas, and in Appendix B.3, we prove estimates on certain smoothing operators $\mathcal{J}_\nu$ that are used Subsection 3.4.

## 1.1 Notation

1. $\mathbb{N}_0 := \mathbb{N} \cup \{0\}$.

2. $\mathbb{R}_+ := (0, +\infty), \mathbb{R}_{\geqslant 0} := [0, \infty)$.

3. $\mathbb{T}^d := \mathbb{R}^d / \mathbb{Z}^d$.

4. $\langle \cdot \rangle := \sqrt{(1 + |\cdot|^2)}$ will denote the usual Japanese bracket.

5. For $X$ a Banach space and $f \in C(\mathbb{R}_+; X)$ we sometimes use
$$f_t := f(t), \qquad \delta_{st} T := f_t - f_s.$$

6. For a function $f \in L^1(\mathbb{R}^d)$ we use the following convention of the Fourier transform: for $\xi \in \mathbb{R}^d$
$$\hat{f}(\xi) := \int_{\mathbb{R}^d} \mathrm{d}x\, e^{-2\pi i \xi \cdot x} f(\xi).$$

    For a function $f \in L^1(\mathbb{T}^d)$ we use the following convention for the coefficients of the Fourier series: for $n \in \mathbb{Z}^d$
$$\hat{f}(n) := \int_{\mathbb{T}^d} \mathrm{d}x\, e^{-2\pi i n \cdot x} f(x).$$

7. We use the following notation for the floor function $\lfloor \cdot \rfloor : \mathbb{R} \to \mathbb{Z}, x \to \lfloor x \rfloor$, taking any real number to the closest integer smaller than or equal to it. When applied to $x \in \mathbb{R}^d$, this is the floor function applied component-wise.

## 2 Summary of the main results

We begin by quickly recalling some well-posedness theory for the linear Schrödinger equation.

DEFINITION 2.1. *A classical solution to equation (1.1) is a function*

$$\varphi \in C^1(\mathbb{R}_+; L^2(\mathbb{R}^d; \mathbb{C})) \cap C(\mathbb{R}_{\geqslant 0}; H^2(\mathbb{R}^d; \mathbb{C}))$$

*such that* $\varphi(0) = \varphi_0$ *and* $\varphi'(t) = H\varphi(t), \forall t \in \mathbb{R}_+$.

By Theorem X.15 of [28] we know that for $V$ satisfying the assumption (1.2) that $H := -\Delta + \varepsilon^{1/2} V$ is self-adjoint on $D(-\Delta) = H^2(\mathbb{R}^d; \mathbb{C})$. Stone's theorem (Theorem 3.24 in [13]) then says that $H$ is the infinitesimal generator of a $\mathscr{C}_0$-group of unitary operators, $S(t)$. By using standard results in the theory of semigroups, one has that for any $\varphi_0 \in H^2(\mathbb{R}^d; \mathbb{C})$, that $\varphi(t) = S(t)\varphi_0 \in C^1(\mathbb{R}_+; L^2(\mathbb{R}^d; \mathbb{C})) \cap C(\mathbb{R}_{\geqslant 0}; H^2(\mathbb{R}^d; \mathbb{C}))$ is the unique classical solution to equation (1.1).

We can then obtain an apriori estimate on the Wigner transform associated to this classical Schrödinger solution. Before doing this, we introduce a representation formula for the Wigner transform that is convenient when working with periodic potentials. This representation uses the Bloch–Floquet–Zak decomposition, introduced by J. Zak in [31] and [32], and which is now a well-known tool in solid state physics (see for instance [27]).

DEFINITION 2.2. *For $\varphi \in \mathscr{S}(\mathbb{R}^d; \mathbb{C})$, $\theta \in \mathbb{R}^d$, $x \in \mathbb{R}^d$, define the **Bloch–Floquet–Zak decomposition of $\varphi$**, or BFZ decomposition of $\varphi$ as:*

$$(\mathscr{U}_{\mathrm{BFZ}}\varphi)_\theta(x) := \tilde{\varphi}(\theta, x) := \sum_{m \in \mathbb{Z}^d} e^{2\pi i \theta \cdot (x-m)} \varphi(x-m). \tag{2.1}$$

We collect some useful properties of the BFZ decomposition in Appendix A. In Appendix B.1 we will use these properties to prove the following representation formula for the Wigner function.

LEMMA 2.3. *Let $\varphi \in \mathscr{S}(\mathbb{R}^d; \mathbb{C})$. Then for for $W_\varphi(x, k)$ defined in (1.3), decomposing the momentum $k = \kappa - \eta$, with $\kappa \in \left(\frac{\mathbb{Z}}{2}\right)^d$ and $\eta \in \left[-\frac{1}{4}, \frac{1}{4}\right]^d$, we have the following representation:*

$$W_\varphi(x, k) = \int_{\mathbb{T}^d} \mathrm{d}y \int_{\mathbb{T}^d} \mathrm{d}\theta\, e^{-4\pi i \theta \cdot x} e^{-4\pi i \kappa \cdot y} \tilde{\varphi}(\eta + \theta, x + y) \tilde{\varphi}^*(\eta - \theta, x - y). \tag{2.2}$$

This can be extended by a density argument to $L^2(\mathbb{R}^d; \mathbb{C})$. To the best of our knowledge, this has not appeared previously in the literature. Noting that the variable $x$ is not just periodic in the overall integrand but individually as an argument of $\tilde{\varphi}$ and $\tilde{\varphi}^*$ (due to the definition of the BFZ transform), we found it useful to introduce the following generalization

DEFINITION 2.4. *Let $z \in \mathbb{T}^d$, $p \in \mathbb{R}^d$ play the role of position variables and let $\eta \in \left[-\frac{1}{4}, \frac{1}{4}\right]^d$, $\kappa \in \left(\frac{\mathbb{Z}}{2}\right)^d$ play the role of momentum variables. Then for $\varphi \in L^2(\mathbb{R}^d; \mathbb{C})$ with associated periodic Bloch decomposition $\tilde{\varphi}$ the Bloch–Wigner function is defined as*

$$\tilde{W}_\varphi(z, p, \eta, \kappa) := \int_{\mathbb{T}^d} \mathrm{d}y \int_{\left[-\frac{1}{2}, \frac{1}{2}\right]^d} \mathrm{d}\theta\, e^{-4\pi i \theta \cdot p} e^{-4\pi i \kappa \cdot y} \tilde{\varphi}(\eta + \theta, z + y) \tilde{\varphi}^*(\eta - \theta, z - y). \tag{2.3}$$

This is well-defined as a consequence of the unitarity of the BFZ transform. We see that for $k = \kappa - \eta$, with $\kappa \in \left(\frac{\mathbb{Z}}{2}\right)^d, \eta \in \left[-\frac{1}{4}, \frac{1}{4}\right)^d$, identifying $[0,1]^d$ with $\mathbb{T}^d$, that

$$W_\varphi(x, k) = \tilde{W}_\varphi(x - \lfloor x \rfloor, x, \eta, \kappa). \tag{2.4}$$

We study the time evolution of the Bloch–Wigner transform associated to a solution of the Schrödinger equation. Denote

$$L^\infty_{\eta, p, \kappa, z} := L^\infty_{\eta, p} l^\infty_\kappa L^\infty_z \left(\left[-\frac{1}{4}, \frac{1}{4}\right)^d \times \mathbb{R}^d \times \left(\frac{\mathbb{Z}}{2}\right)^d \times \mathbb{T}^d\right) \tag{2.5}$$

and

$$L^\infty_{\eta, p, \kappa} L^2_z := L^\infty_{\eta, p} l^\infty_\kappa \left(\left(\left[-\frac{1}{4}, \frac{1}{4}\right)^d \times \mathbb{R}^d \times \left(\frac{\mathbb{Z}}{2}\right)^d\right); L^2_z(\mathbb{T}^d)\right). \tag{2.6}$$

We prove in Appendix B.2 the following

PROPOSITION 2.5. *For $\varphi \in C(\mathbb{R}_{\geqslant 0}; H^2(\mathbb{R}^d; \mathbb{C})) \cap C^1(\mathbb{R}_+; L^2(\mathbb{R}^d; \mathbb{C}))$ satisfying the Schrödinger equation with initial data $\varphi_0$, one has that $\tilde{W}_\varphi \in C^1(\mathbb{R}_+; L^\infty_{\eta, p, \kappa, z})$. Furthermore $(\kappa - \eta) \cdot (\nabla_p + \nabla_z) \tilde{W}_\varphi \in C(\mathbb{R}_+, L^\infty_{\eta, p, \kappa, z})$ and $\tilde{W}_\varphi$ satisfies*

$$\partial_t \tilde{W}_\varphi = -4\pi(\kappa - \eta) \cdot \nabla_p \tilde{W}_\varphi - 4\pi(\kappa - \eta) \cdot \nabla_z \tilde{W}_\varphi + i\varepsilon^{1/2} Q \tilde{W}_\varphi \tag{2.7}$$

*with initial data $\tilde{W}_{\varphi_0}$, where*

$$Q\tilde{W}_\varphi(z, p, \eta, \kappa) = \sum_{n \in \mathbb{Z}^d} e^{2\pi i n \cdot z} \hat{V}(n) \left[\tilde{W}_\varphi\left(z, p, \eta, \kappa + \frac{n}{2}\right) - \tilde{W}_\varphi\left(z, p, \eta, \kappa - \frac{n}{2}\right)\right] \tag{2.8}$$

In order to study the scaling limit, we define the rescaled Bloch–Wigner transform to be

$$\tilde{W}^\varepsilon_\varphi(t, z, p, \eta, \kappa) := \tilde{W}_\varphi\left(\frac{t}{\varepsilon}, z, \frac{p}{\varepsilon}, \eta, \kappa\right), \tag{2.9}$$

with initial data

$$\tilde{W}^\varepsilon_{\varphi_{0,\varepsilon}}(z, p, \eta, \kappa) = \tilde{W}_{\varphi_{0,\varepsilon}}\left(z, \frac{p}{\varepsilon}, \eta, \kappa\right). \tag{2.10}$$

*Remark* 2.6. As a followup to Remark 1.1, in this chapter we will focus on initial data $\varphi_{0,\varepsilon}$ for (1.1) that are uniformly bounded in $L^2(\mathbb{R}^d; \mathbb{C})$ so that by Proposition 2.5, expression (2.10) will be uniformly bounded in $L^\infty_{\eta, p, \kappa, z}$. For instance, one can pick $\varphi_{0,\varepsilon} = \varphi_0$, in which case the limit in $L^\infty_{\eta, p, \kappa, z}$ is trivial. However, the equation for the evolution remains the same, and we will see that even in the setting of initial data converging to something trivial macroscopically, for certain observables, there are challenges in proving convergence of the rescaled Bloch–Wigner functions. We will shortly make another related remark after stating our main theorem.

One can compute the time evolution of the rescaled Bloch–Wigner transform to be

$$\partial_t \tilde{W}^\varepsilon_\varphi = -4\pi\varepsilon^{-1}(\kappa - \eta) \cdot \nabla_z \tilde{W}^\varepsilon_\varphi - 4\pi(\kappa - \eta) \cdot \nabla_p \tilde{W}^\varepsilon_\varphi + i\varepsilon^{-1/2} Q \tilde{W}^\varepsilon_\varphi. \tag{2.11}$$

We work in a moving coordinate frame via

$$U^\varepsilon(t,z,p,\eta,\kappa) := \tilde{W}^\varepsilon_\varphi(t, z + 4\pi\varepsilon^{-1}(\kappa-\eta)t, p, \eta, \kappa), \tag{2.12}$$

and study the Fourier transform of this in the periodic variable $z$ via

$$T^\varepsilon(t,\xi,p,\eta,\kappa) := \mathscr{F}_z U^\varepsilon(t,\xi,p,\eta,\kappa). \tag{2.13}$$

We will use the notation

$$T^{\varepsilon,\eta}(t,\xi,p,\kappa) := T^\varepsilon(t,\xi,p,\eta,\kappa)$$

One also has that

$$T^{\varepsilon,\eta}_0(\xi,p,\kappa) = T^\varepsilon_0(\xi,p,\eta,\kappa) = \mathscr{F}_z \tilde{W}^\varepsilon_{\varphi_0,\varepsilon}(z,p,\eta,\kappa). \tag{2.14}$$

The time evolution of $T^\varepsilon$ can be computed using equations (2.11)-(2.13). One has that

$$\partial_t T^\varepsilon(t,\xi,p,\eta,\kappa) = -4\pi(\kappa-\eta)\cdot\nabla_p T^\varepsilon(t,\xi,p,\eta,\kappa) + Q^\varepsilon_t T^\varepsilon(t,\xi,p,\eta,\kappa), \tag{2.15}$$

where

$$A^{\kappa-\eta} G(\xi,p,\eta,\kappa) := -4\pi(\kappa-\eta)\cdot\nabla_p G(\xi,p,\eta,\kappa),$$

and

$$Q^\varepsilon_t G(\xi,p,\eta,\kappa) := i\varepsilon^{-1/2} \sum_{n\in\mathbb{Z}^d} e^{4\pi^2 i\varepsilon^{-1} n\cdot(2\kappa-2\eta+n-\xi)t} \hat{V}(n) G\left(\xi-n, p, \eta, \kappa+\frac{n}{2}\right)$$

$$- i\varepsilon^{-1/2} \sum_{n\in\mathbb{Z}^d} e^{4\pi^2 i\varepsilon^{-1} n\cdot(2\kappa-2\eta-n+\xi)t} \hat{V}(n) G\left(\xi-n, p, \eta, \kappa-\frac{n}{2}\right).$$

Note that the dynamics of (2.15) does not affect $\eta$. To make sense of the limit of equation (2.15), as $\varepsilon \to 0$, we will use some of the machinery introduced in [1]. To this end, we introduce the following scale: for $m \geq 0$ define the Banach space

$$E_m := \left\{ \psi \in L^1_p l^1_\kappa l^2_\xi : \sum_{|\beta|\leq m} \int_{\mathbb{R}^d} dp \sum_{\kappa\in\left(\frac{\mathbb{Z}}{2}\right)^d} \langle\kappa\rangle^m \|D^\beta_p \psi(\xi,p,\kappa)\|_{l^2_\xi} < \infty \right\}, \tag{2.16}$$

with corresponding dual spaces $E_{-m} := E^*_m$. In particular

$$E_0 = L^1_p l^1_\kappa l^2_\xi\left(\mathbb{R}^d \times \left(\frac{\mathbb{Z}}{2}\right)^d \times \mathbb{Z}^d\right), \qquad E_{-0} = L^\infty_p l^\infty_\kappa l^2_\xi\left(\mathbb{R}^d \times \left(\frac{\mathbb{Z}}{2}\right)^d \times \mathbb{Z}^d\right).$$

This is motivated by the structure of the unbounded transport operator $A^{\kappa-\eta}$ (which worsens regularity in $p$ and summability in $k$), and by the following apriori bound on $T^{\varepsilon,\eta}$, which follows from Proposition 2.7, and equations (2.9), (2.12) and (2.13)

$$\operatorname*{esssup}_{\eta\in\left[-\frac{1}{4},\frac{1}{4}\right)^d} \|T^{\varepsilon,\eta}\|_{L^\infty([0,T];E_{-0})} \leq C. \tag{2.17}$$

Next, we fix

$$\delta \in \left(0, \frac{1}{d+3}\right) \tag{2.18}$$

and define

$$\mathscr{A}_\eta := \left\{\eta\in\left[\frac{1}{4},\frac{1}{4}\right]^d : |2n\cdot\eta - a|^{-1} \leq c_\delta(\eta)^{\frac{1}{1-\delta}} \langle n\rangle^{d+2} \,\forall a\in\mathbb{Z}, n\in\mathbb{Z}^d\setminus\{0\}\right\} \tag{2.19}$$

where $c_\delta \in L^1_\eta\left(\left[-\frac{1}{4}, \frac{1}{4}\right)^d\right)$ is an integrable function that will be constructed explicitly in Lemma 3.2. Lemma 3.2 also asserts that $\mathscr{A}_\eta$ is a full measure subset of $\left[\frac{1}{4}, \frac{1}{4}\right)^d$.

Let

$$Y^{\eta,*}\psi(\xi, p, \kappa) := \frac{i}{4\pi^2} \sum_{n \in \mathbb{Z}^d \setminus \{0\}} |\hat{V}(n)|^2 \left( \frac{\psi(\xi, p, \kappa) - \psi(\xi, p, \kappa - n)\mathbb{I}_{n \perp \xi}}{n \cdot (2\kappa - 2\eta - \xi - n)} \right) \quad (2.20)$$

$$- \frac{i}{4\pi^2} \sum_{n \in \mathbb{Z}^d \setminus \{0\}} |\hat{V}(n)|^2 \left( \frac{\psi(\xi, p, \kappa + n)\mathbb{I}_{n \perp \xi} - \psi(\xi, p, \kappa)}{n \cdot (2\kappa - 2\eta + \xi + n)} \right). \quad (2.21)$$

Note that $Y^{\eta,*}\psi(0, p, \kappa) = 0$ for any $\psi \in E_0$.

DEFINITION 2.7. *Let $\tau > 0$. A function $f \in C([0, \tau]; E_{-0})$ is a weak solution to the linear Boltzmann equation in $[0, \tau]$ with initial condition $f_0 \in E_{-0}$ if*

$$\langle f(\tau), \varphi(\tau) \rangle - \langle f_0, \varphi(0) \rangle = \int_0^\tau dt \langle f(t), (\partial_t + A^{\kappa - \eta, *} + Y^{\eta, *})\varphi(t) \rangle \quad (2.22)$$

*for all $\varphi \in C^1((0, \tau); E_0) \cap C((0, \tau); E_1) \cap C([0, \tau]; E_0)$.*

We are now able to state our main result, which is the following

THEOREM 2.8. *Assume that we have initial data $\varphi_{0,\varepsilon}$ for the Schrödinger equation (1.1) such that $T_0^{\varepsilon,\eta}(\xi, p, \kappa)$ defined in (2.14) converges weakly-$*$ in $E_{-0}$ to a limit $T_0^\eta(\xi, p, \kappa)$. Then, for any $T > 0$, a.e. $\eta \in \left[-\frac{1}{4}, \frac{1}{4}\right)^d$ one has that $T^{\varepsilon,\eta}(t, \xi, p, \kappa)$ converges weakly-$*$ in $L^\infty([0, T]; E_{-0})$ to a limit $T^\eta(t, \xi, p, \kappa)$ and $T^\eta$ is the unique weak solution of equation (2.22) in $[0, T]$ with initial data $T_0^\eta$.*

Note that this is not a claim that the limiting behavior of the rescaled observables can be described by a linear Boltzmann equation.

The proof of this Theorem is contained in Section 3. A key novelty of the proof, and our approach here, is the use of the sewing lemma from rough path theory in avoiding diagrammatic estimates of Duhamel iterates of equation (3.3).

*Remark* 2.9. Recalling equation (1.4), we were interested in weak limits of

$$W^\varepsilon(t, p, k) = W\left(\frac{t}{\varepsilon}, \frac{p}{\varepsilon}, k\right) = \tilde{W}_\varphi\left(\frac{t}{\varepsilon}, \left[\!\left[\frac{p}{\varepsilon}\right]\!\right], \frac{p}{\varepsilon}, \eta, \kappa\right) = \tilde{W}^\varepsilon\left(t, \left[\!\left[\frac{p}{\varepsilon}\right]\!\right], p, \eta, \kappa\right)$$

i.e. weak limits of the rescaled Bloch–Wigner transform with $z = \frac{p}{\varepsilon}$. Our intuition was that $z$, as a "fast variable" on the torus, would cause a homogenization effect, leading to an average over $z$ in the limit. We chose as a first step to not rescale $z$ in equation (2.9), and to just examine the zero Fourier mode after taking the limit (which is 0). In Section 4 we will be more precise about this intuition, when we heuristically use a stationary phase argument to rewrite the expression for the observables in terms of $T^\varepsilon$, evaluated at $\xi = 0$.

We mention two important shortcomings of this result

1. As mentioned above, Theorem 2.8 does not identify the limiting behavior of the rescaled observables, which involve integrating over $\eta$. It leaves open the possibility that when we integrate in $\eta$ and then pass to the limit, we could have concentration on a zero measure set of $\eta$'s for the collision term. We report partial progress on this in Section 4.

2. Furthermore, we stress once more that we have results in $E_{-0}$, since we work with initial data for which we have an *a-priori* bound and uniform estimates on scales $E_m$. However, since we do not expect to have a family of initial data that converge to a nontrivial limit in $E_{-0}$, this is a weak statement at the moment. However we believe that the method of proof we will demonstrate here can be adapted to other scales of Banach spaces in the future, once an appropriate *a-priori* bound has been identified. We show that even in our simpler case, there is an obstruction to proving convergence for the observables.

## 3 Proof of Theorem 2.8

In Subsection 3.1 we use the Duhamel principle to derive a difference equation (3.3) for the time increments of $T^\varepsilon$, and will then show that it has the structure of a rough difference equation. Subsection 3.2 introduces some useful lemmas that will then be used in Subsection 3.3 to prove uniform in $\varepsilon$-estimates for the operators arising in equation (3.3). Finally in Subsection 3.4, we will use the sewing lemma to estimate also the remainder uniformly in $\varepsilon$, and this will allow us to pass to the limit and characterise the limit equation, thus proving the theorem. Applying the sewing lemma in this context will require certain ideas from the theory of unbounded rough drivers introduced in [1].

### 3.1 Deriving a rough difference equation

By using Proposition 2.5 and testing equation (2.15) against functions in $E_2$, we have that for every $F \in E_2$ that

$$\langle \partial_t T_t^{\varepsilon,\eta}, F \rangle = \langle A^{\kappa-\eta} T_t^{\varepsilon,\eta}, F \rangle + \langle Q_t^{\varepsilon,\eta} T_t^{\varepsilon,\eta}, F \rangle$$

where $\langle \cdot, \cdot \rangle$ means we are integrating and summing in $p, \kappa$ and $\xi$, and $A^{\kappa-\eta} = -4\pi(\kappa - \eta) \cdot \nabla_p$. We write $T^{\varepsilon,\eta}$ to indicate that we fix $\eta$, noting that it is just a parameter in the equation and does not get changed by the dynamics. In terms of the adjoint operators $A^{\kappa-\eta,*}$ and $Q_t^{\varepsilon,\eta,*}$ this is

$$\langle \partial_t T_t^{\varepsilon,\eta}, F \rangle = \langle T_t^{\varepsilon,\eta}, A^{\kappa-\eta,*} F \rangle + \langle T_t^{\varepsilon,\eta}, Q_t^{\varepsilon,\eta,*} F \rangle \tag{3.1}$$

where $A^{\kappa-\eta,*} = -A^{\kappa-\eta}$ and

$$Q_u^{\varepsilon,\eta,*} F(\xi, p, \kappa) = i\varepsilon^{-1/2} \sum_{n \in \mathbb{Z}^d} e^{4\pi^2 i \varepsilon^{-1} n \cdot (2\kappa - 2\eta - n - \xi)u} \hat{V}(n) F\left(\xi + n, p, \kappa - \frac{n}{2}\right)$$

$$- i\varepsilon^{-1/2} \sum_{n \in \mathbb{Z}^d} e^{4\pi^2 i \varepsilon^{-1} n \cdot (2\kappa - 2\eta + n + \xi)u} \hat{V}(n) F\left(\xi + n, p, \kappa + \frac{n}{2}\right)$$

Then integrating in time, one has by the fundamental theorem of calculus that

$$\int_s^t du \langle \partial_t T_u^{\varepsilon,\eta}, F \rangle = \langle \delta_{st} T^{\varepsilon,\eta}, F \rangle \tag{3.2}$$

and using this and equation (3.1) once more, one has that the first term on the right hand side is

$$\int_s^t du \langle T_u^{\varepsilon,\eta}, A^{\kappa-\eta,*} F \rangle = \int_s^t du \langle T_s^{\varepsilon,\eta}, A^{\kappa-\eta,*} F \rangle + \int_s^t du \int_s^u dv \langle T_v^{\varepsilon,\eta}, A^{\kappa-\eta,*} A^{\kappa-\eta,*} F \rangle$$

$$+ \int_s^t du \int_s^u dv \langle T_v^{\varepsilon,\eta}, Q_v^{\varepsilon,\eta,*} A^{\kappa-\eta,*} F \rangle$$

Finally, by using (3.2) and equation (3.1) twice more, the second term on the right hand side is

$$\int_s^t du \langle T_u^\varepsilon, Q_u^{\varepsilon,\eta,*} F \rangle = \int_s^t du \langle T_s^\varepsilon, Q_u^{\varepsilon,\eta,*} F \rangle + \int_s^t du \int_s^u dv \langle T_v^\varepsilon, A^{\kappa-\eta,*} Q_u^{\varepsilon,\eta,*} F \rangle$$

$$+ \int_s^t du \int_s^u dv \langle T_v^\varepsilon, Q_v^{\varepsilon,\eta,*} Q_u^{\varepsilon,\eta,*} F \rangle$$

$$= \int_s^t du \langle T_s^\varepsilon, Q_u^{\varepsilon,\eta,*} F \rangle + \int_s^t du \int_s^u dv \langle T_v^\varepsilon, A^{\kappa-\eta,*} Q_u^{\varepsilon,\eta,*} F \rangle$$

$$+ \int_s^t du \int_s^u dv \langle T_s^\varepsilon, Q_v^{\varepsilon,\eta,*} Q_u^{\varepsilon,\eta,*} F \rangle + \int_s^t du \int_s^u dv \int_s^v dw \langle T_w^\varepsilon, A^{\kappa-\eta,*} Q_v^{\varepsilon,\eta,*} Q_u^{\varepsilon,\eta,*} F \rangle$$

$$+ \int_s^t du \int_s^u dv \int_s^v dw \langle T_w^\varepsilon, Q_w^{\varepsilon,\eta,*} Q_v^{\varepsilon,\eta,*} Q_u^{\varepsilon,\eta,*} F \rangle$$

Hence, overall, one has that

$$\langle \delta_{st} T^{\varepsilon,\eta}, F \rangle = \langle T_s^{\varepsilon,\eta}, \mathbb{A}_{st}^{\kappa-\eta,*} F \rangle + \langle T_s^{\varepsilon,\eta}, \mathbb{X}_{st}^{1,\varepsilon,\eta,*} F \rangle + \langle T_s^{\varepsilon,\eta}, \mathbb{X}_{st}^{2,\varepsilon,\eta,*} F \rangle + \langle T_{st}^{\varepsilon,\eta,\natural}, F \rangle \tag{3.3}$$

where

$$\mathbb{A}_{st}^{\kappa-\eta,*} F(\xi, p, \kappa) = 4\pi(t-s)(\kappa-\eta) \cdot \nabla_p F(\xi, p, \kappa)$$

$$\mathbb{X}_{st}^{1,\varepsilon,*} F(\xi, p, \kappa) = \int_s^t du Q_u^{\varepsilon,\eta,*} F(\xi, p, \kappa)$$

and

$$\mathbb{X}_{st}^{2,\varepsilon,\eta,*} F(\xi, p, \kappa) = \int_s^t du \int_s^u dv Q_v^{\varepsilon,\eta,*} Q_u^{\varepsilon,\eta,*} F(\xi, p, \kappa)$$

and where $T_{st}^{\varepsilon,\eta,\natural}$ is, for any fixed $s, t$, the unique linear functional in $E_{-2}$ such that for all $F \in E_2$ we have

$$\langle T_{st}^{\varepsilon,\eta,\natural}, F \rangle = \int_s^t du \int_s^u dv \langle T_v^{\varepsilon,\eta}, A^{\kappa-\eta,*} A^{\kappa-\eta,*} F \rangle \tag{3.4}$$

$$+ \int_s^t du \int_s^u dv \langle T_v^{\varepsilon,\eta}, (Q_v^{\varepsilon,\eta,*} A^{\kappa-\eta,*} + A^{\kappa-\eta,*} Q_u^{\varepsilon,\eta,*}) F \rangle \tag{3.5}$$

$$+ \int_s^t du \int_s^u dv \int_s^v dw \langle T_w^\varepsilon, (A^{\kappa-\eta,*} + Q_w^{\varepsilon,\eta,*}) Q_v^{\varepsilon,\eta,*} Q_u^{\varepsilon,\eta,*} F \rangle \tag{3.6}$$

In order to see that it the right hand side indeed defines a linear functional on $E_2$, we need to bound it. This will be done in Lemma 3.11.

We now compute $\mathbb{X}^{1,\varepsilon,*}_{st}$ and $\mathbb{X}^{2,\varepsilon,\eta,*}_{st}$ explicitly. Consider

$$\mathbb{X}^{1,\varepsilon,*}_{st} F(\xi, p, \kappa) = \int_s^t du\, Q^{\varepsilon,\eta,*}_u F(\xi, p, \kappa)$$

$$= i\varepsilon^{-1/2} \int_s^t du \sum_{n \in \mathbb{Z}^d} e^{4\pi^2 i\varepsilon^{-1} n \cdot (2\kappa - 2\eta - n - \xi)u} \hat{V}(n) F\left(\xi + n, p, \kappa - \frac{n}{2}\right)$$

$$- i\varepsilon^{-1/2} \int_s^t du \sum_{n \in \mathbb{Z}^d} e^{4\pi^2 i\varepsilon^{-1} n \cdot (2\kappa - 2\eta + n + \xi)u} \hat{V}(n) F\left(\xi + n, p, \kappa + \frac{n}{2}\right)$$

$$= i\varepsilon^{-1/2} \sum_{n \in \mathbb{Z}^d \setminus \{0\}} \hat{V}(n) \int_s^t du \left[ e^{ia'_1 u} F\left(\xi + n, p, \kappa - \frac{n}{2}\right) - e^{ia'_2 u} F\left(\xi + n, p, \kappa + \frac{n}{2}\right) \right]$$

where

$$a'_1 = 4\pi^2 \varepsilon^{-1} n \cdot (2\kappa - 2\eta - n - \xi) \tag{3.7}$$

$$a'_2 = 4\pi^2 \varepsilon^{-1} n \cdot (2\kappa - 2\eta + n + \xi) \tag{3.8}$$

Next, consider

$$\mathbb{X}^{2,\varepsilon,\eta,*}_{st} F(\xi, p, \kappa) = \int_s^t du \int_s^u dv\, Q^{\varepsilon,\eta,*}_v Q^{\varepsilon,\eta,*}_u F(\xi, p, \kappa)$$

$$= \int_s^t du \int_s^u dv\, i\varepsilon^{-1/2} \sum_{n \in \mathbb{Z}^d} e^{ia'_1 v} \hat{V}(n) Q^{\varepsilon,\eta,*}_u F\left(\xi + n, p, \kappa - \frac{n}{2}\right)$$

$$- i\varepsilon^{-1/2} \int_s^t du \int_s^u dv \sum_{n \in \mathbb{Z}^d} e^{ia'_2 v} \hat{V}(n) Q^{\varepsilon,\eta,*}_u F\left(\xi + n, p, \kappa + \frac{n}{2}\right)$$

We have that

$$Q^{\varepsilon,\eta,*}_u F\left(\xi + n, p, \kappa - \frac{n}{2}\right) =$$

$$i\varepsilon^{-1/2} \sum_{n' \in \mathbb{Z}^d} e^{4\pi^2 i\varepsilon^{-1} n' \cdot (2\kappa - 2n - 2\eta - n' - \xi)u} \hat{V}(n') F\left(\xi + n + n', p, \kappa - \frac{n}{2} - \frac{n'}{2}\right)$$

$$- i\varepsilon^{-1/2} \sum_{n' \in \mathbb{Z}^d} e^{4\pi^2 i\varepsilon^{-1} n' \cdot (2\kappa - 2\eta + n' + \xi)u} \hat{V}(n') F\left(\xi + n + n', p, \kappa - \frac{n}{2} + \frac{n'}{2}\right)$$

and

$$Q^{\varepsilon,\eta,*}_u F\left(\xi + n, p, \kappa + \frac{n}{2}\right) =$$

$$i\varepsilon^{-1/2} \sum_{n' \in \mathbb{Z}^d} e^{4\pi^2 i\varepsilon^{-1} n' \cdot (2\kappa - 2\eta - n' - \xi)u} \hat{V}(n') F\left(\xi + n + n', p, \kappa + \frac{n}{2} - \frac{n'}{2}\right)$$

$$- i\varepsilon^{-1/2} \sum_{n' \in \mathbb{Z}^d} e^{4\pi^2 i\varepsilon^{-1} n' \cdot (2\kappa + 2n - 2\eta + n' + \xi)u} \hat{V}(n') F\left(\xi + n + n', p, \kappa + \frac{n}{2} + \frac{n'}{2}\right)$$

Hence
$$\mathbb{X}_{st}^{2,\varepsilon,\eta,*}F(\xi,p,\kappa) = -\varepsilon^{-1} \sum_{n,n' \in \mathbb{Z}^d \setminus \{0\}} \hat{V}(n)\hat{V}(n') \int_s^t du \int_s^u dv$$

$$\left[ e^{ia_1'v} e^{ib_1'u} F\left(\xi + n + n', p, \kappa - \frac{n}{2} - \frac{n'}{2}\right) - e^{ia_1'v} e^{ib_2'u} F\left(\xi + n + n', p, \kappa - \frac{n}{2} + \frac{n'}{2}\right) \right]$$

$$+ \varepsilon^{-1} \sum_{n,n' \in \mathbb{Z}^d \setminus \{0\}} \hat{V}(n)\hat{V}(n') \int_s^t du \int_s^u dv$$

$$\left[ e^{ia_2'v} e^{ib_3'u} F\left(\xi + n + n', p, \kappa + \frac{n}{2} - \frac{n'}{2}\right) - e^{ia_2'v} e^{ib_4'u} F\left(\xi + n + n', p, \kappa + \frac{n}{2} + \frac{n'}{2}\right) \right]$$

where
$$b_1' = 4\pi^2 \varepsilon^{-1} n' \cdot (2\kappa - 2n - 2\eta - n' - \xi) \tag{3.9}$$
$$b_2' = 4\pi^2 \varepsilon^{-1} n' \cdot (2\kappa - 2\eta + n' + \xi) \tag{3.10}$$
$$b_3' = 4\pi^2 \varepsilon^{-1} n' \cdot (2\kappa - 2\eta - n' - \xi) \tag{3.11}$$
$$b_4' = 4\pi^2 \varepsilon^{-1} n' \cdot (2\kappa + 2n - 2\eta + n' + \xi) \tag{3.12}$$

Note that
$$a_1' + b_1' = 0 \Leftrightarrow n' \cdot (2\kappa - 2n - 2\eta - n' - \xi) + n \cdot (2\kappa - 2\eta - n - \xi) = 0$$
$$\Leftrightarrow (n + n') \cdot (2\kappa - 2\eta - n - \xi) - n' \cdot (n + n') = 0$$
$$\Leftrightarrow (n + n') \cdot (2\kappa - 2\eta - n - n' - \xi) = 0$$

For this to be 0 for almost every $\eta \in \left[-\frac{1}{4}, \frac{1}{4}\right)^d$, we need $n + n' = 0$ and
$$a_1' + b_2' = 0 \Leftrightarrow n \cdot (2\kappa - 2\eta - n - \xi) + n' \cdot (2\kappa - 2\eta + n' + \xi) = 0$$
$$\Leftrightarrow (n + n') \cdot (2\kappa - 2\eta) + (n' - n) \cdot \xi + |n'|^2 - |n|^2$$

For this to be 0 for almost every $\eta \in \left[-\frac{1}{4}, \frac{1}{4}\right)^d$, we need $n + n' = 0$ and $n \perp \xi$. We argue similarly for $(a_2', b_3')$ and $(a_2', b_4')$. We now introduce the notation: $a, b \in \mathbb{R} \setminus \{0\}$,

$$\varphi_{st}(a,b) := \int_s^t du \int_s^u dr e^{iau} e^{ibr} \tag{3.13}$$

We can thus decompose $\mathbb{X}_{st}^{2,\varepsilon,\eta,*} = \mathbb{Y}_{st}^{\varepsilon,\eta,*} + \mathbb{Z}_{st}^{\varepsilon,\eta,*}$ where

$$\mathbb{Y}_{st}^{\varepsilon,\eta,*} F(\xi,p,\kappa) :=$$
$$-\varepsilon^{-1} \sum_{n \in \mathbb{Z}^d \setminus \{0\}} |\hat{V}(n)|^2 \varphi_{st}(-a_1', a_1')[F(\xi,p,\kappa) - F(\xi,p,\kappa - n)\mathbb{I}_{n \perp \xi}]$$
$$+\varepsilon^{-1} \sum_{n \in \mathbb{Z}^d \setminus \{0\}} |\hat{V}(n)|^2 \varphi_{st}(-a_2', a_2')[F(\xi,p,\kappa + n)\mathbb{I}_{n \perp \xi} - F(\xi,p,\kappa)]$$

and
$$\mathbb{Z}_{st}^{\varepsilon,\eta,*}F(\xi,p,\kappa):=$$

$$-\varepsilon^{-1}\sum_{n,n'\in\mathbb{Z}^d\setminus\{0\}}\hat{V}(n)\hat{V}(n')\varphi_{st}(b_1',a_1')F\left(\xi+n+n',p,\kappa-\frac{n}{2}-\frac{n'}{2}\right)\mathbb{I}_{a_1'\neq b_1'}$$

$$+\varepsilon^{-1}\sum_{n,n'\in\mathbb{Z}^d\setminus\{0\}}\hat{V}(n)\hat{V}(n')\varphi_{st}(b_2',a_1')F\left(\xi+n+n',p,\kappa-\frac{n}{2}+\frac{n'}{2}\right)\mathbb{I}_{a_1'\neq b_2'}$$

$$+\varepsilon^{-1}\sum_{n,n'\in\mathbb{Z}^d\setminus\{0\}}\hat{V}(n)\hat{V}(n')\varphi_{st}(b_3',a_2')F\left(\xi+n+n',p,\kappa+\frac{n}{2}-\frac{n'}{2}\right)\mathbb{I}_{a_2'\neq b_3'}$$

$$-\varepsilon^{-1}\sum_{n,n'\in\mathbb{Z}^d\setminus\{0\}}\hat{V}(n)\hat{V}(n')\varphi_{st}(b_4',a_2')F\left(\xi+n+n',p,\kappa+\frac{n}{2}+\frac{n'}{2}\right)\mathbb{I}_{a_2'\neq b_4'}$$

## 3.2 Some useful lemmas

We will collect here some lemmas that will be useful in estimating terms that show up when attempting to bound the rough drivers on a scale. The first lemma deals with integrals involving two exponentials. We have

LEMMA 3.1. *Let* $\varphi_{st}(a,b)=\int_s^t du\int_s^u dr e^{iau}e^{ibr}$, $a,b\in\mathbb{R}\setminus\{0\}$. *Then we have for* $\gamma\in\left(0,\frac{1}{2}\right)$, *and* $a+b\neq 0$

$$|\varphi_{st}(a,b)|\lesssim\frac{|t-s|^{2\gamma}}{|b|}\left(\frac{1}{|a|^{1-2\gamma}}+\frac{1}{|a+b|^{1-2\gamma}}\right)$$

*and for a bound that is symmetric in a and b one has for* $a+b\neq 0$

$$|\varphi_{st}(a,b)|\lesssim|t-s|^{2\gamma}\left(\frac{1}{|a|^{1-\gamma}|b|^{1-\gamma}}+\frac{1}{|b|^{1/2}|a|^{1-\gamma}|a+b|^{\frac{1-2\gamma}{2}}}\right)$$

$$+(t-s)^{2\gamma}\left(\frac{1}{|a|^{1/2}|b|^{1-\gamma}|a+b|^{\frac{1-2\gamma}{2}}}+\frac{1}{|b|^{1/2}|a|^{1/2}|a+b|^{1-2\gamma}}\right)$$

*and, when* $a+b=0$, *one has*

$$|\varphi_{st}(a,b)|\lesssim\frac{|t-s|}{|a|}$$

PROOF. First notice that for $k\in\mathbb{R}\setminus\{0\}$ we have that

$$\left|\int_s^t e^{iku}du\right|\leqslant|t-s|,\quad\left|\int_s^t e^{iku}du\right|=\left|\frac{e^{ikt}-e^{iks}}{ik}\right|\lesssim\frac{1}{|k|}$$

which by interpolation gives

$$\left|\int_s^t e^{iku}du\right|\lesssim\frac{|t-s|^{2\gamma}}{|k|^{1-2\gamma}}$$

Now, in the first case we have that

$$\varphi_{st}(a,b) = \int_s^t du \int_s^u dr\, e^{iau} e^{ibr} = \int_s^t du \left( e^{iau} \frac{e^{ibu} - e^{ibs}}{ib} \right)$$

$$= \frac{1}{ib} \int_s^t du\, e^{i(a+b)u} - \frac{e^{ibs}}{ib} \int_s^t du\, e^{iau}$$

and applying the triangle inequality and the above interpolated estimate to each term, we get

$$|\varphi_{st}(a,b)| \leq \frac{1}{|b|} \left( \left| \int_s^t e^{i(a+b)u} du \right| + \left| \int_s^t e^{iau} du \right| \right)$$

$$\lesssim \frac{|t-s|^{2\gamma}}{|b|} \left( \frac{1}{|a|^{1-2\gamma}} + \frac{1}{|a+b|^{1-2\gamma}} \right)$$

Now, noticing that one can write

$$\varphi_{st}(a,b) = \int_s^t du \int_s^u dr\, e^{iau} e^{ibr} = \int_s^t dr \int_r^t du\, e^{iau} e^{ibr} = \frac{e^{iat}}{ia} \int_s^t dr\, e^{ibr} - \frac{1}{ia} \int_s^t dr\, e^{i(a+b)r}$$

Using the triangle inequality and the interpolated estimate as before, one has

$$|\varphi_{st}(a,b)| \lesssim \frac{|t-s|^{2\gamma}}{|a|} \left( \frac{1}{|b|^{1-2\gamma}} + \frac{1}{|a+b|^{1-2\gamma}} \right)$$

Interpolating between the two bounds obtained, one has

$$|\varphi_{st}(a,b)| \lesssim |t-s|^{2\gamma} \left( \frac{1}{|a|^{1-\gamma}|b|^{1-\gamma}} + \frac{1}{|b|^{1/2}|a|^{1-\gamma}|a+b|^{\frac{1-2\gamma}{2}}} \right)$$

$$+ (t-s)^{2\gamma} \left( \frac{1}{|a|^{1/2}|b|^{1-\gamma}|a+b|^{\frac{1-2\gamma}{2}}} + \frac{1}{|b|^{1/2}|a|^{1/2}|a+b|^{1-2\gamma}} \right)$$

In the second case, we only have that

$$\varphi_{st}(a,b) = \int_s^t du \int_s^u dr\, e^{ia(u-r)} = \int_s^t du\, e^{iau} \int_s^u dr\, e^{-iar} = -\int_s^t \frac{1 - e^{ia(u-s)}}{ia} du$$

from which we obtain the claim. □

Next we have a lemma that allows us to deal with small divisors. Within our problem, after applying the lemma above, we will encounter terms where $a$ and $b$ are of the form $cn \cdot (\kappa - \eta)$ for integer $n$ and $\kappa$ and $\eta \in \left[ -\frac{1}{2}, \frac{1}{2} \right]^d$. So in order to bound $\frac{1}{|a|}$, one will need to bound $|a|$ away from 0.

LEMMA 3.2. *Let $\delta_0 = \frac{1}{d+3}$. For any $\delta \in (0, \delta_0)$ there exists $c_\delta(\eta) \in L^1\left( \left[ -\frac{1}{2}, \frac{1}{2} \right]^d \right)$ such that almost surely*

$$|n \cdot \eta - a|^{-1} \leq c_\delta(\eta)^{\frac{1}{1-\delta}} \langle n \rangle^{d+2} < \infty, \qquad \forall a \in \mathbb{Z}, n \in \mathbb{Z}^d \setminus \{0\}$$

PROOF. Fix $\delta \in (0, \delta_0)$ and define

$$c'_\delta(\eta) = \sum_{n \in \mathbb{Z}^d \setminus \{0\}} \sum_{a: |a| \leq 2\sqrt{d}|n|} \langle n \rangle^{-d-1-\delta} |n \cdot \eta - a|^{\delta-1}$$

We know that
$$\int_{[-\frac{1}{2},\frac{1}{2}]^d} |n\cdot\eta - a|^{\delta-1}\mathrm{d}\eta \leq C < \infty$$

uniformly in $n$ and $a$: $|a| \leq 2\sqrt{d}|n|$, since
$$\int_{[-\frac{1}{2},\frac{1}{2}]^d} |n\cdot\eta - a|^{\delta-1}\mathrm{d}\eta \leq \int_{B_{\sqrt{d}}(0)} \mathrm{d}\eta |n\cdot\eta - a|^{\delta-1}\mathrm{d}\eta$$

For any rotation matrix $R$, one can define $\eta' := R^{-1}\eta$ to get that this is
$$= \int_{B_{\sqrt{d}}(0)} |n\cdot R\eta' - a|^{\delta-1}\mathrm{d}\eta' = \int_{B_{\sqrt{d}}(0)} |R^T n\cdot\eta' - a|^{\delta-1}\mathrm{d}\eta'$$

Now, one can choose $R \equiv R(n)$ such that $R^T n = (|n|, 0, \ldots, 0)$. This then becomes
$$= \int_{B_{\sqrt{d}}(0)} ||n|\eta'_1 - a|^{\delta-1}\mathrm{d}\eta' = (2\sqrt{d})^{d-1}\int_{-\sqrt{d}}^{\sqrt{d}} \mathrm{d}\eta' ||n|\eta' - a|^{\delta-1}$$

by a change of variables this is
$$= \frac{(2\sqrt{d})^{d-1}}{|n|}\int_{-|n|\sqrt{d}-a}^{|n|\sqrt{d}-a} \mathrm{d}\eta' |\eta'|^{\delta-1}$$

Now, since $|a| \leq 2\sqrt{d}|n|$ in the definition of $c_\delta$, one can estimate
$$\max\{||n|\sqrt{d} - a|, ||n|\sqrt{d} + a|\} \leq 3|n|\sqrt{d}$$

And so the above integral gets estimated as
$$\leq \frac{2(2\sqrt{d})^{d-1}}{|n|}\int_0^{3|n|\sqrt{d}} \mathrm{d}\eta' |\eta'_1|^{\delta-1} \leq \frac{2(2\sqrt{d})^{d-1}}{|n|}\left[\int_0^1 \eta^{\delta-1}\mathrm{d}\eta + \int_1^{3|n|\sqrt{d}} \mathrm{d}\eta\right]$$
$$\leq \frac{C_1(\delta)}{|n|} + C_2(d) \leq C_{\delta,d}$$

for $n \in \mathbb{Z}^d \setminus \{0\}$. Hence we have that
$$\int_{[-\frac{1}{2},\frac{1}{2}]^d} c'_\delta(\eta)\mathrm{d}\eta = \int_{[-\frac{1}{2},\frac{1}{2}]^d} \sum_{n\in\mathbb{Z}^d\setminus\{0\}} \sum_{a\in\mathbb{Z}:|a|\leq 2\sqrt{d}|n|} \langle n\rangle^{-d-1-\delta}|n\cdot\eta - a|^{\delta-1}\mathrm{d}\eta$$

by the dominated convergence and Tonelli theorems, this is
$$= \sum_{n\in\mathbb{Z}^d\setminus\{0\}} \sum_{a\in\mathbb{Z}:|a|\leq 2\sqrt{d}|n|} \langle n\rangle^{-d-1-\delta}\int_{[-\frac{1}{2},\frac{1}{2}]^d} |n\cdot\eta - a|^{\delta-1}\mathrm{d}\eta$$
$$\leq C_{d,\delta} \sum_{n\in\mathbb{Z}^d\setminus\{0\}} \langle n\rangle^{-d-1-\delta} \sum_{a\in\mathbb{Z}:|a|\leq 2\sqrt{d}|n|} 1$$
$$\lesssim C \sum_{n\in\mathbb{Z}^d\setminus\{0\}} \langle n\rangle^{-d-1-\delta}\langle n\rangle \leq C \sum_{n\in\mathbb{Z}^d\setminus\{0\}} \langle n\rangle^{-d-\delta} < \infty$$

Hence we have that $c_\delta(\eta)$ is finite a.e., and hence, for almost every $\eta$, by the fact that $c_\delta(\eta)$ consists of positive terms one has

$$\langle n \rangle^{-d-1-\delta} |n \cdot \eta - a|^{\delta-1} \leq c'_\delta(\eta)$$

$$\Rightarrow \langle n \rangle^{d+1+\delta} |n \cdot \eta - a|^{1-\delta} \geq c'_\delta(\eta)^{-1}$$

$$\Rightarrow |n \cdot \eta - a| \geq c'_\delta(\eta)^{-1/1-\delta} \langle n \rangle^{-(d+1+\delta)/1-\delta}$$

$$\Rightarrow |n \cdot \eta - a|^{-1} \leq c'_\delta(\eta)^{1/1-\delta} \langle n \rangle^{(d+1+\delta)/1-\delta}$$

which for $0 < \delta < \frac{1}{d+3}$

$$|n \cdot \eta - a|^{-1} \leq c'_\delta(\eta)^{\frac{1}{1-\delta}} \langle n \rangle^{d+2}$$

In the case that $|a| > 2\sqrt{d}|n|$ we have by the reverse triangle inequality that

$$|a - n \cdot \eta| \geq |a| - |n \cdot \eta| \geq |a| - \sqrt{d}|n| \geq \sqrt{d}|n| \geq 1$$

$$\Rightarrow |n \cdot \eta - a|^{-1} \leq 1$$

Defining $c_\delta(\eta) = \max\{c'_\delta(\eta), 1\}$ we have the claim. □

REMARK 3.3. Notice that $\int_{\left[-\frac{1}{2}, \frac{1}{2}\right]^d} c_\delta(\eta) \mathrm{d}\eta \xrightarrow{\delta \to 0} \infty$. On the right hand side of the expression in the statement of the lemma, one has $c_\delta(\eta)^{1/1-\delta}$, which is not in $L^1_\eta\left(\left[-\frac{1}{2}, \frac{1}{2}\right]^d\right)$. Clearly, there can never be an $L^1$ function on the right hand side, because one knows that $\eta^{-1}$ is not integrable between 0 and 1.

The above lemma says that a.e $\eta \in \left[-\frac{1}{4}, \frac{1}{4}\right]^d$ belongs to $\mathscr{A}_\eta$. (Recall Definition 2.19). We next have the following estimates on the double exponentials that show up in the operators of the rough difference equation:

LEMMA 3.4. *We have that* $\forall n \in \mathbb{Z}^d \setminus \{0\}, l \in \mathbb{Z}^d, 0 \leq s < t, \eta \in \mathscr{A}_\eta$ *and any* $\gamma \in (0, 1)$

$$\left| \varepsilon^{-1/2} \int_s^t \mathrm{d}u\, e^{4\pi^2 i \varepsilon^{-1} n \cdot (l - 2\eta) u} \right| \lesssim c(\eta)^{\frac{1-\gamma}{1-\delta}} \varepsilon^{1/2-\gamma} |t-s|^\gamma \langle n \rangle^{(d+2)(1-\gamma)}$$

PROOF. We have by the same proof techniques in Lemma 3.1 that

$$\left| \int_s^t \mathrm{d}u\, e^{4\pi^2 i \varepsilon^{-1} n \cdot (l - 2\eta) u} \right| \lesssim \frac{|t-s|^\gamma}{\varepsilon^{-1+\gamma} |n \cdot (l - 2\eta)|^{1-\gamma}}$$

and by Lemma 3.2 and the fact that $\eta \in \mathscr{A}_\eta$, this is

$$\leq \varepsilon^{1-\gamma} |t-s|^\gamma c(\eta)^{\frac{1-\gamma}{1-\delta}} \langle n \rangle^{(d+2)(1-\gamma)}$$

Hence

$$\left| \varepsilon^{-1/2} \int_s^t \mathrm{d}u\, e^{4\pi^2 i \varepsilon^{-1} n \cdot (l - 2\eta) u} \right| \lesssim c(\eta)^{\frac{1-\gamma}{1-\delta}} \varepsilon^{1/2-\gamma} |t-s|^\gamma \langle n \rangle^{(d+2)(1-\gamma)}$$

□

Note that when $\gamma > \delta$ one can integrate the expression on the right hand side over $\eta$ since $c(\eta) \in L^1$ and $\frac{1-\gamma}{1-\delta} < 1$.

LEMMA 3.5. *We have that* $\forall n \in \mathbb{Z}^d \setminus \{0\}, l \in \mathbb{Z}^d, 0 \leq s < u < t, \eta \in \mathcal{A}_\eta$

$$\left| \varepsilon^{-1} \int_s^t du \int_s^u dv\, e^{4\pi^2 i \varepsilon^{-1} n \cdot (l - 2\eta)(u-v)} \right| \lesssim (t-s) c(\eta)^{\frac{1}{1-\delta}} \langle n \rangle^{d+2}$$

PROOF. As in case 2 of Lemma 3.1 we set $a = 4\pi^2 \varepsilon^{-1} n \cdot (l - 2\eta)$ and see that

$$\left| \varepsilon^{-1} \int_s^t \int_s^u e^{ia(u-r)} dr\, du \right| \lesssim \frac{\varepsilon^{-1}(t-s)}{|a|}$$

and by plugging in the value of $a$ and using lemma 3.2 we get that

$$\left| \varepsilon^{-1} \int_s^t \int_s^u e^{ia(u-r)} dr\, du \right| \lesssim (t-s) c(\eta)^{\frac{1}{1-\delta}} \langle n \rangle^{d+2}$$

□

LEMMA 3.6. *We have that* $\forall n, n' \in \mathbb{Z}^d \setminus \{0\}, l, l' \in \mathbb{Z}^d, 0 \leq s < u < t, \eta \in \mathcal{A}_\eta, \gamma \in (0,1)$ *and in the case* $n \cdot (l - 2\eta) + n' \cdot (l' - 2\eta) \neq 0$,

$$\left| \varepsilon^{-1} \int_s^t \int_s^u e^{4\pi^2 i \varepsilon^{-1} n \cdot (l-2\eta) u} e^{4\pi^2 i \varepsilon^{-1} n' \cdot (l'-2\eta) v} dv\, du \right| \lesssim \varepsilon^{1-2\gamma} |t-s|^{2\gamma} c(\eta)^{\frac{2-2\gamma}{1-\delta}} (\langle n \rangle \langle n' \rangle)^{2(d+2)}$$

PROOF. Considering the $a + b \neq 0$ case of Lemma 3.1 we set $a = 4\pi^2 \varepsilon^{-1} n \cdot (l - 2\eta)$ and $b = 4\pi^2 \varepsilon^{-1} n' \cdot (l' - 2\eta)$ and see that when $a + b \neq 0$,

$$\left| \varepsilon^{-1} \int_s^t \int_s^u e^{iau} e^{ibv} dv\, du \right| \lesssim \varepsilon^{-1} |t-s|^{2\gamma} \left( \frac{1}{|a|^{1-\gamma} |b|^{1-\gamma}} + \frac{1}{|b|^{1/2} |a|^{1-\gamma} |a+b|^{\frac{1-2\gamma}{2}}} \right)$$

$$+ \varepsilon^{-1} |t-s|^{2\gamma} \left( \frac{1}{|a|^{1/2} |b|^{1-\gamma} |a+b|^{\frac{1-2\gamma}{2}}} + \frac{1}{|b|^{1/2} |a|^{1/2} |a+b|^{1-2\gamma}} \right)$$

and by Lemma 3.2 this is

$$\lesssim \varepsilon^{1-2\gamma} |t-s|^{2\gamma} c(\eta)^{\frac{2-2\gamma}{1-\delta}} \langle n \rangle^{(d+2)(1-\gamma)} \langle n' \rangle^{(d+2)(1-\gamma)}$$

$$+ \varepsilon^{1-2\gamma} |t-s|^{2\gamma} c(\eta)^{\frac{2-2\gamma}{1-\delta}} \langle n' \rangle^{(d+2)/2} \langle n \rangle^{(d+2)(1-\gamma)} \langle n + n' \rangle^{(d+2)(1-2\gamma)/2}$$

$$+ \varepsilon^{1-2\gamma} |t-s|^{2\gamma} c(\eta)^{\frac{2-2\gamma}{1-\delta}} \langle n \rangle^{(d+2)/2} \langle n' \rangle^{(d+2)(1-\gamma)} \langle n + n' \rangle^{(d+2)(1-2\gamma)/2}$$

$$+ \varepsilon^{1-2\gamma} |t-s|^{2\gamma} c(\eta)^{\frac{2-2\gamma}{1-\delta}} \langle n \rangle^{(d+2)/2} \langle n' \rangle^{(d+2)/2} \langle n + n' \rangle^{(d+2)(1-2\gamma)}$$

combining these, one has that this is

$$\lesssim \varepsilon^{1-2\gamma}|t-s|^{2\gamma}c(\eta)^{\frac{2-2\gamma}{1-\delta}}\langle n\rangle^{d+2}\langle n'\rangle^{d+2}\langle n+n'\rangle^{d+2}$$

$$\lesssim \varepsilon^{1-2\gamma}|t-s|^{2\gamma}c(\eta)^{\frac{2-2\gamma}{1-\delta}}\langle n\rangle^{2(d+2)}\langle n'\rangle^{2(d+2)}$$

□

## 3.3 Uniform operator norm and naive remainder estimates

This Subsection will be devoted to proving the following: For any $m \in \mathbb{N}_0, \eta \in \mathcal{A}_\eta, \gamma \in \left(\frac{1}{3}, \frac{1}{2}\right), \kappa \in \left(\frac{\mathbb{Z}}{2}\right)^d, \eta \in \left[-\frac{1}{4}, \frac{1}{4}\right]^d, 0 \le s < t, \delta$ as in (2.18), we have for any $\psi \in E_m$,

$$\|\mathbb{X}^{1,\varepsilon,\eta,*}_{st}\psi\|_{E_m} \lesssim \varepsilon^{1/2-\gamma}|t-s|^\gamma\|\psi\|_{E_m}$$

$$\|\mathbb{Y}^{\varepsilon,\eta,*}_{st}\psi\|_{E_m} \lesssim (t-s)\|\psi\|_{E_m}$$

$$\|\mathbb{Z}^{\varepsilon,\eta,*}_{st}\psi\|_{E_m} \lesssim \varepsilon^{1-2\gamma}|t-s|^{2\gamma}\|\psi\|_{E_m}$$

and for $\psi \in E_{m+1}$,

$$\|\mathbb{A}^{\kappa-\eta,*}_{st}\psi\|_{E_m} \lesssim |t-s|\|\psi\|_{E_{m+1}}$$

In the above bounds, the constants implicit in $\lesssim$ could depend on $d, V, \gamma, m, \delta, \eta$ but not on $\varepsilon$.

We will also prove the following naive bound on the remainder

$$\|T^{\varepsilon,\eta,\natural}_{st}\|_{E_{-2}} \lesssim \varepsilon^{-3/2}|t-s|^2$$

The uniform in $\varepsilon$-bounds on the operators $\mathbb{X}^{1,\varepsilon,\eta,*}_{st}, \mathbb{Y}^{\varepsilon,\eta,*}_{st}, \mathbb{Z}^{\varepsilon,\eta,*}_{st}, \mathbb{A}^{\kappa-\eta,*}_{st}$ combined with the sewing lemma will improve the bound on the remainder to also be uniform in $\varepsilon$. This will be the content of the next section. This will then allow us to use a compactness argument to pass to the limit $\varepsilon \to 0$.

LEMMA 3.7. *For any $m \in \mathbb{N}_0, \eta \in \mathcal{A}_\eta, \psi \in E_m, 0 \le s < t$, we have*

$$\|\mathbb{X}^{1,\varepsilon,\eta,*}_{st}\psi\|_{E_m} \lesssim c(\eta)^{\frac{1-\gamma}{1-\delta}}\varepsilon^{1/2-\gamma}|t-s|^\gamma\|\psi\|_{E_m} \qquad (3.14)$$

PROOF. Recall that

$$\mathbb{X}^{1,\varepsilon,\eta,*}_{st,\eta}\psi(\xi,p,\kappa) =$$

$$i\varepsilon^{-1/2}\sum_{n\in\mathbb{Z}^d\setminus\{0\}}\hat{V}(n)\int_s^t du\left[e^{ia'_1 u}F\left(\xi+n,p,\kappa-\frac{n}{2}\right) - e^{ia'_2 u}F\left(\xi+n,p,\kappa+\frac{n}{2}\right)\right]$$

Hence, if we want to see where $\mathbb{X}^{1,\varepsilon,\eta,*}_{st}$ maps $\varphi \in E_m$ we first consider for $\beta: |\beta| \le m$

$$\|D^\beta_p \mathbb{X}^{1,\varepsilon,\eta,*}_{st}\psi(\xi,p,\kappa)\|_{l^2_\xi} = \left(\sum_{\xi\in\mathbb{Z}^d}\left|D^\beta_p \mathbb{X}^{1,\varepsilon,\eta,*}_{st}\psi(\xi,p,\kappa)\right|^2\right)^{1/2}$$

To this end, we use Young's inequality to get

$$|a-b|^2 = |a|^2 + |b|^2 - 2a\cdot b \leqslant |a|^2 + |b|^2 + 2\left(\frac{|a|^2}{2} + \frac{|b|^2}{2}\right) = 2|a|^2 + 2|b|^2$$

this allows us to estimate

$$\sum_{\xi \in \mathbb{Z}^d} \left|D_p^\beta \mathbb{X}_{st}^{1,\varepsilon,\eta,*}\psi(\xi,p,\kappa)\right|^2$$

$$\leq 2\sum_{\xi \in \mathbb{Z}^d} \left|\sum_{n \in \mathbb{Z}^d\setminus\{0\}} \hat{V}(n)\left(\varepsilon^{-1/2}\int_s^t e^{ia_1'u}du\right)D_p^\beta\psi\left(\xi+n,p,\kappa-\frac{n}{2}\right)\right|^2$$

$$+ 2\sum_{\xi \in \mathbb{Z}^d} \left|\sum_{n \in \mathbb{Z}^d\setminus\{0\}} \hat{V}(n)\left(\varepsilon^{-1/2}\int_s^t e^{ia_2'u}du\right)D_p^\beta\psi\left(\xi+n,p,\kappa+\frac{n}{2}\right)\right|^2$$

$$= A + B$$

and we shall estimate just the first term, and the second term can be estimated the same way. By Cauchy–Schwarz

$$A \lesssim \sum_{\xi \in \mathbb{Z}^d} \left(\sum_{n \in \mathbb{Z}^d\setminus\{0\}} |\hat{V}(n)|^2 \langle n\rangle^{2M}\right)$$

$$\left(\sum_{n \in \mathbb{Z}^d\setminus\{0\}} \langle n\rangle^{-2M}\left|\varepsilon^{-1/2}\int_s^t e^{ia_1'u}du\right|^2 \left|D_p^\beta\psi\left(\xi+n,p,\kappa-\frac{n}{2}\right)\right|^2\right)$$

which by Lemma 3.4 and the expression for $a_1'$ is, for $\eta \in \mathscr{A}_\eta$

$$\lesssim c(\eta)^{\frac{2(1-\gamma)}{1-\delta}} \varepsilon^{1-2\gamma}|t-s|^{2\gamma}$$

$$\sum_{\xi \in \mathbb{Z}^d}\sum_{n \in \mathbb{Z}^d\setminus\{0\}} \langle n\rangle^{-2M}\langle n\rangle^{(d+2)(1-\gamma)}\left|D_p^\beta\psi(p,\xi+n,k-n)\right|^2$$

By using Tonelli to change the order of summation, this is

$$\lesssim c(\eta)^{\frac{2(1-\gamma)}{1-\delta}} \varepsilon^{1-2\gamma}|t-s|^{2\gamma} \sum_{n \in \mathbb{Z}^d\setminus\{0\}} \langle n\rangle^{-2M}\langle n\rangle^{(d+2)}\left\|D_p^\beta\psi\left(\cdot,p,\kappa-\frac{n}{2}\right)\right\|_{l_\xi^2}^2$$

So, we have that

$$\sum_{\kappa \in \left(\frac{\mathbb{Z}}{2}\right)^d} \langle\kappa\rangle^m \|D_p^\beta \mathbb{X}_{st}^{1,\varepsilon,\eta,*}\psi(p,\xi,\kappa)\|_{l_\xi^2} \lesssim \sum_{\kappa \in \left(\frac{\mathbb{Z}}{2}\right)^d} \langle\kappa\rangle^m A^{1/2} + \sum_{\kappa \in \left(\frac{\mathbb{Z}}{2}\right)^d} \langle\kappa\rangle^m B^{1/2}$$

We estimate

$$\sum_{\kappa \in \left(\frac{\mathbb{Z}}{2}\right)^d} \langle\kappa\rangle^m A^{1/2} \lesssim \|V\|_{H^M(\mathbb{T}^d)} c(\eta)^{\frac{1-\gamma}{1-\delta}} \varepsilon^{1/2-\gamma}|t-s|^\gamma$$

$$\sum_{\kappa \in \left(\frac{\mathbb{Z}}{2}\right)^d} \langle\kappa\rangle^m \left(\sum_{n \in \mathbb{Z}^d\setminus\{0\}} \langle n\rangle^{-2M}\langle n\rangle^{(d+2)}\left\|D_p^\beta\psi\left(\cdot,p,\kappa-\frac{n}{2}\right)\right\|_{l_\xi^2}^2\right)^{1/2}$$

Since the $l^2$-norm of a sequence is upper bounded by its $l^1$-norm, we get that this is

$$\lesssim c(\eta)^{\frac{1-\gamma}{1-\delta}} \varepsilon^{1/2-\gamma} |t-s|^\gamma$$

$$\sum_{\kappa \in \left(\frac{\mathbb{Z}}{2}\right)^d} \sum_{n \in \mathbb{Z}^d \setminus \{0\}} \langle \kappa \rangle^m \langle n \rangle^{-M} \langle n \rangle^{(d+2)/2} \left\| D_p^\beta \psi\left(\xi, p, \kappa - \frac{n}{2}\right) \right\|_{l_\xi^2}$$

By using Tonelli twice more, we have for $M$ large enough

$$\lesssim c(\eta)^{\frac{1-\gamma}{1-\delta}} \varepsilon^{1/2-\gamma} |t-s|^\gamma \sum_{\kappa \in \left(\frac{\mathbb{Z}}{2}\right)^d} \|D_p^\beta \psi(\cdot, p, \kappa)\|_{l_\xi^2}$$

$$\sum_{n \in \mathbb{Z}^d \setminus \{0\}} \left\langle \kappa + \frac{n}{2} \right\rangle^m \langle n \rangle^{-M} \langle n \rangle^{(d+2)/2}$$

Let $M' = M - \frac{d+2}{2}$. Then

$$\sum_{n \in \mathbb{Z}^d \setminus \{0\}} \left\langle \kappa + \frac{n}{2} \right\rangle^m \langle n \rangle^{-M} \langle n \rangle^{(d+2)/2} = \sum_{n \in \mathbb{Z}^d \setminus \{0\}} \left\langle \kappa + \frac{n}{2} \right\rangle^m \langle n \rangle^{-M'}$$

and for $M'$ large enough, this is

$$\leqslant C_m \sum_{n \in \mathbb{Z}^d \setminus \{0\}} (\langle \kappa \rangle^m + \langle n \rangle^m) \langle n \rangle^{-M'} \leqslant C_{m,M} \langle \kappa \rangle^m$$

we have

$$\sum_{\kappa \in \left(\frac{\mathbb{Z}}{2}\right)^d} \langle \kappa \rangle^m A^{1/2} \lesssim \|V\|_{H^M(\mathbb{T}^d)} c(\eta)^{\frac{1-\gamma}{1-\delta}} \varepsilon^{1/2-\gamma} |t-s|^\gamma \sum_{\kappa \in \left(\frac{\mathbb{Z}}{2}\right)^d} \|D_p^\beta \psi(\cdot, p, \kappa)\|_{l_\xi^2} \langle \kappa \rangle^m$$

and similarly for $\sum_{\kappa \in \left(\frac{\mathbb{Z}}{2}\right)^d} \langle \kappa \rangle^m B^{1/2}$. Hence

$$\sum_{\kappa \in \left(\frac{\mathbb{Z}}{2}\right)^d} \langle \kappa \rangle^m \|D_p^\beta \mathbb{X}_{st}^{1,\varepsilon,\eta,*} \psi(\xi, p, \kappa)\|_{l_\xi^2}$$

$$\lesssim c(\eta)^{\frac{1-\gamma}{1-\delta}} \varepsilon^{1/2-\gamma} |t-s|^\gamma \sum_{\kappa \in \left(\frac{\mathbb{Z}}{2}\right)^d} \|D_p^\beta \psi(\xi, p, \kappa)\|_{l_\xi^2} \langle \kappa \rangle^m$$

and so integrating in $p$ and summing over $\beta \colon |\beta| \leqslant m$, we have

$$\|\mathbb{X}_{st}^{1,\varepsilon,*} \psi\|_{E_m} \lesssim c(\eta)^{\frac{1-\gamma}{1-\delta}} \varepsilon^{1/2-\gamma} |t-s|^\gamma \|\psi\|_{E_m} \qquad \square$$

LEMMA 3.8. *For $\kappa \in \left(\frac{\mathbb{Z}}{2}\right)^d, \eta \in \left[-\frac{1}{4}, \frac{1}{4}\right]^d, m \in \mathbb{N}_0, \psi \in E_{m+1}, 0 \leqslant s < t$ we have*

$$\|\mathbb{A}_{st}^{\kappa-\eta,*} \psi\|_{E_m} \lesssim |t-s| \|\psi\|_{E_{m+1}} \tag{3.15}$$

PROOF. We bound the operator $\mathbb{A}_{st}^{\kappa-\eta,*}\psi(\xi,p,\kappa) = 4\pi(t-s)(\kappa-\eta)\cdot\nabla_p\psi(\xi,p,\kappa)$ in the scale as follows, by first considering the quantity

$$\sum_\xi \left|D_p^\beta \mathbb{A}_{st}^{\kappa-\eta,*}\psi\right|^2 = 16\pi^2(t-s)^2 \sum_\xi \left|D_p^\beta (\kappa-\eta)\cdot\nabla_p\psi\right|^2$$

$$= 16\pi^2|t-s|^2 \sum_\xi \left|D_p^\beta(\kappa-\eta)\cdot\nabla_p\psi(\xi,p,\kappa)\right|^2$$

$$= 16\pi^2|t-s|^2 \sum_\xi \left|D_p^\beta \sum_{j=1}^d (\kappa-\eta)_j D_p^{e_j}\psi(\xi,p,\kappa)\right|^2$$

$$= 16\pi^2|t-s|^2 \sum_\xi \left|\sum_{j=1}^d (\kappa-\eta)_j D_p^{\beta+e_j}\psi(\xi,p,\kappa)\right|^2$$

$$\lesssim_d |t-s|^2 \langle\kappa\rangle^2 \sum_\xi \sum_{j=1}^d \left|D_p^{\beta+e_j}\psi(\xi,p,\kappa)\right|^2$$

$$= |t-s|^2 \langle\kappa\rangle^2 \sum_{j=1}^d \sum_\xi \left|D_p^{\beta+e_j}\psi(\xi,p,\kappa)\right|^2$$

Hence by subadditivity of the square root

$$4\pi(t-s)\sum_\kappa \langle\kappa\rangle^m \left(\sum_\xi \left|D_p^\beta(\kappa-\eta)\cdot\nabla_p\psi(\xi,p,\kappa)\right|^2\right)^{1/2}$$

$$\lesssim |t-s|\sum_{j=1}^d \sum_\kappa \langle\kappa\rangle^{m+1}\left(\sum_\xi \left|D_p^{\beta+e_j}\psi(\xi,p,\kappa)\right|^2\right)^{1/2}$$

Hence

$$\|\mathbb{A}_{st}^{\kappa-\eta,*}\psi\|_{E_m} = \sum_{|\beta|\le m}\int_{\mathbb{R}^d} dp \sum_\kappa \langle\kappa\rangle^m \left(\sum_\xi \left|D_p^\beta \mathbb{A}_{st}^{\kappa-\eta,*}\psi(\xi,p,\kappa)\right|^2\right)^{1/2}$$

$$\lesssim |t-s|\sum_{|\beta|\le m}\sum_{j=1}^d \int_{\mathbb{R}^d} dp \sum_\kappa \langle\kappa\rangle^{m+1}\left(\sum_\xi \left|D_p^{\beta+e_j}\psi(\xi,p,\kappa)\right|^2\right)^{1/2}$$

$$\lesssim |t-s|\sum_{|\beta|\le m+1}\int_{\mathbb{R}^d} dp \sum_\kappa \langle\kappa\rangle^{m+1}\|D_p^\beta\psi(\cdot,p,\kappa)\|_{l_\xi^2} \lesssim 4\pi|t-s|\|\psi\|_{E_{m+1}}$$

So

$$\|\mathbb{A}_{st}^{\kappa-\eta,*}\psi\|_{E_m} \lesssim |t-s|\|\psi\|_{E_{m+1}}$$

$\square$

Recall that we had

$$\mathbb{X}_{st}^{2,\varepsilon,\eta,*} = \mathbb{Y}_{st}^{\varepsilon,\eta,*} + \mathbb{Z}_{st}^{\varepsilon,\eta,*}$$

We now estimate the two operators, and will see that $\mathbb{Y}_{st}^{\varepsilon,\eta,*}$ will have a contribution in the limit $\varepsilon \to 0$ while $\mathbb{Z}_{st}^{\varepsilon,\eta,*}$ will vanish.

LEMMA 3.9. *For all $m \in \mathbb{N}_0, \eta \in \mathcal{A}_\eta, \psi \in E_m, 0 \leq s < t$ we have*

$$\|\mathbb{Y}_{st}^{\varepsilon,\eta,*}\psi\|_{E_m} \lesssim (t-s)c(\eta)^{\frac{1}{1-\delta}}\|\psi\|_{E_m} \tag{3.16}$$

PROOF. We recall

$$\mathbb{Y}_{st}^{\varepsilon,\eta,*}F(\xi,p,\kappa) :=$$

$$-\varepsilon^{-1} \sum_{n \in \mathbb{Z}^d \setminus \{0\}} |\hat{V}(n)|^2 \int_s^t du \int_s^u dv\, e^{ia_1'(v-u)}[F(\xi,p,\kappa) - F(\xi,p,\kappa-n)\mathbb{I}_{n \perp \xi}]$$

$$+\varepsilon^{-1} \sum_{n \in \mathbb{Z}^d \setminus \{0\}} |\hat{V}(n)|^2 \int_s^t du \int_s^u dv\, e^{ia_2'(v-u)}[F(\xi,p,\kappa+n)\mathbb{I}_{n \perp \xi} - F(\xi,p,\kappa)]$$

We would like to estimate how $\mathbb{Y}_{st}^{\varepsilon,\eta,*}$ maps on the scale, to this end we first compute

$$\|D_p^\beta \mathbb{Y}_{st}^{\varepsilon,\eta,*}\psi(\cdot,p,\kappa)\|_{l_\xi^2} = \left(\sum_{\xi \in \mathbb{Z}^d} \left|D_p^\beta \mathbb{Y}_{st}^{\varepsilon,\eta,*}\psi(\xi,p,\kappa)\right|^2\right)^{1/2}$$

We have that

$$\sum_{\xi \in \mathbb{Z}^d} \left|D_p^\beta \mathbb{Y}_{st}^{\varepsilon,*}\psi(\xi,p,\kappa)\right|^2$$

$$\lesssim \sum_{\xi \in \mathbb{Z}^d} \left|\sum_{n \in \mathbb{Z}^d \setminus \{0\}} |\hat{V}(n)|^2 \left(\varepsilon^{-1}\int_s^t \int_s^u e^{ia_1'(v-u)}dvdu\right)D_p^\beta \psi(\xi,p,\kappa)\right|^2$$

$$+ \sum_{\xi \in \mathbb{Z}^d} \left|\sum_{n \in \mathbb{Z}^d \setminus \{0\}} |\hat{V}(n)|^2 \left(\varepsilon^{-1}\int_s^t \int_s^u e^{ia_1'(v-u)}dvdu\right)D_p^\beta \psi(\xi,p,\kappa-n)\mathbb{I}_{n \perp \xi}\right|^2$$

$$+ \sum_{\xi \in \mathbb{Z}^d} \left|\sum_{n \in \mathbb{Z}^d \setminus \{0\}} |\hat{V}(n)|^2 \left(\varepsilon^{-1}\int_s^t \int_s^u e^{ia_2'(v-u)}dvdu\right)D_p^\beta \psi(\xi,p,\kappa+n)\mathbb{I}_{n \perp \xi}\right|^2$$

$$+ \sum_{\xi \in \mathbb{Z}^d} \left|\sum_{n \in \mathbb{Z}^d \setminus \{0\}} |\hat{V}(n)|^2 \left(\varepsilon^{-1}\int_s^t \int_s^u e^{ia_2'(v-u)}dvdu\right)D_p^\beta \psi(\xi,p,\kappa)\right|^2$$

$$= A_\mathbb{Y} + B_\mathbb{Y} + C_\mathbb{Y} + D_\mathbb{Y}$$

We have

$$\sum_{\kappa \in \left(\frac{\mathbb{Z}}{2}\right)^d} \langle\kappa\rangle^m \|D_p^\beta \mathbb{Y}_{st}^{\varepsilon,\eta,*}\psi(\cdot,p,\kappa)\|_{l_\xi^2} \lesssim \sum_{\kappa \in \left(\frac{\mathbb{Z}}{2}\right)^d} \langle\kappa\rangle^m \left(A_\mathbb{Y}^{1/2} + B_\mathbb{Y}^{1/2} + C_\mathbb{Y}^{1/2} + D_\mathbb{Y}^{1/2}\right)$$

We will see how to estimate the first two terms, and the last two are estimated analogously. Beginning with the $A_{\mathbb{Y}}$ term we have by Cauchy Schwartz

$$A_{\mathbb{Y}} = \sum_{\xi \in \mathbb{Z}^d} \left| \sum_{n \in \mathbb{Z}^d \setminus \{0\}} |\hat{V}(n)|^2 \left( \varepsilon^{-1} \int_s^t \int_s^u e^{ia_1'(v-u)} \mathrm{d}v \mathrm{d}u \right) D_p^\beta \psi(\xi, p, \kappa) \right|^2$$

$$\lesssim \sum_{\xi \in \mathbb{Z}^d} \sum_{n \in \mathbb{Z}^d \setminus \{0\}} |\hat{V}(n)|^4 \langle n \rangle^{2M}$$

$$\sum_{n \in \mathbb{Z}^d \setminus \{0\}} \langle n \rangle^{-2M} \left( \varepsilon^{-1} \int_s^t \int_s^u e^{ia_1'(v-u)} \mathrm{d}v \mathrm{d}u \right)^2 |D_p^\beta \psi(\xi, p, \kappa)|^2$$

Now, since $V \in \cap_{M \geqslant 0} H^M(\mathbb{T}^d)$ we have that

$$\sum_{n \in \mathbb{Z}^d \setminus \{0\}} |\hat{V}(n)|^4 \langle n \rangle^{2M} = \sum_{n \in \mathbb{Z}^d \setminus \{0\}} |\hat{V}(n)|^2 \langle n \rangle^{2M} |\hat{V}(n)|^2$$

$$\leqslant \|\hat{V}\|_{l^\infty(\mathbb{Z}^d)}^2 \|V\|_{H^M(\mathbb{T}^d)}^2 < \infty$$

so by lemma 3.5 and

$$A_{\mathbb{Y}} \lesssim (t-s)^2 c(\eta)^{\frac{2}{1-\delta}} \sum_{\xi \in \mathbb{Z}^d} \sum_{n \in \mathbb{Z}^d \setminus \{0\}} \langle n \rangle^{-2M} \langle n \rangle^{2d+4} |D_p^\beta \psi(\xi, p, \kappa)|^2$$

which, for $M$ large enough is

$$\lesssim (t-s)^2 c(\eta)^{\frac{2}{1-\delta}} \|D_p^\beta \psi(\cdot, p, \kappa)\|_{l_\xi^2}^2$$

Hence

$$\sum_{\kappa \in \left(\frac{\mathbb{Z}}{2}\right)^d} \langle \kappa \rangle^m A_{\mathbb{Y}}^{1/2} \lesssim (t-s) c(\eta)^{\frac{1}{1-\delta}} \sum_{\kappa \in \left(\frac{\mathbb{Z}}{2}\right)^d} \langle \kappa \rangle^m \|D_p^\beta \psi(\cdot, p, \kappa)\|_{l_\xi^2}$$

Similarly, for the $B_{\mathbb{Y}}$ term, we have

$$B_{\mathbb{Y}} = \sum_{\xi \in \mathbb{Z}^d} \left| \sum_{n \in \mathbb{Z}^d \setminus \{0\}} |\hat{V}(n)|^2 \left( \varepsilon^{-1} \int_s^t \int_s^u e^{ia_1'(v-u)} \mathrm{d}v \mathrm{d}u \right) D_p^\beta \psi(\xi, p, \kappa - n) \mathbb{I}_{n \perp \xi} \right|^2$$

$$\lesssim \sum_{\xi \in \mathbb{Z}^d} \sum_{n \in \mathbb{Z}^d \setminus \{0\}} |\hat{V}(n)|^4 \langle n \rangle^{2M}$$

$$\sum_{n \in \mathbb{Z}^d \setminus \{0\}} \langle n \rangle^{-2M} \left( \varepsilon^{-1} \int_s^t \int_s^u e^{ia_1'(v-u)} \mathrm{d}v \mathrm{d}u \right)^2 |D_p^\beta \psi(\xi, p, \kappa - n)|^2 \mathbb{I}_{n \perp \xi}$$

again using Lemma 3.5 and the fact that $V \in \cap_{M \geqslant 0} H^M(\mathbb{T}^d)$ we have

$$\lesssim (t-s)^2 c(\eta)^{\frac{2}{1-\delta}} \sum_{\xi \in \mathbb{Z}^d} \sum_{n \in \mathbb{Z}^d \setminus \{0\}} \langle n \rangle^{-2M} \langle n \rangle^{2d+4} |D_p^\beta \psi(p, \xi, \kappa - n)|^2 \mathbb{I}_{n \perp \xi}$$

Hence
$$\sum_{\kappa \in \left(\frac{\mathbb{Z}}{2}\right)^d} \langle \kappa \rangle^m B_{\mathbb{Y}}^{1/2} \lesssim (t-s) c(\eta)^{\frac{1}{1-\delta}} \sum_{\kappa \in \left(\frac{\mathbb{Z}}{2}\right)^d} \sum_{n \in \mathbb{Z}^d \setminus \{0\}} \langle n \rangle^{-M} \langle n \rangle^{d+2}$$
$$\langle \kappa \rangle^m \| D_p^\beta \psi(\cdot, p, \kappa - n) \|_{l_\xi^2}$$

Using Tonelli twice the RHS is
$$= (t-s) c(\eta)^{\frac{1}{1-\delta}} \sum_{n \in \mathbb{Z}^d \setminus \{0\}} \sum_{\kappa \in \left(\frac{\mathbb{Z}}{2}\right)^d} \langle n \rangle^{-M} \langle n \rangle^{d+2} \langle \kappa + n \rangle^m \| D_p^\beta \psi(\cdot, p, \kappa) \|_{l_\xi^2}$$

$$= (t-s) c(\eta)^{\frac{1}{1-\delta}} \sum_{\kappa \in \left(\frac{\mathbb{Z}}{2}\right)^d} \| D_p^\beta \psi(\cdot, p, \kappa) \|_{l_\xi^2} \sum_{n \in \mathbb{Z}^d \setminus \{0\}} \langle n \rangle^{-M} \langle n \rangle^{d+2} \langle \kappa + n \rangle^m$$

and as we did in the estimate for $\mathbb{X}_{st}^{1,\varepsilon,*}$ this can be bounded by
$$\lesssim (t-s) c(\eta)^{\frac{1}{1-\delta}} \sum_{\kappa \in \left(\frac{\mathbb{Z}}{2}\right)^d} \| D_p^\beta \psi(\cdot, p, \kappa) \|_{l_\xi^2} \langle \kappa \rangle^m$$

Estimating the $C$ and $D$ terms analogously and integrating in $p$ and summing over $\beta$: $|\beta| \leq m$ we have that
$$\| \mathbb{Y}_{st}^{\varepsilon,\eta,*} \psi \|_{E_m} \lesssim (t-s) c(\eta)^{\frac{1}{1-\delta}} \| \psi \|_{E_m}$$
$\square$

LEMMA 3.10. *For all $m \in \mathbb{N}_0, \eta \in \mathcal{A}_\eta, \psi \in E_m, 0 \leq s < t$ one has*
$$\| \mathbb{Z}_{st}^{\varepsilon,\eta,*} \psi \|_{E_m} \lesssim \varepsilon^{1-2\gamma} |t-s|^{2\gamma} c(\eta)^{\frac{(2-2\gamma)}{1-\delta}} \| \psi \|_{E_m} \qquad (3.17)$$

PROOF. Recall
$$\mathbb{Z}_{st}^{\varepsilon,\eta,*} F(\xi, p, \kappa) :=$$

$$- \varepsilon^{-1} \sum_{n,n' \in \mathbb{Z}^d \setminus \{0\}} \hat{V}(n) \hat{V}(n') \int_s^t du \int_s^u dv\, e^{ia_1' v} e^{ib_1' u} F\left(\xi + n + n', p, \kappa - \frac{n}{2} - \frac{n'}{2}\right) \mathbb{I}_{a_1' \neq b_1'}$$

$$+ \varepsilon^{-1} \sum_{n,n' \in \mathbb{Z}^d \setminus \{0\}} \hat{V}(n) \hat{V}(n') \int_s^t du \int_s^u dv\, e^{ia_1' v} e^{ib_2' u} F\left(\xi + n + n', p, \kappa - \frac{n}{2} + \frac{n'}{2}\right) \mathbb{I}_{a_1' \neq b_2'}$$

$$+ \varepsilon^{-1} \sum_{n,n' \in \mathbb{Z}^d \setminus \{0\}} \hat{V}(n) \hat{V}(n') \int_s^t du \int_s^u dv\, e^{ia_2' v} e^{ib_3' u} F\left(\xi + n + n', p, \kappa + \frac{n}{2} - \frac{n'}{2}\right) \mathbb{I}_{a_2' \neq b_3'}$$

$$- \varepsilon^{-1} \sum_{n,n' \in \mathbb{Z}^d \setminus \{0\}} \hat{V}(n) \hat{V}(n') \int_s^t du \int_s^u dv\, e^{ia_2' v} e^{ib_4' u} F\left(\xi + n + n', p, \kappa + \frac{n}{2} + \frac{n'}{2}\right) \mathbb{I}_{a_2' \neq b_4'}$$

As usual, we first compute
$$\| D_p^\beta \mathbb{Z}_{st}^{\varepsilon,\eta,*} \psi(\cdot, p, \kappa) \|_{l_\xi^2} = \left( \sum_{\xi \in \mathbb{Z}^d} \left| D_p^\beta \mathbb{Z}_{st}^{\varepsilon,\eta,*} \psi(\xi, p, \kappa) \right|^2 \right)^{1/2}$$

We have that

$$\sum_{\xi \in \mathbb{Z}^d} |D_p^\beta \mathbb{Z}_{st}^{\varepsilon,\eta,*} \psi(\xi,p,\kappa)|^2 \lesssim \sum_{\xi \in \mathbb{Z}^d} |D_p^\beta A_Z|^2 + |D_p^\beta B_Z|^2 + |D_p^\beta C_Z|^2 + |D_p^\beta D_Z|^2$$

And this will imply that

$$\sum_{\kappa \in \left(\frac{\mathbb{Z}}{2}\right)^d} \langle \kappa \rangle^m \|D_p^\beta \mathbb{Z}_{st}^{\varepsilon,*} \psi(\cdot,p,\kappa)\|_{l_\xi^2}$$

$$\lesssim \sum_{\kappa \in \left(\frac{\mathbb{Z}}{2}\right)^d} \langle \kappa \rangle^m \left( \|D_p^\beta A_Z\|_{l_\xi^2} + \|D_p^\beta B_Z\|_{l_\xi^2} + \|D_p^\beta C_Z\|_{l_\xi^2} + \|D_p^\beta D_Z\|_{l_\xi^2} \right)$$

We will estimate the first term, and the rest are estimated analogously. We have by using the Cauchy–Schwartz inequality that

$$\|D_p^\beta A_Z\|_{l_\xi^2}^2 = \sum_{\xi \in \mathbb{Z}^d} |D_p^\beta A_Z|^2 \lesssim \sum_{\xi \in \mathbb{Z}^d} \left( \sum_{n,n' \in \mathbb{Z}^d \setminus \{0\}} |\hat{V}(n)\hat{V}(n')|^2 \langle n \rangle^{2M} \langle n' \rangle^{2M} \right)$$

$$\sum_{n,n' \in \mathbb{Z}^d \setminus \{0\}} (\langle n \rangle \langle n' \rangle)^{-2M} \left( \varepsilon^{-1} \int_s^t \int_s^u e^{ia_1' v} e^{ib_1' u} dv du \right)^2$$

$$\left| D_p^\beta \psi\left( \xi + n + n_1, p, \kappa - \frac{n}{2} - \frac{n'}{2} \right) \right|^2 \mathbb{I}_{n+n' \neq 0}$$

Since we have $V \in \cap_{M \geq 0} H^M(\mathbb{T}^d)$, one has that

$$\sum_{n,n' \in \mathbb{Z}^d \setminus \{0\}} |\hat{V}(n)\hat{V}(n')|^2 \langle n \rangle^{2M} \langle n' \rangle^{2M} = \|V\|_{H^M(\mathbb{T}^d)}^4 < \infty$$

So

$$\|D_p^\beta A_Z\|_{l_\xi^2}^2 \lesssim \sum_{\xi \in \mathbb{Z}^d} \sum_{n,n' \in \mathbb{Z}^d \setminus \{0\}} (\langle n \rangle \langle n' \rangle)^{-2M} \left( \varepsilon^{-1} \int_s^t \int_s^u e^{ia_1' v} e^{ib_1' u} dv du \right)^2$$

$$\left| D_p^\beta \psi\left( \xi + n + n', p, \kappa - \frac{n}{2} - \frac{n'}{2} \right) \right|^2 \mathbb{I}_{n+n' \neq 0}$$

and due to the presence of $\mathbb{I}_{n+n' \neq 0}$, by Lemma 3.6 we have

$$\lesssim \varepsilon^{2-4\gamma} |t-s|^{4\gamma} c(\eta)^{\frac{2 \cdot (2-2\gamma)}{1-\delta}}$$

$$\sum_{\xi \in \mathbb{Z}^d} \sum_{n,n' \in \mathbb{Z}^d \setminus \{0\}} (\langle n \rangle \langle n' \rangle)^{-2M+2(d+2)} \left| D_p^\beta \psi\left( \xi + n + n', p, \kappa - \frac{n}{2} - \frac{n'}{2} \right) \right|^2 \mathbb{I}_{n+n' \neq 0}$$

$$\lesssim \varepsilon^{2-4\gamma} |t-s|^{4\gamma} c(\eta)^{\frac{2 \cdot (2-2\gamma)}{1-\delta}}$$

$$\sum_{n,n' \in \mathbb{Z}^d \setminus \{0\}} (\langle n \rangle \langle n' \rangle)^{-2M+2(d+2)} \left\| D_p^\beta \psi\left( \cdot, p, \kappa - \frac{n}{2} - \frac{n'}{2} \right) \right\|_{l_\xi^2}^2 \mathbb{I}_{n+n' \neq 0}$$

This implies that

$$\sum_{\kappa \in \left(\frac{\mathbb{Z}}{2}\right)^d} \langle \kappa \rangle^m \|D_p^\beta A_Z\|_{l_\xi^2} \lesssim \varepsilon^{1-2\gamma} |t-s|^{2\gamma} c(\eta)^{\frac{(2-2\gamma)}{1-\delta}}$$

$$\sum_{\kappa \in \left(\frac{\mathbb{Z}}{2}\right)^d} \langle \kappa \rangle^m \sum_{n,n' \in \mathbb{Z}^d \setminus \{0\}} (\langle n \rangle \langle n' \rangle)^{-M+(d+2)} \left\|D_p^\beta \psi\left(\cdot, p, \kappa - \frac{n}{2} - \frac{n'}{2}\right)\right\|_{l_\xi^2} \mathbb{I}_{n+n' \neq 0}$$

By using Tonelli twice we have that this is

$$\lesssim \varepsilon^{1-2\gamma} |t-s|^{2\gamma} c(\eta)^{\frac{(2-2\gamma)}{1-\delta}} \sum_{\kappa \in \left(\frac{\mathbb{Z}}{2}\right)^d} \|D_p^\beta \psi(p,\xi,\kappa)\|_{l_\xi^2}$$

$$\sum_{n,n' \in \mathbb{Z}^d \setminus \{0\}} (\langle n \rangle \langle n' \rangle)^{-M+(d+2)} \left\langle \kappa + \frac{n}{2} + \frac{n'}{2} \right\rangle^m \mathbb{I}_{n+n' \neq 0}$$

which for $M$ large enough is bounded by

$$\lesssim \varepsilon^{1-2\gamma} |t-s|^{2\gamma} c(\eta)^{\frac{(2-2\gamma)}{1-\delta}} \sum_{\kappa \in \left(\frac{\mathbb{Z}}{2}\right)^d} \|D_p^\beta \psi(\cdot, p, \kappa)\|_{l_\xi^2} \langle \kappa \rangle^m$$

And hence, by arguing similarly for the other terms $B_Z, C_Z$ and $D_Z$ one has that

$$\|\mathbb{Z}_{st}^{\varepsilon,\eta,*} \psi\|_{E_m} \lesssim \varepsilon^{1-2\gamma} |t-s|^{2\gamma} c(\eta)^{\frac{(2-2\gamma)}{1-\delta}} \|\psi\|_{E_m}$$

□

LEMMA 3.11. $T_{st}^{\varepsilon,\eta,\natural}$ defined by equations (3.4), (3.5) and (3.6) is a bounded linear functional on $E_2$, and we have the naive bound

$$\|T_{st}^{\varepsilon,\eta,\natural}\|_{E_{-2}} \lesssim \varepsilon^{-3/2} |t-s|^2 \tag{3.18}$$

PROOF. The proof reduces to showing that for $m \in \{0,1\}$,

$$\|A^{\kappa-\eta,*}\|_{\mathscr{L}(E_{m+1} \to E_m)} \lesssim 1$$

and

$$\|Q_t^{\varepsilon,\eta,*}\|_{\mathscr{L}(E_m \to E_m)} \lesssim \varepsilon^{-1/2}$$

As an example, take the first term, for $F \in E_2$, one has

$$|\langle T_t^{\varepsilon,\eta}, A^{\kappa-\eta,*} A^{\kappa-\eta,*} F \rangle| \leq \|T_t^{\varepsilon,\eta}\|_{E_{-0}} \|A^{\kappa-\eta,*} A^{\kappa-\eta,*} F\|_{E_0}$$

and

$$\|A^{\kappa-\eta,*} F\|_{E_0} \leq \|A^{\kappa-\eta,*}\|_{\mathscr{L}(E_2 \to E_0)} \|A^{\kappa-\eta,*}\|_{\mathscr{L}(E_1 \to E_0)} \|F\|_{E_1} \lesssim \|F\|_{E_1} \lesssim \|F\|_{E_2} < \infty$$

The bound for $A^{\kappa-\eta,*}$ is an immediate consequence of Lemma 3.15. We now compute for $Q_t^{\varepsilon,\eta,*}$: Let $m \in \mathbb{N}_0$, we then have

$$\|Q_t^{\varepsilon,\eta,*} F\|_{E_m} = \sum_{|\beta| \leq m} \int_{\mathbb{R}^d} dp \sum_{\kappa \in \left(\frac{\mathbb{Z}}{2}\right)^d} \langle \kappa \rangle^m \left( \sum_{\xi \in \mathbb{Z}^d} \left|D_p^\beta Q_t^{\varepsilon,\eta,*} F(\xi,p,\kappa)\right|^2 \right)^{1/2}$$

and $\sum_{\xi \in \mathbb{Z}^d} |D_p^\beta Q_t^{\varepsilon,\eta,*} F(\xi,p,\kappa)|^2$ can be naively estimated as

$$\lesssim \varepsilon^{-1} \sum_{\xi \in \mathbb{Z}^d} \left| \sum_{n \in \mathbb{Z}^d} e^{4\pi^2 i \varepsilon^{-1} n \cdot (2\kappa - 2\eta - n - \xi) u} \hat{V}(n) D_p^\beta F\left(\xi + n, p, \kappa - \frac{n}{2}\right) \right|^2$$

$$+ \varepsilon^{-1} \sum_{\xi \in \mathbb{Z}^d} \left| \sum_{n \in \mathbb{Z}^d} e^{4\pi^2 i \varepsilon^{-1} n \cdot (2\kappa - 2\eta + n + \xi) u} \hat{V}(n) D_p^\beta F\left(\xi + n, p, \kappa + \frac{n}{2}\right) \right|^2 = c_1 + c_2$$

and we estimate just the first term by Cauchy–Schwartz, the second is estimated the same way.

$$c_1 \lesssim \varepsilon^{-1} \sum_{\xi \in \mathbb{Z}^d} \left( \sum_{n \in \mathbb{Z}^d} |\hat{V}(n)|^2 \langle n \rangle^{2M} \right) \left( \sum_{n \in \mathbb{Z}^d} \langle n \rangle^{-2M} (D_p^\beta F)^2 \left(\xi + n, p, \kappa - \frac{n}{2}\right) \right)$$

$$\lesssim \varepsilon^{-1} \|V\|_{H^M(\mathbb{T}^d)}^2 \sum_{\xi \in \mathbb{Z}^d} \sum_{n \in \mathbb{Z}^d} \langle n \rangle^{-2M} (D_p^\beta F)^2 \left(\xi + n, p, \kappa - \frac{n}{2}\right)$$

by Tonelli's theorem this is

$$\lesssim \varepsilon^{-1} \|V\|_{H^M(\mathbb{T}^d)}^2 \sum_{n \in \mathbb{Z}^d} \langle n \rangle^{-2M} \sum_{\xi \in \mathbb{Z}^d} (D_p^\beta F)^2 \left(\xi + n, p, \kappa - \frac{n}{2}\right)$$

$$= \varepsilon^{-1} \|V\|_{H^M(\mathbb{T}^d)}^2 \sum_{n \in \mathbb{Z}^d} \langle n \rangle^{-2M} \left\| D_p^\beta F\left(\cdot, p, \kappa - \frac{n}{2}\right) \right\|_{l_\xi^2}^2$$

So using that the $l^2$-norm of a sequence is upper bounded by its $l^1$-norm,

$$\sum_{\kappa \in \left(\frac{\mathbb{Z}}{2}\right)^d} \langle \kappa \rangle^m \left( \sum_{\xi \in \mathbb{Z}^d} |D_p^\beta Q_t^{\varepsilon,\eta,*} F(\xi,p,\kappa)|^2 \right)^{1/2} \lesssim \varepsilon^{-1/2} \|V\|_{H^{2d}(\mathbb{T}^d)}$$

$$\sum_{\kappa \in \left(\frac{\mathbb{Z}}{2}\right)^d} \langle \kappa \rangle^m \sum_{n \in \mathbb{Z}^d} \langle n \rangle^{-M} \left\| D_p^\beta F\left(\cdot, p, \kappa - \frac{n}{2}\right) \right\|_{l_\xi^2}$$

and using Tonelli's theorem, this is

$$\lesssim \varepsilon^{-1/2} \|V\|_{H^{2d}(\mathbb{T}^d)} \sum_{\kappa \in \left(\frac{\mathbb{Z}}{2}\right)^d} \|D_p^\beta F(\cdot, p, \kappa)\|_{l_\xi^2} \sum_{n \in \mathbb{Z}^d} \left\langle \kappa + \frac{n}{2} \right\rangle^m \langle n \rangle^{-M}$$

$$\lesssim \varepsilon^{-1/2} \|V\|_{H^{2d}(\mathbb{T}^d)} \sum_{\kappa \in \left(\frac{\mathbb{Z}}{2}\right)^d} \langle \kappa \rangle^m \|D_p^\beta F(\cdot, p, \kappa)\|_{l_\xi^2}$$

Hence

$$\|Q_t^{\varepsilon,\eta,*} F\|_{E_m} \lesssim \varepsilon^{-1/2} \|V\|_{H^{2d}(\mathbb{T}^d)} \|F\|_{E_m}$$

So we have the claim that

$$\|Q_t^{\varepsilon,\eta,*}\|_{\mathscr{L}(E_m \to E_m)} \lesssim \varepsilon^{-1/2}$$

Hence by chaining the above estimates together with the fact that each time integral gives a $|t-s|$ in the bound, we have proved the claim. □

### 3.4 Passing to the limit

#### 3.4.1 Uniform bound on the remainder

At this stage, we have established that for every $\eta \in \mathscr{A}_\eta$, there exists a family of solutions $\{T^{\varepsilon,\eta}\}_{\varepsilon \in (0,1)}$ in $L^\infty([0,T], E_{-0})$ which satisfy for all $\psi \in E_2$

$$\langle \delta T^{\varepsilon,\eta}_{st}, \psi \rangle = \langle T^{\varepsilon,\eta}_s, \mathbb{A}^{\kappa-\eta,*}_{st} \psi \rangle + \langle T^{\varepsilon,\eta}_s, \mathbb{X}^{1,\varepsilon,\eta,*}_{st} \psi \rangle + \langle T^{\varepsilon,\eta}_s, \mathbb{X}^{2,\varepsilon,\eta,*}_{st} \psi \rangle + \langle T^{\varepsilon,\eta,\natural}_{st}, \psi \rangle$$

where $T^{\varepsilon,\eta,\natural}_{st}$ is an $E_{-2}$ valued $3\gamma$-Hölder map which for any $\psi: \|\psi\|_{E_2} \leqslant 1$ has the property that $\langle T^{\varepsilon,\eta,\natural}_{st}, \psi \rangle \leqslant \varepsilon^{-3/2} |t-s|^2$.

We now use the machinery of unbounded rough drivers introduced in [1] and [10] to obtain a bound on this remainder term that is uniform in $\varepsilon$, which will allow us to pass to the limit, and the bounds on the drivers will allow us to do so. Recall that from estimates (3.14), (3.15), (3.16), (3.17) we have for $\eta \in \mathscr{A}_\eta$, uniformly in $\varepsilon$, and for any $m \in \mathbb{N}_0$ that

$$\|\mathbb{X}^{1,\varepsilon,\eta,*}_{st} \psi\|_{E_m} \lesssim |t-s|^\gamma \|\psi\|_{E_m}$$
$$\|\mathbb{A}^{\kappa-\eta,*}_{st} \psi\|_{E_m} \lesssim |t-s| \|\psi\|_{E_{m+1}}$$
$$\|\mathbb{Z}^{\varepsilon,\eta,*}_{st} \psi\|_{E_m} \lesssim |t-s|^{2\gamma} \|\psi\|_{E_m}$$
$$\|\mathbb{Y}^{\varepsilon,\eta,*}_{st} \psi\|_{E_m} \lesssim |t-s| \|\psi\|_{E_m}$$

In these bounds, the constant $c$ in $\lesssim$ depends on $\eta, \delta, \gamma, V, d$ but is uniform in $\varepsilon$. We then define for $\psi \in E_1$,

$$\langle T^{\varepsilon,\eta,\sharp}_{st}, \psi \rangle := \langle \delta T^{\varepsilon,\eta}_{st}, \psi \rangle - \langle T^{\varepsilon,\eta}_s, \mathbb{X}^{1,\varepsilon,\eta,*}_{st} \psi \rangle$$

and for $\psi \in E_2$ this is equivalent to saying

$$\langle T^{\varepsilon,\eta,\sharp}_{st}, \psi \rangle = \langle T^{\varepsilon,\eta}_s, (\mathbb{A}^{\kappa-\eta,*}_{st} + \mathbb{X}^{2,\varepsilon,\eta,*}_{st}) \psi \rangle + \langle T^{\varepsilon,\eta,\natural}_{st}, \psi \rangle$$

Here is the main point : the theory of rough drivers allows us to convert the uniform in $\varepsilon$-bound for the drivers to a uniform in $\varepsilon$-bound for the remainder. We first introduce some machinery that will be useful in this endeavour. Define the smoothing operators, for $\nu \in (0,1)$

$$\mathcal{J}_\nu \varphi(p, \kappa) := e^{-\nu^{1/2} \langle \kappa \rangle} (\varphi *_p \phi_\nu) \tag{3.19}$$

where

$$\phi_\nu(p) = \nu^{-d/2} \phi\left(\frac{p}{\nu^{1/2}}\right) \tag{3.20}$$

for some mollifier $\phi \in C^\infty_c(\mathbb{R}^d): \int dx \phi(x) = 1$. We have that for $(m,n) \in \{(1,1), (1,2), (2,2)\}$

$$\|\mathcal{J}_\nu\|_{\mathscr{L}(E_m \to E_n)} \lesssim \nu^{-(n-m)} \tag{3.21}$$

and

$$\|\mathcal{J}_\nu - \mathrm{Id}\|_{\mathscr{L}(E_2 \to E_1)} \lesssim \nu^{1/2} \tag{3.22}$$

See Lemmas B.4 and B.5 for proofs of these claims. Now, we can prove the following

LEMMA 3.12. *One has that for any $\varepsilon > 0$, $\eta \in \mathscr{A}_\eta$, $\gamma \in \left(\frac{1}{3}, \frac{1}{2}\right)$ and $0 \leqslant s < t$ that*

$$\|T^{\varepsilon,\eta,\natural}_{st}\|_{E_{-2}} \lesssim C_\gamma C_T |t-s|^{3\gamma} \tag{3.23}$$

*uniformly in ε.*

PROOF. Assume $\|\psi\|_{E_2} \leq 1$. Let $0 < \frac{\min\{1, L\}}{2} \leq |I| \leq \min\{1, L\}$ where $L > 0$ will be defined below in the argument. We compute the increments for the remainder term : for $0 \leq s < u < t \leq |I|$, we have

$$\langle \delta T^{\varepsilon, \eta, \natural}_{sut}, \psi \rangle = \langle T^{\varepsilon, \eta, \natural}_{st}, \psi \rangle - \langle T^{\varepsilon, \eta, \natural}_{su}, \psi \rangle - \langle T^{\varepsilon, \eta, \natural}_{ut}, \psi \rangle$$

using equation (3.3), this is

$$= \langle \delta T^{\varepsilon, \eta}_{st}, \psi \rangle - \langle T^{\varepsilon, \eta}_s, (\mathbb{A}^{\kappa-\eta, *}_{st} + \mathbb{X}^{1, \varepsilon, \eta, *}_{st} + \mathbb{X}^{2, \varepsilon, \eta, *}_{st}) \psi \rangle$$

$$- \langle \delta T^{\varepsilon, \eta}_{su}, \psi \rangle + \langle T^{\varepsilon, \eta}_s, (\mathbb{A}^{\kappa-\eta, *}_{su} + \mathbb{X}^{1, \varepsilon, \eta, *}_{su} + \mathbb{X}^{2, \varepsilon, \eta, *}_{su}) \psi \rangle$$

$$- \langle \delta T^{\varepsilon, \eta}_{ut}, \psi \rangle + \langle T^{\varepsilon, \eta}_u, \mathbb{A}^{\kappa-\eta, *}_{ut} + \mathbb{X}^{1, \varepsilon, \eta, *}_{ut} + \mathbb{X}^{2, \varepsilon, \eta, *}_{ut} \psi \rangle$$

By adding and subtracting $\langle T^{\varepsilon, \eta}_s, (\mathbb{A}^{\kappa-\eta, *}_{ut} + \mathbb{X}^{1, \varepsilon, \eta, *}_{ut} + \mathbb{X}^{2, \varepsilon, \eta, *}_{ut}) \psi \rangle$ and the Chen relations, this is

$$= \langle \delta T^{\varepsilon}_{su}, (\mathbb{A}^{\kappa-\eta, *}_{ut} + \mathbb{X}^{1, \varepsilon, \eta, *}_{ut} + \mathbb{X}^{2, \varepsilon, \eta, *}_{ut}) \psi \rangle - \langle T^{\varepsilon, \eta}_s, \mathbb{X}^{1, \eta, *}_{su} \mathbb{X}^{1, \eta, *}_{ut} \psi \rangle$$

adding and subtracting the smoothing operators 3.19 to the transport term, this is

$$= \langle T^{\varepsilon, \eta, \sharp}_{su}, \mathbb{X}^{1, \varepsilon, \eta, *}_{ut} \psi \rangle + \langle \delta T^{\varepsilon, \eta}_{su}, (\mathbb{I} \pm \mathcal{J}_\nu) \mathbb{A}^{\kappa-\eta, *}_{ut} \psi \rangle + \langle \delta T^{\varepsilon, \eta}_{su}, \mathbb{X}^{2, \varepsilon, \eta, *}_{ut} \psi \rangle$$

since $\mathbb{X}^{1, \varepsilon, \eta, *}_{ut} \psi \in E_2$ this is

$$= \langle T^{\varepsilon, \eta}_s, \mathbb{A}^{\kappa-\eta, *}_{su} \mathbb{X}^{1, \varepsilon, \eta, *}_{ut} \psi \rangle + \langle T^{\varepsilon, \eta}_s, \mathbb{X}^{2, \varepsilon, \eta, *}_{su} \mathbb{X}^{1, \varepsilon, \eta, *}_{ut} \psi \rangle + \langle T^{\varepsilon, \eta, \natural}_{su}, \mathbb{X}^{1, \varepsilon, \eta, *}_{ut} \psi \rangle$$

$$+ \langle T^{\varepsilon, \eta}_s, (\mathbb{A}^{\kappa-\eta, *}_{su} + \mathbb{X}^{1, \varepsilon, \eta, *}_{su} + \mathbb{X}^{2, \varepsilon, \eta, *}_{su}) \mathcal{J}_\nu \mathbb{A}^{\kappa-\eta, *}_{ut} \psi \rangle + \langle T^{\varepsilon, \eta, \natural}_{su}, \mathcal{J}_\nu \mathbb{A}^{\kappa-\eta, *}_{ut} \psi \rangle$$

$$+ \langle \delta T^{\varepsilon, \eta}_{su}, (\mathbb{I} - \mathcal{J}_\nu) \mathbb{A}^{\kappa-\eta, *}_{ut} \psi \rangle$$

$$+ \langle T^{\varepsilon, \eta}_s, (\mathbb{A}^{\kappa-\eta, *}_{su} + \mathbb{X}^{1, \varepsilon, \eta, *}_{su} + \mathbb{X}^{2, \varepsilon, \eta, *}_{su}) \mathbb{X}^{2, \varepsilon, \eta, *}_{ut} \psi \rangle + \langle T^{\varepsilon, \eta, \natural}_{su}, \mathbb{X}^{2, \varepsilon, \eta, *}_{ut} \psi \rangle$$

$$= I_1 + \cdots + I_{12}$$

Denote $C_T := \|T^{\varepsilon, \eta}\|_{L^\infty([0, T]; E_{-0})}$. We have that

$$|I_1| + |I_2| \leq C_T (\|\mathbb{A}^{\kappa-\eta, *}_{su} \mathbb{X}^{1, \varepsilon, \eta, *}_{ut}\|_{\mathscr{L}(E_2 \to E_1)} + \|\mathbb{X}^{2, \varepsilon, \eta, *}_{su} \mathbb{X}^{1, \varepsilon, \eta, *}_{ut}\|_{\mathscr{L}(E_2 \to E_2)})$$

$$|I_4| + |I_5| \lesssim C_T (\|\mathbb{A}^{\kappa-\eta, *}_{su} \mathcal{J}_\nu \mathbb{A}^{\kappa-\eta, *}_{ut}\|_{\mathscr{L}(E_2 \to E_0)} + \|\mathbb{X}^{1, \varepsilon, \eta, *}_{su} \mathcal{J}_\nu \mathbb{A}^{\kappa-\eta, *}_{ut}\|_{\mathscr{L}(E_2 \to E_1)})$$

$$|I_6| + |I_9| \lesssim C_T (\|\mathbb{X}^{2, \varepsilon, \eta, *}_{su} \mathcal{J}_\nu \mathbb{A}^{\kappa-\eta, *}_{ut}\|_{\mathscr{L}(E_2 \to E_1)} + \|\mathbb{A}^{\kappa-\eta, *}_{su} \mathbb{X}^{2, \varepsilon, \eta, *}_{ut}\|_{\mathscr{L}(E_2 \to E_1)})$$

$$|I_{10}| + |I_{11}| \lesssim C_T (\|\mathbb{X}^{1, \varepsilon, \eta, *}_{su} \mathbb{X}^{2, \varepsilon, \eta, *}_{ut}\|_{\mathscr{L}(E_2 \to E_2)} + \|\mathbb{X}^{2, \varepsilon, \eta, *}_{su} \mathbb{X}^{2, \varepsilon, \eta, *}_{ut}\|_{\mathscr{L}(E_2 \to E_2)})$$

$$|I_8| \lesssim C_T \|(\mathbb{I} - \mathcal{J}_\nu) \mathbb{A}^{\kappa-\eta, *}_{ut}\|_{\mathscr{L}(E_2 \to E_0)}$$

$$|I_3| + |I_{12}| \lesssim \|T^{\varepsilon, \eta, \natural}_{su}\|_{E_{-2}} (\|\mathbb{X}^{1, \varepsilon, \eta, *}_{ut}\|_{\mathscr{L}(E_2 \to E_2)} + \|\mathbb{X}^{2, \varepsilon, \eta, *}_{ut}\|_{\mathscr{L}(E_2 \to E_2)})$$

$$|I_7| \lesssim \|T^{\varepsilon, \eta, \natural}_{su}\|_{E_{-2}} \|\mathcal{J}_\nu \mathbb{A}^{\kappa-\eta, *}_{ut}\|_{\mathscr{L}(E_2 \to E_2)}$$

And using the $\varepsilon$-independent bounds on the drivers, and properties (3.21) and (3.22) of the smoothing operator $\mathcal{J}_\nu$, we have

$$|I_1|+|I_2|+|I_4|+|I_5| \lesssim C_T(|t-s|^{1+\gamma}+|t-s|^{3\gamma}+|t-s|^2+|t-s|^{1+\gamma})$$

$$|I_6|+|I_9|+|I_{10}|+|I_{11}| \lesssim C_T(|t-s|^{1+2\gamma}+|t-s|^{1+2\gamma}+|t-s|^{3\gamma}+|t-s|^{4\gamma})$$

$$|I_3|+|I_7|+|I_{12}| \lesssim \|T_{su}^{\varepsilon,\eta,\natural}\|_{E_{-2}}(|t-s|^\gamma+\nu^{-1}|t-s|+|t-s|^{2\gamma})$$

$$|I_8| \lesssim C_T \nu^{1/2}|t-s|$$

Pick $\nu = \frac{|t-s|^{2\gamma}}{|I|^{2\gamma}} \in (0,1)$, so $\nu^{-1}|t-s| = |I|^{2\gamma}|t-s|^{1-2\gamma} \leqslant |I|$, to get that

$$|I_1|+|I_2|+|I_4|+|I_5|+|I_6|+|I_8|+|I_9|+|I_{10}|+|I_{11}| \lesssim C_T|t-s|^{3\gamma}$$

$$|I_3|+|I_7|+|I_{12}| \lesssim \|T_{su}^{\varepsilon,\eta,\natural}\|_{E_{-2}}|I|^\gamma$$

Hence

$$|\langle \delta T_{sut}^{\varepsilon,\eta,\natural},\psi\rangle| \lesssim C_T|t-s|^{3\gamma}+\varepsilon^{-3/2}|t-s|^2|I|^\gamma$$

Hence

$$\|\delta T_{sut}^{\varepsilon,\eta,\natural}\|_{E_{-2}} \lesssim C_T|t-s|^{3\gamma}+\varepsilon^{-3/2}|t-s|^2|I|^\gamma$$

By the sewing lemma (see Lemma 2.2 of [10] or Lemma 4.2 of [15]), setting $C_\gamma = \zeta(3\gamma-1)$ one has

$$\|T_{st}^{\varepsilon,\eta,\natural}\|_{E_{-2}} \lesssim C_\gamma(C_T|t-s|^{3\gamma}+\varepsilon^{-3/2}|t-s|^2|I|^\gamma)$$

Now the $\lesssim$ means there is a constant $c_{\eta,\delta,\gamma,d,V}$ depending on $V,\gamma,d,\eta$ but not on $\varepsilon$ such that

$$\|T_{st}^{\varepsilon,\eta,\natural}\|_{E_{-2}} \leqslant c_{\eta,\delta,\gamma,d,V} C_\gamma(C_T|t-s|^{3\gamma}+\varepsilon^{-3/2}|t-s|^2|I|^\gamma)$$

Choosing $L$ such that $c_{\eta,\gamma,d,V}C_\gamma L^\gamma < \frac{1}{2}$ and choosing $I$ such that $0<|I|\leqslant \min\{1,L\}$ one has

$$\|T_{st}^{\varepsilon,\eta,\natural}\|_{E_{-2}} \leqslant c_{\eta,\delta,\gamma,d,V}C_\gamma C_T|t-s|^{3\gamma}+\frac{1}{2}\varepsilon^{-3/2}|t-s|^2$$

Now if we iterate through the whole procedure once more we get

$$\|T_{st}^{\varepsilon,\eta,\natural}\|_{E_{-2}} \leqslant \left(1+\frac{1}{2}\right)c_{\eta,\delta,\gamma,d,V}C_\gamma C_T|t-s|^{3\gamma}+\frac{1}{4}\varepsilon^{-3/2}|t-s|^2$$

and after $n$ iterations we would get

$$\|T_{st}^{\varepsilon,\eta,\natural}\|_{E_{-2}} \leqslant \left(\sum_{i=0}^n 2^{-i}\right)c_{\eta,\delta,\gamma,d,V}C_\gamma C_T|t-s|^{3\gamma}+\frac{1}{2^{n+1}}\varepsilon^{-3/2}|t-s|^2$$

By sending $n\to\infty$ we get that

$$\|T_{st}^{\varepsilon,\eta,\natural}\|_{E_{-2}} \lesssim c_{\eta,\delta,\gamma,d,V}C_\gamma C_T|t-s|^{3\gamma}$$

And hence we get a bound that is independent of $\varepsilon$. This concludes the proof. □

### 3.4.2 The limiting equation

Recall equations (2.20) and (2.21). We have the following

LEMMA 3.13. $\|(\mathbb{Y}_{st}^{\varepsilon,\eta,*} - \mathbb{Y}_{st}^{\eta,*})\psi\|_{E_0} \to 0$

PROOF. Recall that

$$\mathbb{Y}_{st}^{\varepsilon,\eta,*}\psi(\xi,p,\kappa) :=$$

$$-\varepsilon^{-1}\sum_{n\in\mathbb{Z}^d\setminus\{0\}}|\hat{V}(n)|^2\int_s^t du\int_s^u dv\, e^{ia_1'(v-u)}[\psi(\xi,p,\kappa) - \psi(\xi,p,\kappa-n)\mathbb{I}_{n\perp\xi}]$$

$$+\varepsilon^{-1}\sum_{n\in\mathbb{Z}^d\setminus\{0\}}|\hat{V}(n)|^2\int_s^t du\int_s^u dv\, e^{ia_2'(v-u)}[\psi(\xi,p,\kappa+n)\mathbb{I}_{n\perp\xi} - \psi(\xi,p,\kappa)]$$

We have that

$$\varepsilon^{-1}\int_s^t du\int_s^u dv\, e^{ia_1'(v-u)} = \varepsilon^{-1}\int_s^t du\,\frac{1-e^{ia_1'(s-u)}}{ia_1'} = \varepsilon^{-1}\left(\frac{(t-s)}{ia_1'} + \frac{e^{ia_1'(s-t)}}{(ia_1')^2} - \frac{1}{(ia_1')^2}\right)$$

since $a_1' = 4\pi^2\varepsilon^{-1}n\cdot(2\kappa-2\eta-n-\xi)$ and $a_2' = 4\pi^2\varepsilon^{-1}n\cdot(2\kappa-2\eta+n+\xi)$, this is

$$= \frac{(t-s)}{4\pi^2 in\cdot(2\kappa-2\eta-n-\xi)} + o(\varepsilon)$$

Hence in the limit as $\varepsilon \to 0$ we have that the first term is

$$\frac{i(t-s)}{4\pi^2}\sum_{n\in\mathbb{Z}^d\setminus\{0\}}|\hat{V}(n)|^2\frac{\psi(\xi,p,\kappa)-\psi(\xi,p,\kappa-n)\mathbb{I}_{n\perp\xi}}{n\cdot(2\kappa-2\eta-\xi-n)}$$

Computing the other terms similarly, we have (2.20) and (2.21). □

Finally, we are ready to prove Theorem 2.8.

PROOF. (of Theorem 2.8) Recall the weak formulation of the rough equation, for $\psi \in E_2: \|\psi\|_{E_2} \leqslant 1$

$$\langle \delta T_{st}^{\varepsilon,\eta},\psi\rangle = \langle T_s^{\varepsilon,\eta}, \mathbb{A}_{st}^{\kappa-\eta,*}\psi\rangle + \langle T_s^{\varepsilon,\eta}, \mathbb{X}_{st}^{1,\varepsilon,\eta,*}\psi\rangle + \langle T_s^{\varepsilon,\eta}, \mathbb{X}_{st}^{2,\varepsilon,\eta,*}\psi\rangle + \langle T_{st}^{\varepsilon,\eta,\natural},\psi\rangle$$

one has for a.e. $\eta \in \left[-\frac{1}{4},\frac{1}{4}\right)^d$ that

$$|\langle \delta T_{st}^{\varepsilon,\eta},\psi\rangle| \lesssim C_T(|t-s| + |t-s|^\gamma + |t-s|^{2\gamma}) + \|T_{st}^{\varepsilon,\eta,\natural}\|_{E_{-2}}$$

which by the uniform apriori bounds is

$$|\langle \delta T_{st}^{\varepsilon,\eta},\psi\rangle| \lesssim |t-s|^\gamma C_T + \|T_{st}^{\varepsilon,\eta,\natural}\|_{E_{-2}} \lesssim |t-s|^\gamma C_T + C_T|t-s|^{3\gamma} \lesssim C_T|t-s|^\gamma$$

uniformly in $\varepsilon$. Hence, for any $\psi \in E_2 : \|\psi\|_{E_2} \leqslant 1$, the family $\langle T^{\varepsilon,\eta}(t), \psi \rangle$ which form a uniformly bounded sequence of $\mathbb{R}$-valued $\gamma$-Hölder paths, are uniformly equicontinuous in time, which means by the theorem of Arzela–Ascoli and standard analysis arguments that there exists a subsequence which converges uniformly to some $\gamma$-Hölder real valued function on $[0, T]$.

Next, we note that by the uniform bound $\|T^{\varepsilon,\eta}(t)\|_{E_{-0}} \lesssim C_T^\eta = \|T^{\varepsilon,\eta}\|_{L^\infty([0,T];E_{-0})}$ and by the Banach–Alaoglu theorem that there exists a subsequence $T^{\varepsilon_j}$ converging weakly-* to a function $T^\eta \in L^\infty([0,T]; E_{-0})$. In particular, for any $\psi \in E_2, \langle T^{\varepsilon_j,\eta}, \psi \rangle \to \langle T^\eta, \psi \rangle \in L^\infty([0,T]; \mathbb{R})$ and by the fact that there exists a uniformly converging subsequence from the previous paragraph, the limit $\langle T^\eta(t), \psi \rangle$ is also a $\gamma$-Hölder path.

Now for $\psi \in E_2$, this allows passing to the limit (along the subsequence) in the term on the left hand side of the rough equation, and the first three terms on the right hand side, and this defines the term $\langle T_{st}^{\eta, \natural}, \psi \rangle$. We have

$$\langle T_s^{\varepsilon_j,\eta}, \mathbb{A}_{st}^{\kappa-\eta,*} \psi \rangle \to \langle T_s^\eta, \mathbb{A}_{st}^* \psi \rangle$$

$$\left|\langle T_s^{\varepsilon_j,\eta}, \mathbb{X}_{st}^{1,\varepsilon_j,\eta,*} \psi \rangle\right| \leqslant \|T_s^{\varepsilon_j,\eta}\|_{E_{-0}} \|\mathbb{X}_{st}^{1,\varepsilon_j,\eta,*} \psi\|_{E_0} \lesssim \varepsilon_j^{1/2-\gamma} \to 0$$

which implies that $\langle T_s^{\varepsilon_j,\eta}, \mathbb{X}_{st}^{1,\varepsilon_j,\eta,*} \psi \rangle \to 0$. For the term

$$\langle T_s^{\varepsilon_j,\eta}, \mathbb{X}_{st}^{2,\varepsilon_j,*} \psi \rangle = \langle T_s^{\varepsilon_j,\eta}, \mathbb{Y}_{st}^{\varepsilon_j,\eta,*} \psi \rangle + \langle T_s^{\varepsilon_j,\eta}, \mathbb{Z}_{st}^{\varepsilon_j,\eta,*} \psi \rangle$$

and one has that

$$\left|\langle T_s^{\varepsilon_j,\eta}, \mathbb{Z}_{st}^{\varepsilon_j,\eta,*} \psi \rangle\right| \leqslant \|T_s^{\varepsilon_j,\eta}\|_{E_{-0}} \|\mathbb{Z}_{st}^{\varepsilon_j,\eta,*} \psi\|_{E_0} \lesssim \varepsilon_j^{1-2\gamma} \to 0$$

which implies that $\langle T_s^{\varepsilon_j,\eta}, \mathbb{Z}_{st}^{\varepsilon_j,\eta,*} \psi \rangle \to 0$ and recalling we have that

$$\langle T_s^{\varepsilon_j,\eta}, \mathbb{Y}_{st}^{\varepsilon_j,\eta,*} \psi \rangle \leqslant \langle T_s^{\varepsilon_j,\eta}, \mathbb{Y}_{st}^{\eta,*} \psi \rangle + \langle T_s^{\varepsilon_j,\eta}, (\mathbb{Y}_{st}^{\varepsilon_j,\eta,*} - \mathbb{Y}_{st}^{\eta,*}) \psi \rangle$$

we see that the first term

$$T_s^{\varepsilon_j,\eta}(\mathbb{Y}_{st}^{\eta,*} \psi) \to T_s^\eta(\mathbb{Y}_{st}^{\eta,*} \psi)$$

and the second term goes to 0, as a consequence of Lemma 3.13.

Hence one has

$$\langle T_{st}^{\eta, \natural}, \psi \rangle = \langle \delta T_{st}^\eta, \psi \rangle - \langle T_s^\eta, \mathbb{A}_{st}^{\kappa-\eta,*} \psi \rangle - \langle T_s^\eta, \mathbb{Y}_{st}^{\eta,*} \psi \rangle = \lim_{j \to \infty} T_{st}^{\varepsilon_j,\eta,\natural}(\psi)$$

and we have

$$\left|\langle T_{st}^{\eta,\natural}, \psi \rangle\right| \lesssim C_T \|\psi\|_{E_2} |t-s|^{3\gamma}$$

as a consequence of the fact that the apriori bounds are uniform in $\varepsilon$. Hence we have the existence of an $E_{-0}$ valued path $T(t)$ satisfying, for all $\psi \in E_2$, the rough difference equation

$$\langle \delta T_{st}^\eta, \psi \rangle = \langle T_s^\eta, \mathbb{A}_{st}^{\kappa-\eta,*} \psi \rangle + \langle T_s^\eta, \mathbb{Y}_{st}^{\eta,*} \psi \rangle + \langle T_{st}^{\eta,\natural}, \psi \rangle$$

Now, since $\mathbb{A}_{st}^{\kappa-\eta,*}$ and $\mathbb{Y}_{st}^{\eta,*}$ are of order $|t-s|$, we have for all $\psi \in E_1$

$$\langle T_t^\eta, \psi \rangle - \langle T_0^\eta, \psi \rangle = \int_0^t ds \langle T_s^\eta, Y^{\eta,*} \psi \rangle \tag{3.24}$$

where $Y^{\eta,*} = \frac{\mathbb{Y}_{st}^{\eta,*}}{(t-s)}$. We can deduce this is actually a weak solution to a linear Boltzmann equation in $E_{-0}$. Finally, by the fact that every subsequence of equation (3.3) has a further subsequence converging to a limit that also satisfies equation (3.24), and by the uniqueness of weak solutions to this equation in $E_{-0}$ (for well-posedness theory for linear Boltzmann equations such as this, see [8])), we have that the entire sequence $T^{\varepsilon,\eta}$ converges to $T^\eta$. This concludes the proof. □

## 4  Observables

In this section, we go back to the study of the observables as in (1.5). We first use a heuristic stationary phase argument to restrict to the term $\xi = 0$. When attempting to use the same strategy using the sewing lemma as in Section 3, we show in Subsection 4.1 that one can prove uniform in $\varepsilon$-estimates for the terms involving the non-resonant drivers from (3.3). In Subsection 4.2, we characterize the obstruction to the convergence of the resonant term.

Recall (1.5). For $F \in \mathscr{S}(\mathbb{R}_x^d \times \mathbb{R}_k^d)$, consider the rescaled observable

$$\int_{\mathbb{R}^{2d}} dx\,dk\, W^\varepsilon(t,x,k) F(x,k) = \int_{\mathbb{R}^{2d}} dx\,dk\, W\left(\frac{t}{\varepsilon}, \frac{x}{\varepsilon}, k\right) F(x,k)$$

Using equations (2.4) and (2.9), this is

$$= \int_{\mathbb{R}^d} dx \int_{[-\frac{1}{4},\frac{1}{4}]^d} d\eta \sum_{\kappa \in \left(\frac{\mathbb{Z}}{2}\right)^d} \tilde{W}_\varphi\left(\frac{t}{\varepsilon}, \left[\!\left[\frac{x}{\varepsilon}\right]\!\right], \frac{x}{\varepsilon}, \eta, \kappa\right) F(x, \kappa - \eta)$$

$$= \int_{\mathbb{R}^d} dp \int_{[-\frac{1}{4},\frac{1}{4}]^d} d\eta \sum_{\kappa \in \left(\frac{\mathbb{Z}}{2}\right)^d} \tilde{W}_\varphi^\varepsilon\left(t, \frac{p}{\varepsilon}, p, \eta, \kappa\right) F(p, \kappa - \eta)$$

Now using equations (2.12) and (2.13), this can be rewritten as

$$= \int_{\mathbb{R}^d} dp \int_{[-\frac{1}{4},\frac{1}{4}]^d} d\eta \sum_{\kappa \in \left(\frac{\mathbb{Z}}{2}\right)^d} U^\varepsilon(t, \varepsilon^{-1} p - 4\pi \varepsilon^{-1}(\kappa - \eta) t, p, \eta, \kappa) F(p, \kappa - \eta)$$

$$= \int_{\mathbb{R}^d} dp \int_{[-\frac{1}{4},\frac{1}{4}]^d} d\eta \sum_{\kappa \in \left(\frac{\mathbb{Z}}{2}\right)^d} \sum_{\xi \in \mathbb{Z}^d} e^{2\pi i \varepsilon^{-1} \xi \cdot (p - 4\pi(\kappa-\eta)t)} T^\varepsilon(t, \xi, p, \eta, \kappa) F(p, \kappa - \eta)$$

Consider the mode $\xi = 0$, which we expect to be the only one contributing to the limit by some stationary phase argument (this is yet to be rigorously proved). Hence we restrict to considering

$$O_t^\varepsilon = \int_{[-\frac{1}{4},\frac{1}{4}]^d} d\eta \langle T_t^\varepsilon, \mathbb{I}_{\xi=0} F \rangle = \int_{[-\frac{1}{4},\frac{1}{4}]^d} d\eta \int_{\mathbb{R}^d} dp \sum_{\kappa \in \left(\frac{\mathbb{Z}}{2}\right)^d} T^\varepsilon(t, 0, p, \eta, \kappa) F(p, \kappa - \eta)$$

Then, equation (3.3) and the fact that $\mathbb{X}^{2,\varepsilon,\eta,*}_{st} = \mathbb{Y}^{\varepsilon,\eta,*}_{st} + \mathbb{Z}^{\varepsilon,\eta,*}_{st}$ gives that

$$\delta O^{\varepsilon}_{st} = \int_{[-\frac{1}{4},\frac{1}{4}]^d} d\eta \langle \delta T^{\varepsilon,\eta}_{st}, \mathbb{I}_{\xi=0} F \rangle$$

$$= \int_{[-\frac{1}{4},\frac{1}{4}]^d} d\eta \langle T^{\varepsilon,\eta}_s, A^{\kappa-\eta,*}_{st} \mathbb{I}_{\xi=0} F \rangle + \int_{[-\frac{1}{4},\frac{1}{4}]^d} d\eta \langle T^{\varepsilon,\eta}_s, \mathbb{X}^{1,\varepsilon,\eta,*}_{st} \mathbb{I}_{\xi=0} F \rangle$$

$$+ \int_{[-\frac{1}{4},\frac{1}{4}]^d} d\eta \langle T^{\varepsilon,\eta}_s, \mathbb{Y}^{\varepsilon,\eta,*}_{st} \mathbb{I}_{\xi=0} F \rangle + \int_{[-\frac{1}{4},\frac{1}{4}]^d} d\eta \langle T^{\varepsilon,\eta}_s, \mathbb{Z}^{\varepsilon,\eta,*}_{st} \mathbb{I}_{\xi=0} F \rangle$$

$$+ \int_{[-\frac{1}{4},\frac{1}{4}]^d} d\eta \langle T^{\varepsilon,\eta,\natural}_{st}, \mathbb{I}_{\xi=0} F \rangle$$

We will now attempt to use the same strategy as the previous section- prove uniform in $\varepsilon$-estimates for the leading order terms, and a naive bound on the remainder. We will show that

$$\left| \int_{[-\frac{1}{4},\frac{1}{4}]^d} d\eta \langle T^{\varepsilon,\eta}_s, \mathbb{X}^{1,\varepsilon,\eta,*}_{st} \mathbb{I}_{\xi=0} F \rangle \right| \lesssim \varepsilon^{1/2-\gamma} |t-s|^{\gamma} \|T^{\varepsilon}_s\|_{L^{\infty}_{\eta,p,\kappa} l^2_{\xi}} \|F\|_{L^1_{p,k}(\mathbb{R}^{2d})} \qquad (4.1)$$

and

$$\left| \int_{[-\frac{1}{4},\frac{1}{4}]^d} d\eta \langle T^{\varepsilon,\eta}_s, \mathbb{Z}^{\varepsilon,\eta,*}_{st} \mathbb{I}_{\xi=0} F \rangle \right| \lesssim |t-s|^{2\gamma} \varepsilon^{1-2\gamma} \|T^{\varepsilon}_s\|_{L^{\infty}_{\eta,p,\kappa} l^2_{\xi}} \|F\|_{L^1_{p,k}(\mathbb{R}^{2d})} \qquad (4.2)$$

and that the terms with $A^{\kappa-\eta,*}_{st}$ and $\mathbb{Y}^{\varepsilon,\eta,*}_{st}$ are uniformly bounded in $\varepsilon$, and have time regularity $|t-s|$, but do not decay as $\varepsilon \to 0$. This is similar to what we did in the case of fixed $\eta$. However, we will see that the limit $\varepsilon \to 0$ cannot be taken for the resonant term without having continuity in the $\eta$-variable for $T^{\varepsilon}$ and any potential limit $T$.

The transport term is the easiest to handle. We have that

$$\int_{[-\frac{1}{4},\frac{1}{4}]^d} d\eta \langle T^{\varepsilon}_s, A^*_{st} \mathbb{I}_{\xi=0} F \rangle = \int_{[-\frac{1}{4},\frac{1}{4}]^d} d\eta \sum_{\kappa \in (\frac{\mathbb{Z}}{2})^d} \int_{\mathbb{R}^d} dp\, T^{\varepsilon}_s(0,p,\eta,\kappa) A^*_{st} F(p,\kappa-\eta)$$

$$= 4\pi(t-s) \int_{[-\frac{1}{4},\frac{1}{4}]^d} d\eta \sum_{\kappa \in (\frac{\mathbb{Z}}{2})^d} \int_{\mathbb{R}^d} dp\, T^{\varepsilon}_s(0,p,\eta,\kappa)(\kappa-\eta) \cdot \nabla_p F(p,\kappa-\eta)$$

Hence

$$\left| \int_{[-\frac{1}{4},\frac{1}{4}]^d} d\eta \langle T^{\varepsilon}_s, A^*_{st} \mathbb{I}_{\xi=0} F \rangle \right|$$

$$\lesssim |t-s| \|T^{\varepsilon}_s\|_{L^{\infty}_{\eta,p,\kappa} l^2_{\xi}} \int_{[-\frac{1}{4},\frac{1}{4}]^d} d\eta \sum_{\kappa \in (\frac{\mathbb{Z}}{2})^d} \int_{\mathbb{R}^d} dp\, |(\kappa-\eta) \cdot \nabla_p F(p,\kappa-\eta)|$$

## 4.1 Uniform bounds on the non-resonant terms

In this section we prove the bounds in equations (4.1) and (4.2). First consider the term with $\mathbb{X}^{1,\varepsilon,*}_{st}$.

$$\int_{[-\frac{1}{4},\frac{1}{4}]^d} d\eta \langle T^{\varepsilon}_s, \mathbb{X}^{1,\varepsilon,*}_{st} \mathbb{I}_{\xi=0} F \rangle$$

We will now see that this can be bounded uniformly in $\varepsilon$. This is

$$= \int_{[-\frac{1}{4},\frac{1}{4}]^d} d\eta \sum_{\kappa \in (\frac{\mathbb{Z}}{2})^d} \int_{\mathbb{R}^d} dp \sum_{\xi \in \mathbb{Z}^d} T_s^\varepsilon(\xi, p, \eta, \kappa) \mathbb{X}_{st}^{1;\varepsilon,*}(\mathbb{I}_{\xi=0} F(p, \kappa - \eta))$$

$$= i\varepsilon^{-1/2} \int_{[-\frac{1}{4},\frac{1}{4}]^d} d\eta \sum_{\kappa \in (\frac{\mathbb{Z}}{2})^d} \int_{\mathbb{R}^d} dp \sum_{\xi \in \mathbb{Z}^d} T_s^\varepsilon(\xi, p, \eta, \kappa)$$

$$\sum_{n \in \mathbb{Z}^d \setminus \{0\}} \hat{V}(n) \int_s^t du \left[ e^{ia_1'u} F\left(p, \kappa - \eta - \frac{n}{2}\right) - e^{ia_2'u} F\left(p, \kappa - \eta + \frac{n}{2}\right) \right] \mathbb{I}_{\xi+n=0}$$

where we recall expressions (3.7) and (3.8). Simplifying, this is

$$= i\varepsilon^{-1/2} \int_{[-\frac{1}{4},\frac{1}{4}]^d} d\eta \sum_{\kappa \in (\frac{\mathbb{Z}}{2})^d} \int_{\mathbb{R}^d} dp \sum_{n \in \mathbb{Z}^d \setminus \{0\}} \hat{V}(n) T_s^\varepsilon(-n, p, \eta, \kappa)$$

$$\left[ F\left(p, \kappa - \eta - \frac{n}{2}\right) - F\left(p, \kappa - \eta + \frac{n}{2}\right) \right] \int_s^t du\, e^{8\pi^2 i \varepsilon^{-1} n \cdot (\kappa - \eta) u}$$

Splitting terms, and taking absolute values, the first term can be bounded by

$$\leqslant \int_{[-\frac{1}{4},\frac{1}{4}]^d} d\eta \sum_{\kappa \in (\frac{\mathbb{Z}}{2})^d} \int_{\mathbb{R}^d} dp$$

$$\left| \sum_{n \in \mathbb{Z}^d \setminus \{0\}} \hat{V}(n) T_s^\varepsilon(-n, p, \eta, \kappa) F\left(p, \kappa - \eta - \frac{n}{2}\right) \varepsilon^{-1/2} \int_s^t du\, e^{8\pi^2 i \varepsilon^{-1} n \cdot (\kappa - \eta) u} \right|$$

using the Hölder's inequality, this is

$$\leqslant \int_{[-\frac{1}{4},\frac{1}{4}]^d} d\eta \sum_{\kappa \in (\frac{\mathbb{Z}}{2})^d} \int_{\mathbb{R}^d} dp \left( \sup_{n \in \mathbb{Z}^d \setminus \{0\}} |\hat{V}(n)| \langle n \rangle^M \right)$$

$$\sum_{n \in \mathbb{Z}^d \setminus \{0\}} \langle n \rangle^{-M} |T_s^\varepsilon(-n, p, \eta, \kappa)| F\left(p, \kappa - \eta - \frac{n}{2}\right) \left( \varepsilon^{-1/2} \int_s^t du\, e^{8\pi^2 i \varepsilon^{-1} n \cdot (\kappa - \eta) u} \right)$$

$$\lesssim_V \|T_s^\varepsilon\|_{L^\infty_{\eta,p,\kappa} l^\infty_\xi} \int_{[-\frac{1}{4},\frac{1}{4}]^d} d\eta \sum_{\kappa \in (\frac{\mathbb{Z}}{2})^d} \int_{\mathbb{R}^d} dp$$

$$\sum_{n \in \mathbb{Z}^d \setminus \{0\}} F\left(p, \kappa - \eta - \frac{n}{2}\right) \langle n \rangle^{-M} \left| \varepsilon^{-1/2} \int_s^t du\, e^{8\pi^2 i \varepsilon^{-1} n \cdot (\kappa - \eta) u} \right|$$

by Lemma 3.4, this is

$$\lesssim_V \|T_s^\varepsilon\|_{L^\infty_{\eta,p,\kappa} l^2_\xi} \int_{[-\frac{1}{4},\frac{1}{4}]^d} d\eta \sum_{\kappa \in (\frac{\mathbb{Z}}{2})^d} \int_{\mathbb{R}^d} dp$$

$$\sum_{n \in \mathbb{Z}^d \setminus \{0\}} \langle n \rangle^{-M} F\left(p, \kappa - \eta - \frac{n}{2}\right) c(\eta)^{\frac{1-\gamma}{1-\delta}} \varepsilon^{\frac{1}{2} - \gamma} |t - s|^\gamma \langle n \rangle^{(d+2)(1-\gamma)}$$

by Tonelli's theorem, this is

$$\lesssim_V \varepsilon^{1/2-\gamma}|t-s|^\gamma \|T_s^\varepsilon\|_{L^\infty_{\eta,p,\kappa}l^2_\xi} \sum_{n\in\mathbb{Z}^d\setminus\{0\}} \langle n\rangle^{-M}\langle n\rangle^{(d+2)(1-\gamma)}$$

$$\int_{\mathbb{R}^d} dp \sum_{\kappa\in(\frac{\mathbb{Z}}{2})^d} \sup_{\eta\in[-\frac{1}{4},\frac{1}{4}]^d} F\left(p,\kappa-\eta-\frac{n}{2}\right) \int_{[-\frac{1}{4},\frac{1}{4}]^d} d\eta c(\eta)^{\frac{1-\gamma}{1-\delta}}$$

$$=\varepsilon^{1/2-\gamma}|t-s|^\gamma \|T_s^\varepsilon\|_{L^\infty_{\eta,p,\kappa}l^2_\xi} \sum_{n\in\mathbb{Z}^d\setminus\{0\}} \langle n\rangle^{-M}\langle n\rangle^{(d+2)(1-\gamma)}$$

$$\int_{\mathbb{R}^d} dp \int_{\mathbb{R}^d} dk F(p,k) \int_{[-\frac{1}{4},\frac{1}{4}]^d} d\eta c(\eta)^{\frac{1-\gamma}{1-\delta}}$$

Now, choosing $\delta<\min\{\gamma,\frac{1}{d+3}\}$, the integral in $\eta$ is finite, since $c(\eta)\in L^1$. The sum over $n$ is finite choosing $M$ large enough and the regularity of $F$ says that the above expression is uniformly bounded in $\varepsilon$, i.e.,

$$\int_{[-\frac{1}{4},\frac{1}{4}]^d} d\eta \langle T_s^\varepsilon, \mathbb{X}_{st}^{1;\varepsilon,*}\mathbb{I}_{\xi=0}F\rangle \lesssim_V \varepsilon^{1/2-\gamma}|t-s|^\gamma \|T_s^\varepsilon\|_{L^\infty_{\eta,p,\kappa}l^2_\xi}\|F\|_{L^1_{p,k}(\mathbb{R}^{2d})}$$

This is equation (4.1). Next we consider the term with $\mathbb{Z}_{st}^{\varepsilon,\eta,*}$

$$\int_{[-\frac{1}{4},\frac{1}{4}]^d} d\eta \langle T_s^{\varepsilon,\eta}, \mathbb{Z}_{st}^{\varepsilon,\eta,*}\mathbb{I}_{\xi=0}F\rangle$$

$$=\int_{[-\frac{1}{4},\frac{1}{4}]^d} d\eta \int_{\mathbb{R}^d} dp \sum_{\kappa\in(\frac{\mathbb{Z}}{2})^d} T_s^\varepsilon(0,p,\eta,\kappa)\mathbb{Z}_{st}^{\varepsilon,\eta,*}(\mathbb{I}_{\xi=0}F(p,\kappa-\eta))$$

where

$$\mathbb{Z}_{st}^{\varepsilon,\eta,*}(\mathbb{I}_{\xi=0}F(p,\kappa-\eta))=$$

$$=-\varepsilon^{-1}\sum_{n,n'\in\mathbb{Z}^d\setminus\{0\}} \hat{V}(n)\hat{V}(n')\varphi_{st}(b_1',a_1')F\left(p,\kappa-\eta-\frac{n}{2}-\frac{n'}{2}\right)\mathbb{I}_{a_1'\neq b_1'}\mathbb{I}_{\xi+n+n'=0}$$

$$+\varepsilon^{-1}\sum_{n,n'\in\mathbb{Z}^d\setminus\{0\}} \hat{V}(n)\hat{V}(n')\varphi_{st}(b_2',a_1')F\left(p,\kappa-\eta-\frac{n}{2}+\frac{n'}{2}\right)\mathbb{I}_{a_1'\neq b_2'}\mathbb{I}_{\xi+n+n'=0}$$

$$+\varepsilon^{-1}\sum_{n,n'\in\mathbb{Z}^d\setminus\{0\}} \hat{V}(n)\hat{V}(n')\varphi_{st}(b_3',a_2')F\left(\kappa-\eta+\frac{n}{2}-\frac{n'}{2}\right)\mathbb{I}_{a_2'\neq b_3'}\mathbb{I}_{\xi+n+n'=0}$$

$$-\varepsilon^{-1}\sum_{n,n'\in\mathbb{Z}^d\setminus\{0\}} \hat{V}(n)\hat{V}(n')\varphi_{st}(b_4',a_2')F\left(\kappa-\eta+\frac{n}{2}+\frac{n'}{2}\right)\mathbb{I}_{a_2'\neq b_4'}\mathbb{I}_{\xi+n+n'=0}$$

Recalling expressions (3.7)-(3.13), for

$$\alpha_1=4\pi^2\varepsilon^{-1}n\cdot(2\kappa-2\eta+n'), \qquad \alpha_2=4\pi^2\varepsilon^{-1}n\cdot(2\kappa-2\eta-n')$$

$$\beta_1=4\pi^2\varepsilon^{-1}n'\cdot(2\kappa-2\eta-n), \qquad \beta_2=4\pi^2\varepsilon^{-1}n'\cdot(2\kappa-2\eta+n)$$

and
$$G_1(p, \kappa, \eta, n, n') = F\left(p, \kappa - \eta - \frac{n}{2} - \frac{n'}{2}\right) - F\left(p, \kappa - \eta - \frac{n}{2} + \frac{n'}{2}\right)$$
$$G_2(p, \kappa, \eta, n, n') = F\left(p, \kappa - \eta + \frac{n}{2} - \frac{n'}{2}\right) - F\left(p, \kappa - \eta + \frac{n}{2} - \frac{n'}{2}\right)$$
one has that
$$\mathbb{Z}_{st}^{\varepsilon,\eta,*}(\mathbb{I}_{\xi=0}F(p, \kappa - \eta)) =$$
$$= -\varepsilon^{-1} \sum_{n,n' \in \mathbb{Z}^d \setminus \{0\}} \hat{V}(n)\hat{V}(n')\varphi_{st}(\beta_1, \alpha_1)G_1(p, \kappa, \eta, n, n')\mathbb{I}_{\alpha_1 \neq \beta_1}$$
$$+ \varepsilon^{-1} \sum_{n,n' \in \mathbb{Z}^d \setminus \{0\}} \hat{V}(n)\hat{V}(n')\varphi_{st}(\beta_2, \alpha_2)G_2(p, \kappa, \eta, n, n')\mathbb{I}_{\alpha_2 \neq \beta_2}$$
$$= O_Z^1 + O_Z^2$$

Both terms can be estimated the same way. We will show how to bound $O_Z^1$. Consider now
$$O_Z^1 = -\varepsilon^{-1} \int_{[-\frac{1}{4}, \frac{1}{4}]^d} d\eta \int_{\mathbb{R}^d} dp \sum_{\kappa \in \left(\frac{\mathbb{Z}}{2}\right)^d} T_s^\varepsilon(0, p, \eta, \kappa) \sum_{n,n' \in \mathbb{Z}^d \setminus \{0\}} \hat{V}(n)\hat{V}(n')$$
$$\varphi_{st}(\beta_1, \alpha_1)G_1(p, \kappa, \eta, n, n')\mathbb{I}_{\alpha_1 \neq \beta_1}$$

So
$$|O_Z^1| \leqslant \varepsilon^{-1} \|T_s^\varepsilon\|_{L^\infty_{\eta,p,\kappa} l^\infty_\xi} \int_{[-\frac{1}{4}, \frac{1}{4}]^d} d\eta \int_{\mathbb{R}^d} dp \sum_{\kappa \in \left(\frac{\mathbb{Z}}{2}\right)^d}$$
$$\left|\sum_{n,n' \in \mathbb{Z}^d \setminus \{0\}} \hat{V}(n)\hat{V}(n')\varphi_{st}(\beta_1, \alpha_1)G_1(p, \kappa, \eta, n, n')\mathbb{I}_{\alpha_1 \neq \beta_1}\right|$$

By Hölder's inequality, this is
$$\leqslant \varepsilon^{-1} \|T_s^\varepsilon\|_{L^\infty_{\eta,p,\kappa} l^2_\xi} \int_{[-\frac{1}{4}, \frac{1}{4}]^d} d\eta \int_{\mathbb{R}^d} dp \sum_{\kappa \in \left(\frac{\mathbb{Z}}{2}\right)^d} \left(\sup_n |\hat{V}(n)|\langle n \rangle^M\right)^2$$
$$\sum_{n,n' \in \mathbb{Z}^d \setminus \{0\}} \langle n \rangle^{-M} \langle n' \rangle^{-M} \varphi_{st}(\beta_1, \alpha_1)|G_1(p, \kappa, \eta, n, n')|\mathbb{I}_{\alpha_1 \neq \beta_1}$$
$$\lesssim \varepsilon^{-1} \|T_s^\varepsilon\|_{L^\infty_{\eta,p,\kappa} l^2_\xi} \int_{\mathbb{R}^d} dp \sum_{\kappa \in \left(\frac{\mathbb{Z}}{2}\right)^d} \left(\sup_n |\hat{V}(n)|\langle n \rangle^M\right)^2$$
$$\sum_{n,n' \in \mathbb{Z}^d \setminus \{0\}} \langle n \rangle^{-M} \langle n' \rangle^{-M} \sup_\eta |G_1(p, \kappa, \eta, n, n')| \int_{[-\frac{1}{4}, \frac{1}{4}]^d} d\eta \varphi_{st}(\beta_1, \alpha_1)\mathbb{I}_{\alpha_1 \neq \beta_1}$$
$$\lesssim_V \varepsilon^{-1} \|T_s^\varepsilon\|_{L^\infty_{\eta,p,\kappa} l^2_\xi} \int_{\mathbb{R}^d} dp \sum_{\kappa \in \left(\frac{\mathbb{Z}}{2}\right)^d} \sup_{\eta \in [-\frac{1}{4}, \frac{1}{4}]^d} F(p, \kappa - \eta)$$
$$\sum_{n,n' \in \mathbb{Z}^d \setminus \{0\}} \langle n \rangle^{-M} \langle n' \rangle^{-M} \int_{[-\frac{1}{4}, \frac{1}{4}]^d} d\eta \varphi_{st}(\beta_1, \alpha_1)\mathbb{I}_{\alpha_1 \neq \beta_1}$$

By Lemma 3.1 we have that

$$\int_{\left[-\frac{1}{4},\frac{1}{4}\right]^d} d\eta |\varphi_{st}(\beta_1,\alpha_1)|\mathbb{I}_{\alpha_1\neq\beta_1}$$

$$\lesssim \frac{|t-s|^{2\gamma}}{\varepsilon^{2\gamma-2}} \int_{\left[-\frac{1}{4},\frac{1}{4}\right]^d} d\eta \left( \frac{1}{|\beta_1|^{1-\gamma}|\alpha_1|^{1-\gamma}} + \frac{1}{|\alpha_1|^{1/2}|\beta_1|^{1-\gamma}|\beta_1+\alpha_1|^{\frac{1-2\gamma}{2}}} \right) \mathbb{I}_{\alpha_1\neq\beta_1}$$

$$+ \frac{|t-s|^{2\gamma}}{\varepsilon^{2\gamma-2}} \int_{\left[-\frac{1}{4},\frac{1}{4}\right]^d} d\eta \left( \frac{\mathbb{I}_{\alpha_1\neq\beta_1}}{|\beta_1|^{1/2}|\alpha_1|^{1-\gamma}|\beta_1+\alpha_1|^{\frac{1-2\gamma}{2}}} + \frac{\mathbb{I}_{\alpha_1\neq\beta_1}}{|\alpha_1|^{1/2}|\beta_1|^{1/2}|\beta_1+\alpha_1|^{1-2\gamma}} \right)$$

If we can show that each of the integrals in $\eta$ can be bounded by $C(d,\gamma)\langle n\rangle^4\langle n'\rangle^4$, we will have that

$$|O_Z^1| \lesssim |t-s|^{2\gamma} \varepsilon^{1-2\gamma} \|T_s^\varepsilon\|_{L^\infty_{\eta,p,\kappa}l^2_\xi} \|F\|_{L^1_{p,k}}$$

which is uniformly bounded in $\varepsilon$, and vanishes as $\varepsilon \to 0$. To show the bound on the integrals in $\eta$, we need to modify the argument used in Lemma 3.2, by noticing that each singular term is integrable when isolated from the others. This will be the content of the following

LEMMA 4.1. *For $i\in\{1,2\}$, we have that for $\alpha+\beta+\sigma = 2-2\gamma, 0<\alpha,\beta<1, 0\leqslant \sigma<1$ that*

$$\int_{\left[-\frac{1}{4},\frac{1}{4}\right]^d} d\eta \frac{1}{|\beta_i|^\alpha|\alpha_i|^\beta|\beta_i+\alpha_i|^\sigma}\mathbb{I}_{\alpha_i\neq\beta_i} \lesssim_{d,\gamma} \langle n\rangle^4\langle n'\rangle^4$$

PROOF. We will prove it for the case $i=1$, the case $i=2$ is the done the same way. Note that $\sigma<\alpha$ and $\sigma<\beta$ in each expression. Using the form of $\alpha_1$ and $\beta_1$, after a change of variables $2\eta \to \eta$, this yields an expression of the form

$$\int_{\left[-\frac{1}{2},\frac{1}{2}\right]^d} d\eta \frac{1}{|n\cdot\eta-k_1|^\alpha |n'\cdot\eta-k_2|^\beta |n\cdot\eta-k_1+n'\cdot\eta-k_2|^\sigma}\mathbb{I}_{n\cdot\eta-k_1+n'\cdot\eta-k_2\neq 0}$$

for some $k_1,k_2\in\mathbb{Z}$, upto some constant factor which is neglected. We break this down into two cases, one where $n,n'$ are collinear and one where they aren't. For the collinear case, assume wlog that $n'=cn, c\in\mathbb{Z}\setminus\{0\}$. Hence $|c|\geqslant 1$. Note also that in the collinear case, we cannot then have $k_2=ck_1$, since if we use the actual form of $k_1$ and $k_2$ from $\alpha_1,\beta_1$ we would have that $k_2=ck_1$ would mean that

$$n'\cdot 2\kappa - n'\cdot n = k_2 \stackrel{!}{=} ck_1 = c(2n\cdot\kappa + n\cdot n') \Leftrightarrow 2cn\cdot\kappa - cn\cdot n = 2cn\cdot\kappa + c^2 n\cdot n$$

$$\Leftrightarrow c^2 = -c \Leftrightarrow c = -1$$

But in this case $n\cdot\eta - k_1 + n'\cdot\eta - k_2 = n\cdot\eta - k_1 - n'\cdot\eta + k_1 = 0$ which violates the non-resonance condition. Hence we have that $k_2 \neq ck_1$ in the collinear case where $n' = cn$. Then writing $k_2 = ck_1 + l$ for some $l \in \mathbb{Z}\setminus\{0\}$ the above expression becomes

$$\int_{[-\frac{1}{2},\frac{1}{2}]^d} d\eta \frac{1}{|n\cdot\eta - k_1|^\alpha |cn\cdot\eta - ck_1 - l|^\beta |n\cdot\eta - k_1 + cn\cdot\eta - ck_1 - l|^\sigma}$$

$$= \int_{[-\frac{1}{2},\frac{1}{2}]^d} d\eta \frac{1}{|n\cdot\eta - k_1|^\alpha |c|^\beta \left|n\cdot\eta - k_1 - \frac{l}{c}\right|^\beta |n\cdot\eta - k_1 + cn\cdot\eta - ck_1 - l|^\sigma}$$

If $c = -1$ then $|n\cdot\eta - k_1 + cn\cdot\eta - ck_1 - l|^\sigma = |l|^\sigma$ and one adapts the steps we show below in the case $c \neq -1$. For $c \neq -1$, this is

$$= \int_{[-\frac{1}{2},\frac{1}{2}]^d} d\eta \frac{1}{|n\cdot\eta - k_1|^\alpha |c|^\beta \left|n\cdot\eta - k_1 - \frac{l}{c}\right|^\beta |1 + c|^\sigma \left|n\cdot\eta - k_1 - \frac{l}{1+c}\right|^\sigma}$$

Let $R$ be a rotation matrix such that $Rn = |n|e_1$. Then changing variables via $\eta' = R^T\eta$ this is

$$\leq \int_{B_{\sqrt{d}}(0)} d\eta' \frac{1}{||n|\eta'_1 - k_1|^\alpha |c|^\beta \left||n|\eta'_1 - k_1 - \frac{l}{c}\right|^\beta |1 + c|^\sigma \left||n|\eta'_1 - k_1 - \frac{l}{1+c}\right|^\sigma}$$

$$\lesssim \int_{-\sqrt{d}}^{\sqrt{d}} d\eta' \frac{1}{||n|\eta'_1 - k_1|^\alpha |c|^\beta \left||n|\eta'_1 - k_1 - \frac{l}{c}\right|^\beta |1 + c|^\sigma \left||n|\eta'_1 - k_1 - \frac{l}{1+c}\right|^\sigma}$$

changing variables once more $\eta = |n|\eta'_1$ and since $|c| \geq 1, |1 + c| \geq 1$ this is

$$\lesssim \frac{1}{|n|} \int_{-|n|\sqrt{d}}^{|n|\sqrt{d}} d\eta \frac{1}{|\eta - k_1|^\alpha \left|\eta - k_1 - \frac{l}{c}\right|^\beta \left|\eta - k_1 - \frac{l}{1+c}\right|^\sigma}$$

We have singularities therefore at $\eta = k_1, \eta = k_1 + \frac{l}{c}, \eta = k_1 + \frac{l}{1+c}$. (In the case $c = -1$ this important point would still be true for the two denominator terms). Since $l \in \mathbb{Z}\setminus\{0\}$ the singularities never occur at the same point. We can therefore rewrite this as

$$\frac{1}{|n||c|^\beta |1 + c|^\sigma} \int_{-|n|\sqrt{d}}^{|n|\sqrt{d}} d\eta \frac{1}{|\eta - k_1|^\alpha |\eta - k_2|^\beta |\eta - k_3|^\sigma}$$

for $k_1 < k_2 < k_3$. We assume that $k_1, k_2, k_3$ lie in $[-|n|\sqrt{d}, |n|\sqrt{d}]$, we would have a constant such that the corresponding term would be upper bounded by, and we could estimate the remaining integral as we will do below. Now let $r = \min\left\{1, \frac{k_2 - k_1}{2}, \frac{k_3 - k_2}{2}\right\}$ and let $S_{k_j} = [k_j - l, k_j + l] \cap [-|n|\sqrt{d}, |n|\sqrt{d}]$ for $j \in \{1, 2, 3\}$ and let $N = [-|n|\sqrt{d}, |n|\sqrt{d}] \setminus \cup_{j=1}^3 S_{k_j}$. On $N$ each of the denominators is greater than $\frac{r}{2}$, so

$$\int_N d\eta \frac{d\eta}{|\eta - k_1|^\alpha |\eta - k_2|^\beta |\eta - k_3|^\sigma} \leq \frac{|N|}{r^{2-2\gamma}} \lesssim_{d,\gamma} \frac{|n|}{r^{2-2\gamma}}$$

On any of the $S_{k_j}$ for instance, $S_{k_1}$ one has that

$$\int_{S_{k_1}} d\eta \frac{1}{|\eta - k_1|^\alpha |\eta - k_2|^\beta |\eta - k_3|^\sigma} \lesssim \frac{1}{r^{\beta+\sigma}} \int_{S_{k_1}} d\eta \frac{1}{|\eta - k_1|^\alpha} \lesssim \frac{1}{r^{2-2\gamma}} \int_{-1}^{1} d\eta \frac{1}{|\eta|^\alpha} \lesssim \frac{1}{r^{2-2\gamma}}$$

Hence overall we have that

$$\frac{1}{|n|} \int_{-|n|\sqrt{d}}^{|n|\sqrt{d}} d\eta \frac{1}{|\eta - k_1|^\alpha |\eta - k_2|^\beta |\eta - k_3|^\sigma} \lesssim \frac{1}{r^{2-2\gamma}}$$

Now plugging in that $k_2 - k_1 = \frac{l}{1+c}$ and $k_3 - k_2 = \frac{l}{c} - \frac{l}{1+c} = \frac{l}{c(1+c)}$ we have that for $c \notin \{0, 1\}$ that $r \geqslant \frac{1}{|c||1+c|}$ hence,

$$\frac{1}{|n|} \int_{-|n|\sqrt{d}}^{|n|\sqrt{d}} d\eta \frac{1}{|\eta - k_1|^\alpha |\eta - k_2|^\beta |\eta - k_3|^\sigma} \lesssim |c|^{4-4\gamma}$$

Finally using that $|c| = \frac{|n'|}{|n|} \leqslant |n'| \leqslant \langle n' \rangle$ we have that this is

$$\lesssim \langle n' \rangle^{4-4\gamma} \lesssim \langle n' \rangle^4 \langle n \rangle^4$$

as claimed. We next consider the non-collinear case. Here, since $n' \neq cn$, we have that $n, n'$ span a two dimensional plane $P_{n,n'}$. We can define two orthogonal vectors on this plane via the Gram–Schmidt procedure

$$u_1 := \frac{n}{\|n\|}, \qquad u_2 := \frac{\tilde{n}}{\|\tilde{n}\|}, \tilde{n} = n' - (n' \cdot u_1) u_1$$

We can then create an orthogonal matrix $R$ of the form

$$R = \begin{pmatrix} u_1^T \\ u_2^T \\ \vdots \\ u_d^T \end{pmatrix}$$

Then defining $s = R\eta \Leftrightarrow \eta = R^T s = s_1 u_1 + s_2 u_2 + \sum_{j=3}^d s_j u_j$ we have that $n \cdot \eta = \|n\| u_1 \cdot \eta = \|n\| s_1$ and since $n' \in \text{span}\{u_1, u_2\}$, $n' \cdot \eta = (n' \cdot u_1) s_1 + (n' \cdot u_2) s_2$. Then

$$\int_{[-\frac{1}{2}, \frac{1}{2}]^d} d\eta \frac{1}{|n \cdot \eta - k_1|^\alpha |n' \cdot \eta - k_2|^\beta |n \cdot \eta - k_1 + n' \cdot \eta - k_2|^\sigma}$$

$$\leqslant \int_{B_{\sqrt{d}}(0)} ds \frac{1}{\|n\| s_1 - k_1|^\alpha |(n' \cdot u_1) s_1 + (n' \cdot u_2) s_2 - k_2|^\beta}$$

$$\frac{1}{\|n\| s_1 - k_1 + (n' \cdot u_1) s_1 + (n' \cdot u_2) s_2 - k_2|^\sigma}$$

$$\lesssim \int_{B_{\sqrt{d}}(0)} ds_2 ds_1 \frac{1}{\|n\| s_1 - k_1|^\alpha |(n' \cdot u_1) s_1 + (n' \cdot u_2) s_2 - k_2|^\beta}$$

$$\frac{1}{\|n\| s_1 - k_1 + (n' \cdot u_1) s_1 + (n' \cdot u_2) s_2 - k_2|^\sigma}$$

where now we are in a 2d ball. Let us define

$$\begin{pmatrix} v_1 \\ v_2 \end{pmatrix} = C\begin{pmatrix} s_1 \\ s_2 \end{pmatrix}, \qquad C = \begin{pmatrix} \|n\| & 0 \\ (n'\cdot u_1) & (n'\cdot u_2) \end{pmatrix}$$

Then $\det C = \|n\|(n'\cdot u_2)$.

$$n'\cdot u_2 = n'\cdot \frac{\tilde{n}}{\|\tilde{n}\|} = n'\cdot \frac{n'}{\|\tilde{n}\|} - (n'\cdot u_1)\frac{(n'\cdot u_1)}{\|\tilde{n}\|} = \frac{\|n'\|^2}{\|\tilde{n}\|} - \frac{\|n'\cdot u_1\|^2}{\|\tilde{n}\|}$$

$$= \frac{\|n'\|^2}{\|\tilde{n}\|} - \frac{\|n'\cdot n\|^2}{\|n\|^2\|\tilde{n}\|} = \frac{\|n'\|^2}{\|\tilde{n}\|} - \frac{\|n'\|^2}{\|\tilde{n}\|}\cos^2(\theta) = \frac{\|n'\|^2}{\|\tilde{n}\|}\sin^2(\theta) = \|n'\||\sin\theta|$$

since by construction $\tilde{n} = n'\sin\theta \Rightarrow \|\tilde{n}\| = \|n'\||\sin\theta|$. Here $\theta$ is the angle between $n$ and $n'$ in $P$. Hence

$$\det C = \|n\|\|n'\||\sin\theta|$$

Hence

$$\int_{B_{\sqrt{d}}(0)} ds_2 ds_1 \frac{1}{\|n\|s_1 - k_1|^\alpha |(n'\cdot u_1)s_1 + (n'\cdot u_2)s_2 - k_2|^\beta}$$

$$\frac{1}{\|n\|s_1 - k_1 + (n'\cdot u_1)s_1 + (n'\cdot u_2)s_2 - k_2|^\sigma}$$

$$\lesssim \frac{1}{\|n\|\|n'\||\sin\theta|}\int_{C(B_{\sqrt{d}}(0))} dv_2 dv_1 \frac{1}{|v_1-k_1|^\alpha |v_2-k_2|^\beta |v_1-k_1+v_2-k_2|^\sigma}$$

$$\lesssim \frac{1}{\|n\|\|n'\||\sin\theta|}\int_{B_{\sqrt{d}(|n|+|n'|)}(0)} dv_2 dv_1 \frac{1}{|v_1-k_1|^\alpha |v_2-k_2|^\beta |v_1-k_1+v_2-k_2|^\sigma}$$

We see that there are singularities at $v_1 = k_1, v_2 = k_2$, but if these singularities lie outside the ball they are harmless, since we then just need to estimate the volume of the ball in 2d and we are done. Hence we can bound the above by

$$\lesssim \frac{1}{\|n\|\|n'\||\sin\theta|}\int_{B_{\sqrt{d}(|n|+|n'|)}(0)} dv_2 dv_1 \frac{1}{|v_1|^\alpha |v_2|^\beta |v_1+v_2|^\sigma}$$

Changing to polar coordinates and writing $R = \sqrt{d}(|n|+|n'|)$, this is

$$\lesssim \frac{1}{\|n\|\|n'\||\sin\theta|}\int_0^R \frac{r}{r^{2-2\gamma}}dr \int_0^{2\pi} d\theta \frac{1}{|\cos\theta|^\alpha |\sin\theta|^\beta |\cos\theta+\sin\theta|^\sigma}$$

$$= \frac{1}{\|n\|\|n'\||\sin\theta|}\frac{R^{2\gamma}}{2\gamma}\int_0^{2\pi} \frac{d\theta}{|\cos\theta|^{1/2}|\sin\theta|^{1-\gamma}|\cos\theta+\sin\theta|^{\frac{1-2\gamma}{2}}}$$

Now the problematic points in $\theta$ are when one of the denominators becomes 0. The important point is they cannot all be 0 at the same $\theta$. $\cos\theta = 0$ at $\theta_1 = \frac{\pi}{2}$ and $\theta_2 = \frac{3\pi}{2}$. $\sin\theta = 0$ at $\theta_3 = 0, \theta_4 = \pi, \theta_5 = 2\pi$ and $\sin\theta + \cos\theta = 0$ at $\theta_6 = \frac{3\pi}{4}, \theta_7 = \frac{7\pi}{4}$. By picking $0 < \delta_\theta \ll 1$ and defining $S_{\theta_j} = (\theta_j + [-\delta, \delta]) \cap [0, 2\pi]$ we can decompose the integral onto each $S_{\theta_j}$ and $C \setminus \cup_{j=1}^7 S_{\theta_j}$. On $C \setminus \cup_{j=1}^7 S_{\theta_j}$ each of the denominators is bounded away from 0 and then the integral can be easily estimated. On each $S_{\theta_j}$, there is only singular contribution. For instance, for $S_{\theta_1}$ we have that there exists a constant $c$ such that $\frac{1}{|\sin\theta|}$, $\frac{1}{|\cos\theta + \sin\theta|} \leqslant c$. Then

$$\int_{S_{\theta_j}} \frac{\mathrm{d}\theta}{|\cos\theta|^{1/2} |\sin\theta|^{1-\gamma} |\cos\theta + \sin\theta|^{\frac{1-2\gamma}{2}}} \lesssim c^{\frac{3}{2}-2\gamma} \int_{\frac{\pi}{2}-\delta_\theta}^{\frac{\pi}{2}+\delta_\theta} \frac{\mathrm{d}\theta}{|\cos\theta|^{1/2}}$$

$$\lesssim c^{\frac{3}{2}-2\gamma} \int_0^{\delta_\theta} \frac{\mathrm{d}\theta}{\theta^{1/2}} \lesssim c^{\frac{3}{2}-2\gamma} \delta_\theta^{1-1/2} \lesssim c^{2-2\gamma} \delta_\theta^{1-\min\{\frac{1}{2}, 1-\gamma, \frac{1}{2}-\gamma\}}$$

and this constant $c$ only depends on $\delta_\theta$. Hence, overall the term is bounded by

$$\lesssim \frac{R^{2\gamma}}{\|n\| \|n'\| |\sin\theta|} \lesssim \frac{\|n\|^{2\gamma-1} + \|n'\|^{2\gamma-1}}{|\sin\theta|} \lesssim \|n\| \|n'\| (\|n\|^{2\gamma-1} + \|n'\|^{2\gamma-1})$$

$$\lesssim \|n\|^{2\gamma} \|n'\| + \|n\| \|n'\|^{2\gamma} \lesssim \|n\| \|n'\| \lesssim \langle n \rangle^4 \langle n' \rangle^4$$

Hence we can conclude. □

## 4.2 The resonant term

Next consider

$$\int_{[-\frac{1}{4},\frac{1}{4}]^d} \mathrm{d}\eta \langle T_s^{\varepsilon,\eta}, \mathbb{Y}_{st}^{\varepsilon,\eta,*} \mathbb{I}_{\xi=0} F \rangle$$

$$= \int_{[-\frac{1}{4},\frac{1}{4}]^d} \mathrm{d}\eta \int_{\mathbb{R}^d} \mathrm{d}p \sum_{\kappa \in (\frac{\mathbb{Z}}{2})^d} T_s^\varepsilon(0, p, \eta, \kappa) \mathbb{Y}_{st}^{\varepsilon,\eta,*} (\mathbb{I}_{\xi=0} F(p, \kappa - \eta))$$

where, we recall that

$$\mathbb{Y}_{st}^{\varepsilon,\eta,*} f(\xi, p, \kappa) =$$

$$-\varepsilon^{-1} \sum_{n \in \mathbb{Z}^d \setminus \{0\}} |\hat{V}(n)|^2 \int_s^t \mathrm{d}u \int_s^u \mathrm{d}v\, e^{ia_1'(v-u)} [f(\xi, p, \kappa) - f(\xi, p, \kappa - n) \mathbb{I}_{n \perp \xi}]$$

$$+\varepsilon^{-1} \sum_{n \in \mathbb{Z}^d \setminus \{0\}} |\hat{V}(n)|^2 \int_s^t \mathrm{d}u \int_s^u \mathrm{d}v\, e^{ia_2'(v-u)} [f(\xi, p, \kappa + n) \mathbb{I}_{n \perp \xi} - f(\xi, p, \kappa)]$$

So, the expression concerned becomes

$$-\varepsilon^{-1}\int_{[-\frac{1}{4},\frac{1}{4}]^d}d\eta\int_{\mathbb{R}^d}dp\sum_{\kappa\in\left(\frac{\mathbb{Z}}{2}\right)^d}T_s^\varepsilon(0,p,\eta,\kappa)\sum_{n\in\mathbb{Z}^d\setminus\{0\}}|\hat{V}(n)|^2\int_s^t du\int_s^u dv\, e^{ic_1(v-u)}$$

$$[F(p,\kappa-\eta)-F(p,\kappa-\eta-n)]$$

$$+\varepsilon^{-1}\int_{[-\frac{1}{4},\frac{1}{4}]^d}d\eta\int_{\mathbb{R}^d}dp\sum_{\kappa\in\left(\frac{\mathbb{Z}}{2}\right)^d}T_s^\varepsilon(0,p,\eta,\kappa)\sum_{n\in\mathbb{Z}^d\setminus\{0\}}|\hat{V}(n)|^2\int_s^t du\int_s^u dv\, e^{ic_2(v-u)}$$

$$[F(p,\kappa-\eta+n)-F(p,\kappa-\eta)]$$

where now

$$c_1=4\pi^2\varepsilon^{-1}n\cdot(2\kappa-2\eta-n),\qquad c_2=4\pi^2\varepsilon^{-1}n\cdot(2\kappa-2\eta+n)$$

Changing variables in the second expression, this is

$$=-\varepsilon^{-1}\int_{[-\frac{1}{4},\frac{1}{4}]^d}d\eta\int_{\mathbb{R}^d}dp\sum_{\kappa\in\left(\frac{\mathbb{Z}}{2}\right)^d}T_s^\varepsilon(0,p,\eta,\kappa)\sum_{n\in\mathbb{Z}^d\setminus\{0\}}|\hat{V}(n)|^2\int_s^t du\int_s^u dv\, e^{ic_1(v-u)}$$

$$[F(p,\kappa-\eta)-F(p,\kappa-\eta-n)]$$

$$-\varepsilon^{-1}\int_{[-\frac{1}{4},\frac{1}{4}]^d}d\eta\int_{\mathbb{R}^d}dp\sum_{\kappa\in\left(\frac{\mathbb{Z}}{2}\right)^d}T_s^\varepsilon(0,p,\eta,\kappa)\sum_{n\in\mathbb{Z}^d\setminus\{0\}}|\hat{V}(-n)|^2\int_s^t du\int_s^u dv\, e^{-ic_1(v-u)}$$

$$[F(p,\kappa-\eta)-F(p,\kappa-\eta-n)]$$

Recalling that $|\hat{V}(n)|^2=|\hat{V}(-n)|^2$, and using that

$$e^{ic_1(v-u)}+e^{-ic_1(v-u)}=2\cos(c_1(v-u))$$

this simplifies to

$$=2\int_{[-\frac{1}{4},\frac{1}{4}]^d}d\eta\int_{\mathbb{R}^d}dp\sum_{\kappa\in\left(\frac{\mathbb{Z}}{2}\right)^d}T_s^\varepsilon(0,p,\eta,\kappa)\sum_{n\in\mathbb{Z}^d\setminus\{0\}}|\hat{V}(n)|^2$$

$$\int_s^t du\int_s^u dv(\varepsilon^{-1}\cos(c_1(v-u)))[F(p,\kappa-\eta-n)-F(p,\kappa-\eta)]$$

We next perform the time integration:

$$\varepsilon^{-1}\int_s^t du\int_s^u dv\cos(c_1(v-u))=-\frac{\varepsilon^{-1}}{c_1}\int_s^t du\sin(c_1(s-u))=\frac{\varepsilon^{-1}}{c_1}\int_s^t du\sin(c_1(u-s))$$

$$=-\frac{\varepsilon^{-1}}{c_1^2}(\cos(c_1(t-s))-1)$$

$$=\frac{\varepsilon^{-1}}{|4\pi^2\varepsilon^{-1}n\cdot(2\kappa-2\eta-n)|^2}(1-\cos(4\pi^2\varepsilon^{-1}n\cdot(2\kappa-2\eta-n)(t-s)))$$

A computation shows that this is an approximation of the delta function on the hyperplane $n\cdot(2\kappa - 2\eta - n) = 0$, upto a constant factor. We can show the concentration on this set as follows: let

$$H_{n,k,\eta}(r) = \left\{(n, k, \eta) \in \mathbb{Z}^d \setminus \{0\} \times \left(\frac{\mathbb{Z}}{2}\right)^d \times \left(-\frac{1}{4}, \frac{1}{4}\right)^d : |n\cdot(2\kappa - 2\eta - n)| \leq r\right\}.$$

Fix $r > 0$. Let $c_1' = \varepsilon c_1$. From the above computations we have that

$$\int_{\left[-\frac{1}{4},\frac{1}{4}\right]^d} d\eta \langle T_s^{\varepsilon,\eta}, \mathbb{Y}_{st}^{\varepsilon,\eta,*} \mathbb{I}_{\xi=0} F \rangle$$

$$= 2(t-s) \int_{\left[-\frac{1}{4},\frac{1}{4}\right]^d} d\eta \int_{\mathbb{R}^d} dp \sum_{\kappa \in \left(\frac{\mathbb{Z}}{2}\right)^d} T_s^{\varepsilon}(0, p, \eta, \kappa) \sum_{n \in \mathbb{Z}^d \setminus \{0\}} |\hat{V}(n)|^2$$

$$\left(\varepsilon^{-1}(t-s) \frac{1 - \cos(\varepsilon^{-1} c_1'(t-s))}{|\varepsilon^{-1} c_1'(t-s)|^2}\right) [F(p, \kappa - \eta - n) - F(p, \kappa - \eta)]$$

$$\left(\mathbb{I}_{H_{n,k,\eta}(r)} + \mathbb{I}_{H_{n,k,\eta}^c(r)}\right).$$

For the term containing $\mathbb{I}_{H_{n,k,\eta}^c(r)}$, we have that $c_1' > \frac{r}{4\pi^2}$. Hence, we can bound that term in absolute value by

$$\lesssim \frac{\varepsilon}{r} \|T_s^\varepsilon\|_{L^\infty_{\eta,p,\kappa,z}} \sum_{n \in \mathbb{Z}^d \setminus \{0\}} |\hat{V}(n)|^2$$

$$\int_{\left[-\frac{1}{4},\frac{1}{4}\right]^d} d\eta \int_{\mathbb{R}^d} dp \sum_{\kappa \in \left(\frac{\mathbb{Z}}{2}\right)^d} |F(p, \kappa - \eta) - F(p, \kappa - \eta - n)|$$

$$\lesssim \frac{\varepsilon}{r} \|T_s^\varepsilon\|_{L^\infty_{\eta,p,\kappa,z}} \|F\|_{L^1(\mathbb{R}^{2d})}$$

which for any positive $r$, in the limit $\varepsilon \to 0$ is $0$.

However, the difficulty in concluding is that on the zero measure set, we do not know continuity of $T^\varepsilon$ in $\eta$, hence we do not know yet if the limit exists.

*Remark* 4.2. Note that

$$4\pi^2 n\cdot(2\kappa - 2\eta - n) = 4\pi^2(2n\cdot(\kappa - \eta) - |n|^2) = 4\pi^2|\kappa - \eta|^2 - 4\pi^2|\kappa - \eta - n|^2$$

In an upcoming work we show how this is related to energy band crossings of the Laplacian, and that one can make a strong statement for observables concentrated away from such momenta.

We have the following image in $d = 2$, of what the problematic zero-measure set looks like

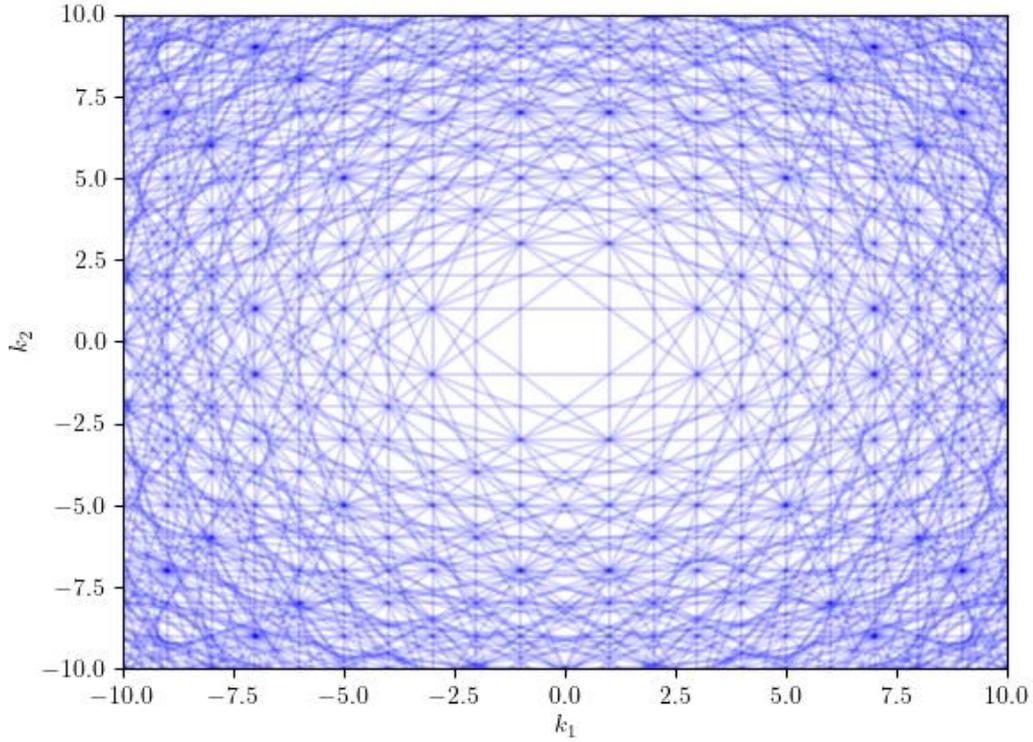

The image is generated purely with straight lines as follows: Since $2\kappa - 2\eta$ are representations for a momenta $k \in \mathbb{R}^d$, we are looking for momenta $k \in \mathbb{R}^d$ such that there exists a non-zero lattice vector $n \in \mathbb{Z}^d \setminus \{0\}$ for which $n \cdot (\kappa - n) = 0$. The diagram is generated by looping over lattice vectors $n$ and drawing a line perpendicular to the line segment $[0, n]$, passing through $n$. Any vector $k$ on this line has the property that $n \cdot k = |n|^2$.

*Example* 4.3. (Single mode potential) Let's demonstrate what the resonant term looks like when the potential has only one mode. Assume and $\hat{V}(n) = 0$ unless $n = \begin{pmatrix} \sigma \\ 0 \end{pmatrix}$ for $\sigma \in \{\pm 1\}$. Then we have an expression of the form

$$\int_{[-\frac{1}{4},\frac{1}{4}]^d} d\eta \int_{\mathbb{R}^d} dp \sum_{\kappa \in (\frac{\mathbb{Z}}{2})^d} \sum_{\sigma \in \{\pm 1\}} T_s^\varepsilon(0, p, \eta, \kappa) \delta_\varepsilon(2\kappa_1 - 2\eta_1 - \sigma)$$

$$\left[ F\left(p, \kappa - \eta - \begin{pmatrix} \sigma \\ 0 \end{pmatrix}\right) - F(p, \kappa - \eta) \right]$$

If we assume $\operatorname{supp} F \subset S_1 \times \mathbb{R}^{d-1}$, where $S_1 = \left(\left[-\frac{1}{2} - \rho, -\frac{1}{2} + \rho\right] \cup \left[\frac{1}{2} - \rho, \frac{1}{2} + \rho\right]\right)^c$, for some $\rho \ll 1$, then

$$\begin{pmatrix} 2 & -2 \\ 1 & -1 \end{pmatrix} \begin{pmatrix} \kappa_1 \\ \eta_1 \end{pmatrix} = \begin{pmatrix} \sigma \\ d \end{pmatrix}, \quad \begin{pmatrix} 2 & -2 \\ 1 & -1 \end{pmatrix} \begin{pmatrix} \kappa_1 \\ \eta_1 \end{pmatrix} = \begin{pmatrix} \sigma \\ d + \sigma \end{pmatrix}$$

has no solution for any $d \in S_1$. One has that for any $\varepsilon' > 0, \sigma \in \{\pm 1\}$, there exists $\varepsilon > 0$ such $\int_{[\frac{\sigma}{2}-\rho,\frac{\sigma}{2}+\rho]^c} \delta_\varepsilon(2\kappa_1 - 2\eta_1 - \sigma) < \varepsilon'$. Then the above term is bounded in absolute value by

$$\lesssim \|T_s^\varepsilon\|_{L^\infty_{\eta,p,\kappa} L^2_\xi} \int_{[-\frac{1}{4},\frac{1}{4}]^d} d\eta \int_{\mathbb{R}^d} dp \sum_{\kappa \in (\frac{\mathbb{Z}}{2})^d} \sum_{\sigma \in \{\pm 1\}} \delta_\varepsilon(2\kappa_1 - 2\eta_1 - \sigma)$$

$$\left| F\left(p, \kappa - \eta - \begin{pmatrix} \sigma \\ 0 \end{pmatrix}\right) - F(p, \kappa - \eta) \right| \left( \mathbb{I}_{\kappa_1 - \eta_1 \in [\frac{\sigma}{2}-\rho, \frac{\sigma}{2}+\rho]} + \mathbb{I}_{\kappa_1 - \eta_1 \notin [\frac{\sigma}{2}-\rho, \frac{\sigma}{2}+\rho]} \right)$$

when $\kappa_1 - \eta_1 \notin \left[\frac{\sigma}{2} - \rho, \frac{\sigma}{2} + \rho\right]$, one can bound this by

$$\lesssim \|T_s^\varepsilon\|_{L^\infty_{\eta,p,\kappa} L^2_\xi} \|F\|_{L^1_p L^\infty_k} \int_{[\frac{\sigma}{2}-\rho,\frac{\sigma}{2}+\rho]^c} \delta_\varepsilon(2\kappa_1 - 2\eta_1 - \sigma)$$

$$\lesssim \varepsilon' \|T_s^\varepsilon\|_{L^\infty_{\eta,p,\kappa} L^2_\xi} \|F\|_{L^1_p L^\infty_k}$$

Since $\varepsilon'$ was arbitrary, this part is 0. On the other hand, for the term

$$\|T_s^\varepsilon\|_{L^\infty_{\eta,p,\kappa} L^2_\xi} \int_{[-\frac{1}{4},\frac{1}{4}]^d} d\eta \int_{\mathbb{R}^d} dp \sum_{\kappa \in (\frac{\mathbb{Z}}{2})^d} \sum_{\sigma \in \{\pm 1\}} \delta_\varepsilon(2\kappa_1 - 2\eta_1 - \sigma)$$

$$\left| F\left(p, \kappa - \eta - \begin{pmatrix} \sigma \\ 0 \end{pmatrix}\right) - F(p, \kappa - \eta) \right| \mathbb{I}_{\kappa_1 - \eta_1 \in [\frac{\sigma}{2}-\rho, \frac{\sigma}{2}+\rho]}$$

one has that $\kappa_1 - \eta_1 \in \left[\frac{\sigma}{2} - \rho, \frac{\sigma}{2} + \rho\right]$ hence both $\kappa_1 - \eta_1 \in S_1^c$ and $\kappa_1 - \eta_1 - \sigma \in S_1^c$. Hence $F\left(p, \kappa - \eta - \begin{pmatrix} \sigma \\ 0 \end{pmatrix}\right) = F(p, \kappa - \eta) = 0$ and also this part is 0. Hence, for any $F \in C_c^\infty(S_1 \times \mathbb{R}^{d-1})$, the entire term is zero, and we expect only trivial transport in the limit for its corresponding observable. For this potential, in the picture above, the problematic zero-measure set reduces to two vertical lines passing through $\begin{pmatrix} 1 \\ 0 \end{pmatrix}$ and $\begin{pmatrix} -1 \\ 0 \end{pmatrix}$ respectively.

## A  The Bloch–Floquet–Zak decomposition

Let us state a few basic properties of the BFZ decomposition:

1. $\tilde{\varphi}(\theta, x)$ is $\mathbb{Z}^d$-periodic in $x$: For $n \in \mathbb{Z}^d$

$$\tilde{\varphi}(\theta, x + n) = \sum_{m \in \mathbb{Z}^d} e^{2\pi i \theta \cdot (x + n - m)} \varphi(x + n - m)$$

$$= \sum_{m' \in \mathbb{Z}^d} e^{2\pi i \theta \cdot (x - m')} \varphi(x - m') = \tilde{\varphi}(\theta, x)$$

Hence, we can identify $\tilde{\varphi}(\theta, x)$ with a complex valued function on $\mathbb{R}^d \times \mathbb{T}^d$. We also note that it is $\mathbb{Z}^d$-quasiperiodic in $\theta$:

$$\tilde{\varphi}(\theta + n, x) = \sum_{m \in \mathbb{Z}^d} e^{2\pi i (\theta + n) \cdot (x - m)} \varphi(x - m)$$

$$= e^{2\pi i n \cdot x} \sum_{m \in \mathbb{Z}^d} e^{2\pi i \theta \cdot (x - m)} \varphi(x - m) = e^{2\pi i n \cdot x} \tilde{\varphi}(\theta, x)$$

By complex conjugation we have that

$$\tilde{\varphi}^*(\theta, x+n) = \tilde{\varphi}^*(\theta, x) \qquad \tilde{\varphi}^*(\theta+n, x) = e^{-2\pi i n \cdot x} \tilde{\varphi}^*(\theta, x)$$

2. Note that the BFZ transform does not commute with complex conjugation. We have that

$$\widetilde{\varphi^*}(\theta, x) := \sum_{m \in \mathbb{Z}^d} e^{2\pi i \theta \cdot (x-m)} \varphi^*(x-m)$$

whereas

$$(\tilde{\varphi})^*(\theta, x) := \sum_{m \in \mathbb{Z}^d} e^{-2\pi i \theta \cdot (x-m)} \varphi^*(x-m)$$

3. We have $\mathbb{Z}^d$-periodicity in the $\theta$ variable for the functions $e^{-2\pi i \theta \cdot x} \tilde{\varphi}(\theta, x)$ (and hence $e^{2\pi i \theta \cdot x} \tilde{\varphi}^*(\theta, x)$) and $|\tilde{\varphi}(\theta, x)|^2$ since

$$e^{-2\pi i (\theta+n) \cdot x} \tilde{\varphi}(\theta+n, x) = e^{-2\pi i (\theta+n) \cdot x} e^{2\pi i n \cdot x} \tilde{\varphi}(\theta, x) = e^{-2\pi i \theta \cdot x} \tilde{\varphi}(\theta, x)$$

and so

$$|\tilde{\varphi}(\theta+n, x)|^2 = \tilde{\varphi}^*(\theta+n, x) \tilde{\varphi}(\theta+n, x)$$

$$\Rightarrow |\tilde{\varphi}(\theta+n, x)|^2 = e^{-2\pi i n \cdot x} \tilde{\varphi}^*(\theta, x) e^{2\pi i n \cdot x} \tilde{\varphi}(\theta, x) = |\tilde{\varphi}(\theta, x)|^2 \tag{A.1}$$

4. For $\varphi \in \mathscr{S}(\mathbb{R}^d; \mathbb{C})$ we can invert the BFZ transform. One has that for all $\varphi \in \mathscr{S}(\mathbb{R}^d; \mathbb{C})$

$$\varphi(x) = \int_{\mathbb{T}^d} e^{-2\pi i \theta \cdot x} \tilde{\varphi}(\theta, x) \mathrm{d}\theta \tag{A.2}$$

This can be seen from the following computation

$$\int_{\mathbb{T}^d} e^{-2\pi i \theta \cdot x} \tilde{\varphi}(\theta, x) \mathrm{d}\theta = \int_{[0,1]^d} e^{-2\pi i \theta \cdot x} \tilde{\varphi}(\theta, x) \mathrm{d}\theta$$

$$= \int_{[0,1]^d} \sum_{m \in \mathbb{Z}^d} e^{-2\pi i \theta \cdot m} \varphi(x-m) \mathrm{d}\theta$$

which by Fubini's theorem is

$$= \sum_{m \in \mathbb{Z}^d} \underbrace{\int_{[0,1]^d} e^{-2\pi i \theta \cdot m} \mathrm{d}\theta}_{\delta_{m,0}} \varphi(x-m) = \varphi(x)$$

Similarly, by complex conjugation,

$$\varphi^*(x) = \int_{\mathbb{T}^d} e^{2\pi i \theta \cdot x} \tilde{\varphi}^*(\theta, x) \mathrm{d}\theta$$

5. If $\varphi \in \mathscr{S}(\mathbb{R}^d; \mathbb{C})$ then $\tilde{\varphi} \in L^2\left(\left[-\frac{1}{2}, \frac{1}{2}\right] \times \mathbb{T}^d\right)$ since

$$\int_{\left[-\frac{1}{2}, \frac{1}{2}\right]^d \times \mathbb{T}^d} \mathrm{d}\theta \mathrm{d}x |\tilde{\varphi}(\theta, x)|^2 = \int_{\mathbb{T}^d \times \mathbb{T}^d} \mathrm{d}\theta \mathrm{d}x \tilde{\varphi}(\theta, x) \tilde{\varphi}^*(\theta, x)$$

$$= \int_{\mathbb{T}^d \times \mathbb{T}^d} \mathrm{d}\theta \mathrm{d}x \sum_{m, m'} e^{2\pi i \theta \cdot (x-m)} \varphi(x-m) e^{-2\pi i \theta \cdot (x-m')} \varphi^*(x-m')$$

$$= \int_{\mathbb{T}^d \times \mathbb{T}^d} \mathrm{d}x \mathrm{d}\theta \sum_{m, m'} e^{-2\pi i \theta \cdot (m-m')} \varphi(x-m) \varphi^*(x-m')$$

We now apply Fubini, since $\sum_m |\varphi(x-m)| \sum_{m'} |\varphi(x-m')| < \infty$, to write this as

$$= \int_{\mathbb{T}^d} \mathrm{d}x \sum_{m, m'} \underbrace{\int_{\mathbb{T}^d} \mathrm{d}\theta e^{-2\pi i \theta \cdot (m-m')}}_{\delta_{m-m'}} \varphi(x-m) \varphi^*(x-m')$$

$$= \int_{\mathbb{T}^d} \mathrm{d}x \sum_m \varphi(x-m) \varphi^*(x-m) = \|\varphi\|^2_{L^2(\mathbb{R}^d)}$$

6. For $\varphi \in \mathscr{S}(\mathbb{R}^d; \mathbb{C})$, we have that $\tilde{\varphi} \in C(\mathbb{R}^d \times \mathbb{T}^d)$: Fix $\theta \in \mathbb{R}^d, x \in \mathbb{T}^d, \varepsilon > 0$. Then we then have that for $\delta_1, \delta_2 \in \mathbb{R}^d$:

$$\tilde{\varphi}(\theta + \delta_1, x + \delta_2) - \tilde{\varphi}(\theta, x) = \sum_{m \in \mathbb{Z}^d} (e^{2\pi i \delta_1 \cdot (x-m)} e^{2\pi i \theta \cdot \delta_2} - 1) e^{2\pi i \theta \cdot (x-m)} \varphi(x-m)$$

Since $\varphi$ is Schwartz, we can define a ball $B$ such that

$$\sum_{m \in \mathbb{Z}^d \cap B^c} |\varphi(x-m)| < \frac{\varepsilon}{4}$$

This then gives us that

$$|\tilde{\varphi}(\theta + \delta, x) - \tilde{\varphi}(\theta, x)| \leq \left| \sum_{m \in \mathbb{Z}^d \cap B} (e^{2\pi i \delta_1 \cdot (x-m)} e^{2\pi i \theta \cdot \delta_2} - 1) e^{2\pi i \theta \cdot (x-m)} \varphi(x-m) \right|$$

$$+ \left| \sum_{m \in \mathbb{Z}^d \cap B^c} (e^{2\pi i \delta_1 \cdot (x-m)} e^{2\pi i \theta \cdot \delta_2} - 1) e^{2\pi i \theta \cdot (x-m)} \varphi(x-m) \right|$$

$$< \left| \sum_{m \in \mathbb{Z}^d \cap B} (e^{2\pi i \delta_1 \cdot (x-m)} e^{2\pi i \theta \cdot \delta_2} - 1) e^{2\pi i \theta \cdot (x-m)} \varphi(x-m) \right| + \frac{\varepsilon}{2}$$

and then choose $\delta_1, \delta_2$ small enough such that the first term is also smaller than $\varepsilon/2$, which can be done, since the sum is finite and the terms are bounded.

7. Defining $\gamma_m u(x) := e^{2\pi i m \cdot x} u(x)$, one can show that the Bloch–Floquet–Zak transform extends to a unitary transformation, also called the Block–Floquet–Zak transform, of $L^2(\mathbb{R}^d; \mathbb{C})$ into $\mathscr{H}_\gamma$, the Hilbert space of $L^2(\mathbb{T}^d)$ valued $\gamma$-equivariant functions, i.e.,

$$\mathscr{H}_\gamma := \{u \in L^2_{\mathrm{loc}}(\mathbb{R}^d; L^2(\mathbb{T}^d; \mathbb{C})) : \tilde{u}(\theta + n, \cdot) = \gamma_n \tilde{u}(\theta, \cdot), \text{ for } \theta \in \mathbb{R}^d, n \in \mathbb{Z}^d\}$$

This is similar to how the Fourier transform extends from $\mathscr{S}(\mathbb{R}^d)$ to a unitary transformation on $L^2(\mathbb{R}^d)$. The above Hilbert space can be endowed with the scalar product

$$\langle u, v \rangle_{\mathscr{H}_\gamma} := \int_{\mathbb{T}^d} d\theta \langle \tilde{u}(\theta, \cdot), \tilde{v}(\theta, \cdot) \rangle_{L^2(\mathbb{T}^d)} = \int_{\mathbb{T}^d} d\theta \int_{\mathbb{T}^d} dx\, \tilde{u}^*(\theta, x) \tilde{v}(\theta, x)$$

and therefore with the norm

$$\|u\|_{\mathscr{H}_\gamma} := \left( \int_{\mathbb{T}^d} d\theta \|\tilde{u}(\theta, \cdot)\|^2_{L^2(\mathbb{T}^d)} \right)^{1/2} = \left( \int_{\mathbb{T}^d} d\theta \int_{\mathbb{T}^d} dx\, |\tilde{u}(\theta, x)|^2 \right)^{1/2}$$

and the inverse BFZ transform is explicitly given by

$$(\mathscr{U}_{\text{BFZ}}^{-1} \tilde{u})(x) = \int_{\mathbb{T}^d} d\theta\, e^{-2\pi i \theta \cdot x} \tilde{u}(\theta, x)$$

See [27] for more details.

Next we have a lemma that describes the regularity of the BFZ transform of a function with the regularity of a classical solution to the Schrödinger equation.

LEMMA A.1. *Let $\varphi \in C(\mathbb{R}; H^2(\mathbb{R}^d; \mathbb{C})) \cap C^1(\mathbb{R}; L^2(\mathbb{R}^d; \mathbb{C}))$, then $\tilde{\varphi} \in C(\mathbb{R}; L^2_{\text{loc}}(\mathbb{R}^d; H^2(\mathbb{T}^d; \mathbb{C}))) \cap C^1(\mathbb{R}; \mathscr{H}_\gamma)$. Furthermore, one has that*

$$\widetilde{\partial_t \varphi}(t) = \partial_t \tilde{\varphi}(t) \tag{A.3}$$

$$\widetilde{\Delta \varphi}(t, \theta, x) = \Delta_x \tilde{\varphi}(t) - 4\pi^2 |\theta|^2 \tilde{\varphi}(t, \theta, x) - 4\pi i \theta \cdot \nabla_x \tilde{\varphi}(t, \theta, x) \tag{A.4}$$

*and*

$$\widetilde{\Delta \varphi^*}(t, \theta, x) = \Delta_x \tilde{\varphi}^*(t, \theta, x) - 4\pi^2 |\theta|^2 \tilde{\varphi}^*(t, \theta, x) + 4\pi i \theta \cdot \nabla_x \tilde{\varphi}^*(t, \theta, x) \tag{A.5}$$

*as functions in $\mathscr{H}_\gamma$.*

PROOF. First, we have that since $U_{\text{BFZ}}$ is a unitary transformation, it is a bounded linear operator from $L^2(\mathbb{R}^d; \mathbb{C})$ to $H_\gamma$, so in particular, it is a continuous linear operator. Hence using the linearity and continuity, one has that

$$\partial_t \tilde{\varphi}(t) = \partial_t U_{\text{BFZ}} \varphi(t) = \lim_{h \to 0} \frac{U_{\text{BFZ}} \varphi(t+h) - U_{\text{BFZ}} \varphi(t)}{h}$$

$$= \lim_{h \to 0} U_{\text{BFZ}} \left( \frac{\varphi(t+h) - \varphi(t)}{h} \right)$$

$$= U_{\text{BFZ}} \left( \lim_{h \to 0} \frac{\varphi(t+h) - \varphi(t)}{h} \right) = U_{\text{BFZ}}(\partial_t \varphi(t)) = \widetilde{\partial_t \varphi}(t)$$

The unitarity of $U_{\text{BFZ}}$ also immediately implies that $\tilde{\varphi} \in C^1(\mathbb{R}; \mathscr{H}_\gamma)$, since $\varphi \in C^1(\mathbb{R}; L^2(\mathbb{R}^d; \mathbb{C}))$. Next, we will show that $\tilde{\varphi} \in C(\mathbb{R}; L^2_{\text{loc}}(\mathbb{R}^d; H^1(\mathbb{T}^d; \mathbb{C})))$. One can proceed by computing that for $\varphi(t) \in H^2(\mathbb{R}^d; \mathbb{C}), \psi \in C_c^\infty(\mathbb{R}^d; \mathbb{C}), j \in \{1, \ldots, d\}$

$$\int_{\mathbb{R}^d} \partial_{x_j} \psi^*(x) \varphi(t, x) dx = -\int_{\mathbb{R}^d} \psi^*(x) v_j(t, x) dx = -\int_{\mathbb{T}^d \times \mathbb{T}^d} (\tilde{\psi})^*(\theta, x) \tilde{v}_j(t, \theta, x) d\theta dx$$

where $v_j(t) = \partial_{x_j}\varphi(t)$ weak is the unique function in $C(\mathbb{R}; L^2(\mathbb{R}^d; \mathbb{C}))$ that satisfies the first identity, and $\tilde{v}_j(t,\theta,x)$ is its BFZ transform. The left hand side can also be written as

$$\int_{\mathbb{R}^d} \partial_{x_j}\psi^*(x)\varphi(t,x)\mathrm{d}x = \int_{\mathbb{T}^d \times \mathbb{T}^d} (\widetilde{\partial_{x_j}\psi})^*(\theta,x)\tilde{\varphi}(t,\theta,x)\mathrm{d}\theta\mathrm{d}x$$

since one has an explicit representation for the BFZ of $\partial_{x_j}\psi$, this is

$$= \int_{\mathbb{T}^d \times \mathbb{T}^d} \tilde{\varphi}(t,\theta,x) \sum_{m \in \mathbb{Z}^d} e^{-2\pi i\theta\cdot(x-m)} \partial_{x_j}\psi^*(x-m)$$

which by the chain rule is

$$= \int_{\mathbb{T}^d \times \mathbb{T}^d} \mathrm{d}\theta \mathrm{d}x \tilde{\varphi}(t,\theta,x) \partial_{x_j}\left(\sum_{m \in \mathbb{Z}^d} e^{-2\pi i\theta\cdot(x-m)} \psi^*(x-m)\right)$$

$$- \int_{\mathbb{T}^d \times \mathbb{T}^d} \mathrm{d}\theta \mathrm{d}x \tilde{\varphi}(t,\theta,x) \sum_{m \in \mathbb{Z}^d} \partial_{x_j}(e^{-2\pi i\theta\cdot(x-m)}) \psi^*(x-m)$$

$$= \int_{\mathbb{T}^d \times \mathbb{T}^d} \tilde{\varphi}(t,\theta,x)(\partial_{x_j}(\tilde{\psi})^*(\theta,x) + 2\pi i\theta_j(\tilde{\psi})^*(\theta,x))\mathrm{d}\theta\mathrm{d}x$$

Hence for all $t \in \mathbb{R}$ one has

$$\int_{\mathbb{T}^d \times \mathbb{T}^d} \tilde{\varphi}(t,\theta,x)\partial_{x_j}(\tilde{\psi})^*(\theta,x)\mathrm{d}\theta\mathrm{d}x =$$

$$-\int_{\mathbb{T}^d \times \mathbb{T}^d} [\tilde{v}_j(t,\theta,x) + 2\pi i\theta_j\tilde{\varphi}(t,\theta,x)](\tilde{\psi})^*(\theta,x)\mathrm{d}\theta\mathrm{d}x$$

Since $\tilde{v}_j(t)$ and $2\pi i\theta_j\tilde{\varphi}(t)$ are functions in $\mathcal{H}_\gamma$ for all $t \in \mathbb{R}, j \in \{1,\ldots,d\}$ one writes

$$\partial_{x_j}\tilde{\varphi}(t) = \tilde{v}_j(t) + 2\pi i\theta_j\tilde{\varphi}(t) \text{ weak on } \mathcal{H}_\gamma$$

One deduces that $\tilde{\varphi} \in C(\mathbb{R}; L^2_{\text{loc},\theta}(\mathbb{R}^d; H^1_x(\mathbb{T}^d; \mathbb{C})))$. Rearranging it, one has

$$\tilde{v}_j(t) = \widetilde{\partial_{x_j}\varphi}(t) = \partial_{x_j}\tilde{\varphi}(t) - 2\pi i\theta_j\tilde{\varphi}(t) \text{ weak on } \mathcal{H}_\gamma \qquad (A.6)$$

Hence

$$\widetilde{\nabla\varphi}(t) = \tilde{v}(t) = \begin{pmatrix} \tilde{v}_1(t) \\ \vdots \\ \tilde{v}_d(t) \end{pmatrix} = \nabla\tilde{\varphi}(t) - 2\pi i\theta\tilde{\varphi}(t) \text{ weak on } \mathcal{H}_\gamma$$

and by conjugating, one has

$$(\widetilde{\nabla\varphi})^*(t) = \tilde{v}^*(t) = \nabla(\tilde{\varphi})^*(t) + 2\pi i\theta(\tilde{\varphi})^*(t) \text{ weak on } \mathcal{H}_\gamma \qquad (A.7)$$

Similarly one can deduce that $\tilde{\varphi} \in C(\mathbb{R}; L^2_{\text{loc},\theta}(\mathbb{R}^d; H^2(\mathbb{T}^d; \mathbb{C})))$. For $j,k \in \{1,\ldots,d\}$ and $\psi \in C^\infty_c(\mathbb{R}^d; \mathbb{C})$

$$\int_{\mathbb{R}^d} \varphi(t,x)\partial^2_{x_j x_k}\psi^*(x)\mathrm{d}x = \int_{\mathbb{R}^d} v_{jk}(t,x)\psi^*(x)\mathrm{d}x = \int_{\mathbb{T}^d \times \mathbb{T}^d} \tilde{v}_{jk}(t,\theta,x)(\tilde{\psi})^*(\theta,x)\mathrm{d}\theta\mathrm{d}x$$

where $v_{jk}(t)$ is the unique function in $L^2(\mathbb{R}^d; \mathbb{C})$ that satisfies the first identity, and $\tilde{v}_{jk}(t)$ is its Bloch–Floquet–Zak transform. The left hand side can also be written as

$$\int_{\mathbb{R}^d} \varphi(t,x) \partial^2_{x_j x_k} \psi^*(x) dx = \int_{\mathbb{T}^d \times \mathbb{T}^d} \tilde{\varphi}(t,\theta,x) \widetilde{(\partial_{x_j} \partial_{x_k} \psi)}^*(\theta, x) d\theta dx$$

$$= \int_{\mathbb{T}^d \times \mathbb{T}^d} d\theta dx \tilde{\varphi}(t,\theta,x) (\partial_{x_j} \widetilde{(\partial_{x_k} \psi)}^*(\theta,x) + 2\pi i \theta_j \widetilde{(\partial_{x_k} \psi)}^*(\theta,x))$$

and using equation (A.7) above, one has

$$= \int_{\mathbb{T}^d \times \mathbb{T}^d} d\theta dx \tilde{\varphi}(t,\theta,x) \times$$

$$\times (\partial^2_{x_j x_k} \tilde{\psi}^*(\theta,x) + 2\pi i \theta_k \partial_{x_j} \tilde{\psi}^*(\theta,x) + 2\pi i \theta_j \partial_{x_k} \tilde{\psi}^*(\theta,x) - 4\pi^2 \theta_j \theta_k \tilde{\psi}^*(\theta,x))$$

Now, we use that

$$\int_{\mathbb{T}^d \times \mathbb{T}^d} d\theta dx \tilde{\varphi}(t,\theta,x) \partial_{x_j} \tilde{\psi}^*(\theta,x) = -\int_{\mathbb{T}^d \times \mathbb{T}^d} d\theta dx (\tilde{v}_j(t) + 2\pi i \theta_j \tilde{\varphi}(t)) \tilde{\psi}^*(\theta,x)$$

To write the above as

$$\int_{\mathbb{T}^d \times \mathbb{T}^d} d\theta dx \tilde{\varphi}(t,\theta,x) \partial^2_{x_j x_k} \tilde{\psi}^*(\theta,x)$$

$$-2\pi i \int_{\mathbb{T}^d \times \mathbb{T}^d} d\theta dx (\theta_k(\tilde{v}_j(t) + 2\pi i \theta_j \tilde{\varphi}(t)) + \theta_j(\tilde{v}_k(t) + 2\pi i \theta_k \tilde{\varphi}(t))) \tilde{\psi}^*(\theta,x)$$

$$-4\pi^2 \int_{\mathbb{T}^d \times \mathbb{T}^d} d\theta dx \theta_j \theta_k \tilde{\varphi}(t,\theta,x) \tilde{\psi}^*(\theta,x)$$

Hence

$$\int_{\mathbb{T}^d \times \mathbb{T}^d} d\theta dx \tilde{\varphi}(t,\theta,x) \partial^2_{x_j x_k} \tilde{\psi}^*(\theta,x) =$$

$$\int_{\mathbb{T}^d \times \mathbb{T}^d} \tilde{v}_{jk}(t,\theta,x) (\tilde{\psi})^*(\theta,x) d\theta dx$$

$$+2\pi i \int_{\mathbb{T}^d \times \mathbb{T}^d} d\theta dx (\theta_k(\tilde{v}_j(t) + 2\pi i \theta_j \tilde{\varphi}(t)) + \theta_j(\tilde{v}_k(t) + 2\pi i \theta_k \tilde{\varphi}(t))) \tilde{\psi}^*(\theta,x)$$

$$+4\pi^2 \int_{\mathbb{T}^d \times \mathbb{T}^d} d\theta dx \theta_j \theta_k \tilde{\varphi}(t,\theta,x) \tilde{\psi}^*(\theta,x)$$

from which one deduces that $\tilde{\varphi} \in C(\mathbb{R}; L^2_{\text{loc}}(\mathbb{R}^d; H^2(\mathbb{T}^d; \mathbb{C})))$ and one has that

$$\partial^2_{x_j x_k} \tilde{\varphi}(t) = \tilde{v}_{jk}(t) + 2\pi i (\theta_k(\tilde{v}_j(t) + 2\pi i \theta_j \tilde{\varphi}(t)) + \theta_j(\tilde{v}_k(t) + 2\pi i \theta_k \tilde{\varphi}(t))) + 4\pi^2 \theta_j \theta_k \tilde{\varphi}(t)$$

weak on $\mathcal{H}_\gamma$. Hence

$$\Delta \tilde{\varphi}(t) = \sum_j (\tilde{v}_{jj}(t) + 2\pi i (\theta_j(\tilde{v}_j(t) + 2\pi i \theta_j \tilde{\varphi}(t)) + \theta_j(\tilde{v}_j(t) + 2\pi i \theta_j \tilde{\varphi}(t))) + 4\pi^2 \theta_j^2 \tilde{\varphi}(t))$$

$$= \sum_j (\tilde{v}_{jj}(t) + 2\pi i (2\theta_j \tilde{v}_j(t) + 4\pi i \theta_j^2 \tilde{\varphi}(t)) + 4\pi^2 \theta_j^2 \tilde{\varphi}(t))$$

$$= \sum_j (\tilde{v}_{jj}(t) + 4\pi i \theta_j \tilde{v}_j(t) - 8\pi^2 i \theta_j^2 \tilde{\varphi}(t) + 4\pi^2 \theta_j^2 \tilde{\varphi}(t))$$

$$= \sum_j (\tilde{v}_{jj}(t) + 4\pi i \theta_j \tilde{v}_j(t) - 4\pi^2 i \theta_j^2 \tilde{\varphi}(t))$$

Finally, plugging in the expression for $\tilde{v}_j(t)$, this is

$$= \sum_j \left( \tilde{v}_{jj}(t) + 4\pi i \theta_j (\partial_{x_j} \tilde{\varphi}(t) - 2\pi i \theta_j \tilde{\varphi}(t)) - 4\pi^2 i \theta_j^2 \tilde{\varphi}(t) \right)$$

$$= \sum_j \left( \tilde{v}_{jj}(t) + 4\pi i \theta_j \partial_{x_j} \tilde{\varphi}(t) + 8\pi^2 \theta_j^2 \tilde{\varphi}(t) - 4\pi^2 i \theta_j^2 \tilde{\varphi}(t) \right)$$

$$= \sum_j \left( \tilde{v}_{jj}(t) + 4\pi i \theta_j \partial_{x_j} \tilde{\varphi}(t) + 4\pi^2 \theta_j^2 \tilde{\varphi}(t) \right)$$

Rewriting it, one has

$$\sum_j \tilde{v}_{jj}(t) = \widetilde{\Delta \varphi}(t) = \Delta_x \tilde{\varphi}(t) - 4\pi^2 |\theta|^2 \tilde{\varphi}(t) - 4\pi i \theta \cdot \nabla_x \tilde{\varphi}(t) \text{ weak on } \mathcal{H}_\gamma$$

which is equation (A.4), and conjugating yields equation (A.5). □

*Remark* A.2. Alternatively, one could compute directly for $\varphi \in \mathscr{S}(\mathbb{R}^d; \mathbb{C})$ that

$$\widetilde{\Delta_x \varphi}(\theta, x) = \sum_{m \in \mathbb{Z}^d} e^{2\pi i \theta \cdot (x-m)} \Delta_x \varphi(x) = \sum_{j=1}^d \sum_{m \in \mathbb{Z}^d} e^{2\pi i \theta \cdot (x-m)} \partial_{x_j x_j}^2 \varphi(x)$$

$$= \sum_{j=1}^d \partial_{x_j} \left( \sum_{m \in \mathbb{Z}^d} e^{2\pi i \theta \cdot (x-m)} \partial_{x_j} \varphi(x) \right) - 2\pi i \sum_{j=1}^d \theta_j \sum_{m \in \mathbb{Z}^d} e^{2\pi i \theta \cdot (x-m)} \partial_{x_j} \varphi(x)$$

$$= \sum_{j=1}^d \partial_{x_j x_j}^2 \left( \sum_{m \in \mathbb{Z}^d} e^{2\pi i \theta \cdot (x-m)} \varphi(x) \right) - 2\pi i \sum_{j=1}^d \theta_j \partial_{x_j} \left( \sum_{m \in \mathbb{Z}^d} e^{2\pi i \theta \cdot (x-m)} \varphi(x) \right)$$

$$- 2\pi i \sum_{j=1}^d \theta_j \partial_{x_j} \left( \sum_{m \in \mathbb{Z}^d} e^{2\pi i \theta \cdot (x-m)} \varphi(x) \right) - 4\pi^2 \sum_{j=1}^d \theta_j^2 \sum_{m \in \mathbb{Z}^d} e^{2\pi i \theta \cdot (x-m)} \varphi(x)$$

$$= \sum_{j=1}^d \partial_{x_j x_j}^2 \tilde{\varphi}(\theta, x) - 4\pi i \sum_{j=1}^d \theta_j \partial_{x_j} \tilde{\varphi}(\theta, x) - 4\pi^2 \sum_{j=1}^d \theta_j^2 \tilde{\varphi}(\theta, x)$$

$$= \Delta_x \tilde{\varphi}(\theta, x) - 4\pi i \theta \cdot \nabla_x \tilde{\varphi}(\theta, x) - 4\pi^2 |\theta|^2 \tilde{\varphi}(\theta, x)$$

and make a density argument using the unitarity of the BFZ transform. One might shorten the computation for the second derivative by recursively using the computation for the first derivative.

## B Proofs of auxilliary results

### B.1 Proof of Lemma 2.3

In the proof of Lemma 2.3 below, we will use several properties of the BFZ transform that we have listed in Appendix A.

PROOF. For $x, k \in \mathbb{R}^d, \varphi \in \mathscr{S}(\mathbb{R}^d; \mathbb{C})$

$$W_\varphi(x, k) = \int_{\mathbb{R}^d} dy \, e^{2\pi i k \cdot y} \varphi\left(x - \frac{y}{2}\right) \varphi^*\left(x + \frac{y}{2}\right)$$

We now regularize this expression. This is

$$= \int_{\mathbb{R}^d} dy \lim_{\varepsilon \to 0} e^{2\pi i k \cdot y} e^{-\pi \varepsilon |y|^2} \varphi\left(x - \frac{y}{2}\right) \varphi^*\left(x + \frac{y}{2}\right)$$

by the dominated convergence theorem, this is

$$= \lim_{\varepsilon \to 0} \int_{\mathbb{R}^d} dy\, e^{2\pi i k \cdot y} e^{-\pi \varepsilon |y|^2} \varphi\left(x - \frac{y}{2}\right) \varphi^*\left(x + \frac{y}{2}\right)$$

Plugging in expression (A.2) and its complex conjugate into this equation, this is

$$= \lim_{\varepsilon \to 0} \int_{\mathbb{R}^d} dy \int_{\mathbb{T}^d} d\theta \int_{\mathbb{T}^d} d\theta'\, e^{2\pi i k \cdot y} e^{-\pi \varepsilon |y|^2}$$

$$e^{-2\pi i \theta \cdot \left(x - \frac{y}{2}\right)} \tilde{\varphi}\left(\theta, x - \frac{y}{2}\right) e^{2\pi i \theta' \cdot \left(x + \frac{y}{2}\right)} \tilde{\varphi}^*\left(\theta', x + \frac{y}{2}\right)$$

$$= \lim_{\varepsilon \to 0} \int_{\mathbb{R}^d} dy \int_{\mathbb{T}^d} d\theta \int_{\mathbb{T}^d} d\theta'\, e^{2\pi i \left(k + \frac{\theta}{2} + \frac{\theta'}{2}\right) \cdot y} e^{-\pi \varepsilon |y|^2} e^{-2\pi i (\theta - \theta') \cdot x} \tilde{\varphi}\left(\theta, x - \frac{y}{2}\right) \tilde{\varphi}^*\left(\theta', x + \frac{y}{2}\right)$$

We decompose $\mathbb{R}^d \ni y = y + m \in [-1,1]^d + (2\mathbb{Z})^d$ to get

$$= \lim_{\varepsilon \to 0} \int_{[-1,1]^d} dy \sum_{m \in (2\mathbb{Z})^d} \int_{\mathbb{T}^d} d\theta \int_{\mathbb{T}^d} d\theta'\, e^{2\pi i \left(k + \frac{\theta}{2} + \frac{\theta'}{2}\right) \cdot (y + m)} e^{-\pi \varepsilon |y + m|^2} e^{-2\pi i (\theta - \theta') \cdot x}$$

$$\tilde{\varphi}\left(\theta, x - \frac{y}{2} - \frac{m}{2}\right) \tilde{\varphi}^*\left(\theta', x + \frac{y}{2} + \frac{m}{2}\right)$$

which, by the $\mathbb{Z}^d$-periodicity of the BFZ transform in the second variable, and a change of variables, is

$$= \lim_{\varepsilon \to 0} \int_{\left[-\frac{1}{2}, \frac{1}{2}\right]^d} dy \sum_{m \in \mathbb{Z}^d} \int_{\mathbb{T}^d} d\theta \int_{\mathbb{T}^d} d\theta'\, e^{2\pi i \left(k + \frac{\theta}{2} + \frac{\theta'}{2}\right) \cdot (2y + 2m)} e^{-\pi \varepsilon |2y + 2m|^2} e^{-2\pi i (\theta - \theta') \cdot x}$$

$$\tilde{\varphi}(\theta, x - y) \tilde{\varphi}^*(\theta', x + y)$$

Since we introduced the regularization, Fubini's theorem says that this is

$$= \lim_{\varepsilon \to 0} \int_{\left[-\frac{1}{2}, \frac{1}{2}\right]^d} dy \int_{\mathbb{T}^d} d\theta \int_{\mathbb{T}^d} d\theta' \sum_{m \in \mathbb{Z}^d} e^{2\pi i (2k + \theta + \theta') \cdot (y + m)} e^{-4\pi \varepsilon |y + m|^2} e^{-2\pi i (\theta - \theta') \cdot x}$$

$$\tilde{\varphi}(\theta, x - y) \tilde{\varphi}^*(\theta', x + y)$$

We let $g(z) = e^{-4\pi \varepsilon |z|^2}$, $h(m) = e^{2\pi i (2k + \theta + \theta') \cdot z} g(z)$. Let $f(z) = \tau_{-y} h(z)$, where $\tau_{-y} F(x) := F(x + y)$. Then $\hat{f}(\xi) = e^{2\pi i y \cdot \xi} \hat{h}(\xi)$ and

$$\hat{h}(\xi) = \int_{\mathbb{R}^d} e^{-2\pi i \xi \cdot z} e^{2\pi i (2k + \theta + \theta') \cdot z} g(z) dz = \hat{g}(\xi - 2k - \theta - \theta')$$

Using Lemma C.1, we have that this is

$$= \frac{1}{(4\varepsilon)^{\frac{d}{2}}} e^{-\frac{\pi |\xi - 2k - \theta - \theta'|^2}{4\varepsilon}}$$

Hence

$$\hat{f}(\xi) = \frac{1}{(4\varepsilon)^{\frac{d}{2}}} e^{-\frac{\pi|\xi - 2k - \theta - \theta'|^2}{4\varepsilon}} e^{2\pi i y \cdot \xi}$$

Putting this together with the Poisson summation formula (Lemma C.2) we get that

$$W_\varphi(x,k) = \lim_{\varepsilon \to 0} \int_{[-\frac{1}{2},\frac{1}{2}]^d} dy \int_{\mathbb{T}^d} d\theta \int_{\mathbb{T}^d} d\theta' \sum_{m \in \mathbb{Z}^d} e^{2\pi i y \cdot m} \frac{e^{-\frac{\pi|m - 2k - \theta - \theta'|^2}{4\varepsilon}}}{(4\varepsilon)^{d/2}}$$

$$e^{-2\pi i (\theta - \theta') \cdot x} \tilde{\varphi}(\theta, x - y) \tilde{\varphi}^*(\theta', x + y)$$

Notice that as $\varepsilon \to 0$, the Gaussian better approximates the delta and we expect only certain terms to survive in the limit. To see this rigorously, we split the sum over $m$ as

$$= \lim_{\varepsilon \to 0} \int_{[-\frac{1}{2},\frac{1}{2}]^d} dy \int_{[-\frac{1}{2},\frac{1}{2}]^d} d\theta \int_{[-\frac{1}{2},\frac{1}{2}]^d} d\theta' \, \tilde{\varphi}(\theta, x - y) \tilde{\varphi}^*(\theta', x + y)$$

$$\sum_{m \in \mathbb{Z}^d} e^{2\pi i y \cdot m} \frac{e^{-\frac{\pi|m - 2k - \theta - \theta'|^2}{4\varepsilon}}}{(4\varepsilon)^{d/2}} e^{-2\pi i (\theta - \theta') \cdot x} \mathbb{I}_{[-1,1]^d + 2k + \theta}(m)$$

$$+ \lim_{\varepsilon \to 0} \int_{[-\frac{1}{2},\frac{1}{2}]^d} dy \int_{[-\frac{1}{2},\frac{1}{2}]^d} d\theta \int_{[-\frac{1}{2},\frac{1}{2}]^d} d\theta' \, \tilde{\varphi}(\theta, x - y) \tilde{\varphi}^*(\theta', x + y)$$

$$\sum_{m \in \mathbb{Z}^d} e^{2\pi i y \cdot m} \frac{e^{-\frac{\pi|m - 2k - \theta - \theta'|^2}{4\varepsilon}}}{(4\varepsilon)^{d/2}} e^{-2\pi i (\theta - \theta') \cdot x} \left(1 - \mathbb{I}_{[-1,1]^d + 2k + \theta}(m)\right)$$

Since $\sum_{m \in \mathbb{Z}^d} \frac{e^{-\frac{\pi|m - 2k - \theta - \theta'|^2}{4\varepsilon}}}{(4\varepsilon)^{d/2}} \left(1 - \mathbb{I}_{[-1,1]^d + 2k + \theta}(m)\right) \leqslant C$ uniformly in $\varepsilon \in (0,1)$ and $\theta, \theta' \in [-\frac{1}{2}, \frac{1}{2}]^d$, the second sum by dominated convergence, is

$$= \int_{[-\frac{1}{2},\frac{1}{2}]^d} dy \int_{[-\frac{1}{2},\frac{1}{2}]^d} d\theta \int_{[-\frac{1}{2},\frac{1}{2}]^d} d\theta' \, \tilde{\varphi}(\theta, x - y) \tilde{\varphi}^*(\theta', x + y)$$

$$\sum_{m \in \mathbb{Z}^d} e^{2\pi i y \cdot m} \lim_{\varepsilon \to 0} \frac{e^{-\frac{\pi|m - 2k - \theta - \theta'|^2}{4\varepsilon}}}{(4\varepsilon)^{d/2}} e^{-2\pi i (\theta - \theta') \cdot x} \left(1 - \mathbb{I}_{[-1,1]^d + 2k + \theta}(m)\right)$$

Let $c = \mathrm{dist}\left(\left\{[-\frac{1}{2},\frac{1}{2}]^d\right\}, ([-1,1]^d)^c\right)$, so $c > 0$, and the above term is bounded in absolute value by

$$\leq \int_{[-\frac{1}{2},\frac{1}{2}]^d} dy \int_{[-\frac{1}{2},\frac{1}{2}]^d} d\theta \int_{[-\frac{1}{2},\frac{1}{2}]^d} d\theta' \, |\tilde{\varphi}(\theta, x - y) \tilde{\varphi}^*(\theta', x + y)|$$

$$\sum_{m \in \mathbb{Z}^d} \underbrace{\lim_{\varepsilon \to 0} \frac{e^{-c\pi/4\varepsilon}}{(4\varepsilon)^{d/2}} \left(1 - \mathbb{I}_{[-1,1]^d + 2k + \theta}(m)\right)}_{=0} = 0$$

For a fixed $k, \theta$ the sum in the first term becomes a finite sum and hence we can use the linearity of the Lebesgue integral to write it as

$$\lim_{\varepsilon \to 0} \int_{\left[-\frac{1}{2},\frac{1}{2}\right]^d} dy \int_{\left[-\frac{1}{2},\frac{1}{2}\right]^d} d\theta \tilde{\varphi}(\theta, x-y) \sum_{m \in \mathbb{Z}^d} e^{2\pi i y \cdot m} \mathbb{I}_{[-1,1]^d + 2k + \theta}(m)$$

$$\int_{\left[-\frac{1}{2},\frac{1}{2}\right]^d} d\theta' \frac{e^{-\frac{\pi|m-2k-\theta-\theta'|^2}{4\varepsilon}}}{(4\varepsilon)^{d/2}} e^{-2\pi i (\theta - \theta') \cdot x} \tilde{\varphi}^*(\theta', x+y)$$

$$= \lim_{\varepsilon \to 0} \int_{\left[-\frac{1}{2},\frac{1}{2}\right]^d} dy \int_{\left[-\frac{1}{2},\frac{1}{2}\right]^d} d\theta \tilde{\varphi}(\theta, x-y) \sum_{m \in \mathbb{Z}^d} e^{2\pi i y \cdot m} \mathbb{I}_{[-1,1]^d + 2k + \theta}(m)$$

$$\int_{\mathbb{R}^d} d\theta' \frac{e^{-\frac{\pi|m-2k-\theta-\theta'|^2}{4\varepsilon}}}{(4\varepsilon)^{d/2}} e^{-2\pi i (\theta - \theta') \cdot x} \tilde{\varphi}^*(\theta', x+y) \mathbb{I}_{\left[-\frac{1}{2},\frac{1}{2}\right]^d}(\theta')$$

We make the following change of variables

$$\bar{\theta} = \frac{(m-2k-\theta-\theta')}{\sqrt{4\varepsilon\pi^{-1}}} \Rightarrow \theta' = m-2k-\theta - \sqrt{4\varepsilon\pi^{-1}}\bar{\theta}, \qquad d\theta' = -(4\varepsilon\pi^{-1})^{d/2} d\bar{\theta}$$

Hence the above expression is

$$= -\lim_{\varepsilon \to 0} \int_{\left[-\frac{1}{2},\frac{1}{2}\right]^d} dy \int_{\left[-\frac{1}{2},\frac{1}{2}\right]^d} d\theta \tilde{\varphi}(\theta, x-y) \sum_{m \in \mathbb{Z}^d} e^{2\pi i y \cdot m} \mathbb{I}_{[-1,1]^d + 2k + \theta}(m) \times$$

$$\times \int_{\mathbb{R}^d} \frac{d\bar{\theta}}{(\pi)^{d/2}} e^{-|\bar{\theta}|^2} e^{-2\pi i \left(2\theta - m + 2k + \sqrt{4\varepsilon\pi^{-1}}\bar{\theta}\right) \cdot x}$$

$$\tilde{\varphi}^*\left(m - 2k - \theta - \sqrt{4\varepsilon\pi^{-1}}\bar{\theta}, x+y\right) \mathbb{I}_{\left[-\frac{1}{2},\frac{1}{2}\right]^d + 2k + \theta - m}\left(-\sqrt{4\varepsilon\pi^{-1}}\bar{\theta}\right)$$

The continuity of $\tilde{\varphi}^*$ implies that $\sup_{\theta, x \in \left[-\frac{1}{2},\frac{1}{2}\right]^d} \tilde{\varphi}^*(\theta, x) \leq C$ and also allows us to use dominated convergence once more, to get that

$$W_\varphi(x,k) = -\int_{\left[-\frac{1}{2},\frac{1}{2}\right]^d} dy \int_{\left[-\frac{1}{2},\frac{1}{2}\right]^d} d\theta \tilde{\varphi}(\theta, x-y) \sum_{m \in \mathbb{Z}^d} e^{2\pi i y \cdot m} \mathbb{I}_{[-1,1]^d + 2k + \theta}(m)$$

$$\int_{\mathbb{R}^d} \frac{d\bar{\theta}}{(\pi)^{d/2}} e^{-|\bar{\theta}|^2} e^{-2\pi i (2\theta - m + 2k) \cdot x} \tilde{\varphi}^*(m - 2k - \theta, x+y) \mathbb{I}_{\left[-\frac{1}{2},\frac{1}{2}\right]^d + 2k + \theta}(m)$$

taking the product of the indicator functions, and using the fact that $\int_{\mathbb{R}^d} dx e^{-|x|^2} = (\pi)^{d/2}$, this is

$$= -\int_{\left[-\frac{1}{2},\frac{1}{2}\right]^d} dy \int_{\left[-\frac{1}{2},\frac{1}{2}\right]^d} d\theta \tilde{\varphi}(\theta, x-y) \sum_{m \in \mathbb{Z}^d} e^{2\pi i y \cdot m} \mathbb{I}_{\left[-\frac{1}{2},\frac{1}{2}\right]^d + 2k + \theta}(m)$$

$$e^{-2\pi i (2\theta - m + 2k) \cdot x} \tilde{\varphi}^*(m - 2k - \theta, x+y)$$

By $\mathbb{Z}^d$-quasiperiodicity of $\varphi^*$ in the first variable, this is

$$= -\int_{[-\frac{1}{2},\frac{1}{2}]^d} dy \int_{[-\frac{1}{2},\frac{1}{2}]^d} d\theta \tilde{\varphi}(\theta, x-y) \sum_{m \in \mathbb{Z}^d} e^{2\pi i y \cdot m} \mathbb{I}_{[-\frac{1}{2},\frac{1}{2}]^d + 2k+\theta}(m)$$

$$e^{-2\pi i(2\theta - m + 2k) \cdot x} \tilde{\varphi}^*(-2k-\theta, x+y) e^{-2\pi i m \cdot (x+y)}$$

$$= -\int_{[-\frac{1}{2},\frac{1}{2}]^d} dy \int_{[-\frac{1}{2},\frac{1}{2}]^d} d\theta \tilde{\varphi}(\theta, x-y) e^{-4\pi i(\theta+k) \cdot x} \tilde{\varphi}^*(-2k-\theta, x+y)$$

$$\sum_{m \in \mathbb{Z}^d} \mathbb{I}_{[-\frac{1}{2},\frac{1}{2}]^d + 2k+\theta}(m)$$

since $\sum_{m \in \mathbb{Z}^d} \mathbb{I}_{[-\frac{1}{2},\frac{1}{2}]^d + 2k+\theta}(m) = 1, \theta$ a.s. this is

$$= -\int_{[-\frac{1}{2},\frac{1}{2}]^d} dy \int_{[-\frac{1}{2},\frac{1}{2}]^d} d\theta \tilde{\varphi}(\theta, x-y) e^{-4\pi i(\theta+k) \cdot x} \tilde{\varphi}^*(-2k-\theta, x+y)$$

by the $\mathbb{Z}^d$-periodicity of the integrands in $y$ and $\theta$, this is

$$= -\int_{\mathbb{T}^d} dy \int_{\mathbb{T}^d} d\theta \tilde{\varphi}(\theta, x-y) e^{-4\pi i(\theta+k) \cdot x} \tilde{\varphi}^*(-2k-\theta, x+y)$$

Now by splitting $k = \kappa + [\![k]\!] \in \left(\frac{\mathbb{Z}}{2}\right)^d + \left[-\frac{1}{4}, \frac{1}{4}\right)^d$ we have

$$W_\varphi(x,k) = -\int_{\mathbb{T}^d} dy \int_{\mathbb{T}^d} d\theta e^{-4\pi i(\theta+\kappa+[\![k]\!]) \cdot x} \tilde{\varphi}(\theta, x-y) \tilde{\varphi}^*(-2\kappa - 2[\![k]\!] - \theta, x+y)$$

using $\mathbb{Z}^d$-quasiperiodicity in the first variable of $\varphi^*$ once more

$$= -\int_{\mathbb{T}^d} dy \int_{\mathbb{T}^d} d\theta e^{-4\pi i(\theta+[\![k]\!]) \cdot x} e^{4\pi i \kappa \cdot y} \tilde{\varphi}(\theta, x-y) \tilde{\varphi}^*(-2[\![k]\!] - \theta, x+y)$$

and finally by shifting by $-[\![k]\!]$ in the $\theta$-variable

$$= -\int_{\mathbb{T}^d} dy \int_{\mathbb{T}^d} d\theta e^{-4\pi i \theta \cdot x} e^{4\pi i \kappa \cdot y} \tilde{\varphi}(-[\![k]\!] + \theta, x-y) \tilde{\varphi}^*(-[\![k]\!] - \theta, x+y)$$

We conclude by a change of variables that

$$W_\varphi(x,k) = \int_{\mathbb{T}^d} dy \int_{\mathbb{T}^d} d\theta e^{-4\pi i \theta \cdot x} e^{-4\pi i \kappa \cdot y} \tilde{\varphi}(-[\![k]\!] + \theta, x+y) \tilde{\varphi}^*(-[\![k]\!] - \theta, x-y) \qquad \square$$

*Remark* B.1. By another change of variables, we see that

$$W_\varphi(x,k) = 2^{2d} \int_{\mathbb{T}^d} dy \int_{\mathbb{T}^d} d\theta e^{2\pi i \theta \cdot x} e^{2\pi i \kappa \cdot y} \tilde{\varphi}\left(-[\![k]\!] - \frac{\theta}{2}, x - \frac{y}{2}\right) \tilde{\varphi}^*\left(-[\![k]\!] + \frac{\theta}{2}, x + \frac{y}{2}\right)$$

and in this form, the similarity to the usual Wigner transform is even more apparent.

## B.2 Proof of Proposition 2.5

Similar to the usual $L^\infty$ estimate on the Wigner function (see [24]) we have an a-priori bound on the Bloch–Wigner function. Recall definitions (2.5) and (2.6).

LEMMA B.2. *For $\varphi \in L^2(\mathbb{R}^d; \mathbb{C})$ one has that $\tilde{W}_\varphi \in L^\infty_{\eta,p,\kappa,z}$*

PROOF. We compute

$$|\tilde{W}_\varphi(z,p,\eta,\kappa)| = \left| \int_{\mathbb{T}^d} dy \int_{\left[-\frac{1}{2},\frac{1}{2}\right]^d} d\theta\, e^{-4\pi i\theta \cdot p} e^{-4\pi i\kappa \cdot y} \tilde{\varphi}(\eta+\theta, z+y)\tilde{\varphi}^*(\eta-\theta, z-y) \right|$$

$$\leq \int_{\mathbb{T}^d} dy \int_{\left[-\frac{1}{2},\frac{1}{2}\right]^d} d\theta\, |\tilde{\varphi}(\eta+\theta, z+y)\tilde{\varphi}^*(\eta-\theta, z-y)|$$

by using Cauchy–Schwartz inequality, this is

$$\leq \left( \int_{\mathbb{T}^d} dy \int_{\left[-\frac{1}{2},\frac{1}{2}\right]^d} d\theta\, |\tilde{\varphi}(\eta+\theta, z+y)|^2 \right)^{1/2} \left( \int_{\mathbb{T}^d} dy \int_{\left[-\frac{1}{2},\frac{1}{2}\right]^d} d\theta\, |\tilde{\varphi}^*(\eta-\theta, z-y)|^2 \right)^{1/2}$$

using equation (A.1), this is

$$= \left( \int_{\mathbb{T}^d} dy \int_{\mathbb{T}^d} d\theta\, |\tilde{\varphi}(\eta+\theta, z+y)|^2 \right)^{1/2} \left( \int_{\mathbb{T}^d} dy \int_{\mathbb{T}^d} d\theta\, |\tilde{\varphi}^*(\eta-\theta, z-y)|^2 \right)^{1/2}$$

Hence we can shift in the $\theta$ and $y$ variables and use the fact that $\mathcal{U}_{\text{BFZ}}$ is a unitary transformation, to have that this is

$$= \left( \int_{\mathbb{T}^d} dy \int_{\mathbb{T}^d} d\theta\, |\tilde{\varphi}(\theta,y)|^2 \right)^{1/2} \left( \int_{\mathbb{T}^d} dy \int_{\mathbb{T}^d} d\theta\, |\tilde{\varphi}^*(\theta,y)|^2 \right)^{1/2} = \|\tilde{\varphi}\|^2_{\mathcal{H}_\gamma} = \|\varphi\|^2_{L^2(\mathbb{R}^d)}$$

□

Hence

$$\tilde{W}_\varphi(z,p,\eta,\kappa) \in L^\infty_{\eta,p,\kappa,z} \subset L^\infty_{\eta,p,\kappa} L^2_z \tag{B.1}$$

We will now consider now $\tilde{W}_\varphi(t,z,p,\eta,\kappa)$ associated to $\varphi(t,x)$, the solution of the Schrödinger equation (1.1).

First, we define a Hermitian form

$$F[\tilde{\varphi}, \tilde{\psi}](z,p,\eta,\kappa) := \int_{\mathbb{T}^d} dy \int_{\left[-\frac{1}{2},\frac{1}{2}\right]^d} d\theta\, e^{-4\pi i\theta \cdot p} e^{-4\pi i\kappa \cdot y} \tilde{\varphi}(\eta+\theta, x+y)\tilde{\psi}^*(\eta-\theta, x-y)$$

and

$$G[\tilde{\varphi}](z,p,\eta,\kappa) := F[\tilde{\varphi}, \tilde{\varphi}](z,p,\eta,\kappa)$$

LEMMA B.3. *$F: \mathcal{H}_\gamma \times \mathcal{H}_\gamma \to L^\infty_{\eta,p,\kappa,z}$ and $G: \mathcal{H}_\gamma \to L^\infty_{\eta,p,\kappa,z}$ are continuous maps.*

PROOF. By picking $\tilde{\varphi}, \tilde{\psi}, \tilde{\varphi}_1, \tilde{\psi}_1 \in \mathcal{H}_f$ one has that

$$F[\tilde{\varphi}, \tilde{\psi}](z, p, \eta, \kappa) - F[\tilde{\varphi}_1, \tilde{\psi}_1](z, p, \eta, \kappa)$$

$$= \int_{\mathbb{T}^d} dy \int_{\left[-\frac{1}{2}, \frac{1}{2}\right]^d} d\theta e^{-4\pi i \theta \cdot p} e^{-4\pi i \kappa \cdot y} \tilde{\varphi}(\eta + \theta, z + y) \tilde{\psi}^*(\eta - \theta, z - y)$$

$$- \int_{\mathbb{T}^d} dy \int_{\left[-\frac{1}{2}, \frac{1}{2}\right]^d} d\theta e^{-4\pi i \theta \cdot p} e^{-4\pi i \kappa \cdot y} \tilde{\varphi}_1(\eta + \theta, z + y) \tilde{\psi}_1^*(\eta - \theta, z - y)$$

$$\pm \int_{\mathbb{T}^d} dy \int_{\left[-\frac{1}{2}, \frac{1}{2}\right]^d} d\theta e^{-4\pi i \theta \cdot p} e^{-4\pi i \kappa \cdot y} \tilde{\varphi}_1(\eta + \theta, z + y) \tilde{\psi}^*(\eta - \theta, z - y)$$

by rearranging the terms, this is

$$= \int_{\mathbb{T}^d} dy \int_{\left[-\frac{1}{2}, \frac{1}{2}\right]^d} d\theta e^{-4\pi i \theta \cdot p} e^{-4\pi i \kappa \cdot y} (\tilde{\varphi} - \tilde{\varphi}_1)(\eta + \theta, z + y) \tilde{\psi}^*(\eta - \theta, z - y)$$

$$+ \int_{\mathbb{T}^d} dy \int_{\left[-\frac{1}{2}, \frac{1}{2}\right]^d} d\theta e^{-4\pi i \theta \cdot p} e^{-4\pi i \kappa \cdot y} \tilde{\varphi}_1(\eta + \theta, z + y)(\tilde{\psi}^* - \tilde{\psi}_1^*)(\eta - \theta, z - y)$$

Hence

$$|F[\tilde{\varphi}, \tilde{\psi}](z, p, \eta, \kappa) - F[\tilde{\varphi}_1, \tilde{\psi}_1](z, p, \eta, \kappa)|$$

$$\leqslant \|(\tilde{\varphi} - \tilde{\varphi}_1)(\eta + \cdot, z + \cdot) \tilde{\psi}^*(\eta - \cdot, z - \cdot)\|_{L^1\left(\left[-\frac{1}{2}, \frac{1}{2}\right]^d \times \mathbb{T}^d\right)}$$

$$+ \|\tilde{\varphi}_1(\eta + \cdot, z + \cdot)(\tilde{\psi}^* - \tilde{\psi}_1^*)(\eta - \cdot, z - \cdot)\|_{L^1\left(\left[-\frac{1}{2}, \frac{1}{2}\right]^d \times \mathbb{T}^d\right)}$$

by the Cauchy–Schwartz inequality and A.1 this is

$$\leqslant \|(\tilde{\varphi} - \tilde{\varphi}_1)(\eta + \cdot, z + \cdot)\|_{L^2(\mathbb{T}^d \times \mathbb{T}^d)} \|\tilde{\psi}^*(\eta - \cdot, z - \cdot)\|_{L^2(\mathbb{T}^d \times \mathbb{T}^d)}$$

$$+ \|\tilde{\varphi}_1(\eta + \cdot, z + \cdot)\|_{L^2(\mathbb{T}^d \times \mathbb{T}^d)} \|(\tilde{\psi}^* - \tilde{\psi}_1^*)(\eta - \cdot, z - \cdot)\|_{L^2(\mathbb{T}^d \times \mathbb{T}^d)}$$

And by the shift invariance of the $L^2$ norm, this is

$$\leqslant \|\tilde{\varphi} - \tilde{\varphi}_1\|_{\mathcal{H}_\gamma} \|\tilde{\psi}^*\|_{\mathcal{H}_\gamma} + \|\tilde{\varphi}_1\|_{\mathcal{H}_\gamma} \|\tilde{\psi}^* - \tilde{\psi}_1^*\|_{\mathcal{H}_f}$$

And one can deduce continuity of the map from here, and also for $G[\tilde{\varphi}]$, by replacing $\tilde{\psi}$ and $\tilde{\psi}_1$ by $\tilde{\varphi}$ and $\tilde{\varphi}_1$ respectively in the above computations. □

PROOF. (of Proposition 2.5) One can compute for $\varphi \in C(\mathbb{R}; H^2(\mathbb{R}^d; \mathbb{C})) \cap C^1(\mathbb{R}; L^2(\mathbb{R}^d; \mathbb{C}))$:

$$\frac{\tilde{W}_\varphi(t+h) - \tilde{W}_\varphi(t)}{h} = \frac{F[\tilde{\varphi}(t+h), \tilde{\varphi}(t+h)] - F[\tilde{\varphi}(t), \tilde{\varphi}(t)]}{h} \pm \frac{F[\tilde{\varphi}(t), \tilde{\varphi}(t+h)]}{h}$$

since $F$ is a Hermitian form by Lemma B.3, this is

$$= F\left[\frac{(\tilde{\varphi}(t+h) - \tilde{\varphi}(t))}{h}, \tilde{\varphi}(t+h)\right] + F\left[\tilde{\varphi}(t), \frac{\tilde{\varphi}(t+h) - \tilde{\varphi}(t)}{h}\right]$$

By the continuity of $F$ one has

$$\partial_t \tilde{W}_\varphi(t) = \lim_{h\to 0} \frac{\tilde{W}_\varphi(t+h) - \tilde{W}_\varphi(t)}{h}$$

$$= F[\partial_t \tilde{\varphi}(t), \tilde{\varphi}(t)] + F[\tilde{\varphi}(t), \partial_t \tilde{\varphi}(t)]$$

Thus, we have

$$\partial_t \tilde{W}_\varphi(t) = F[\partial_t \tilde{\varphi}(t), \tilde{\varphi}(t)] + F[\tilde{\varphi}(t), \partial_t \tilde{\varphi}(t)] \tag{B.2}$$

Similarly, by using Lemma A.5, one has that,

$$\partial_{z_j} \tilde{W}_\varphi(t) = F[\partial_{z_j} \tilde{\varphi}(t), \tilde{\varphi}(t)] + F[\tilde{\varphi}(t), \partial_{z_j} \tilde{\varphi}(t)]$$

and

$$\partial_{z_i z_j} \tilde{W}_\varphi(t) = F[\partial_{z_i z_j} \tilde{\varphi}(t), \tilde{\varphi}(t)] + F[\tilde{\varphi}(t), \partial_{z_i z_j} \tilde{\varphi}(t)]$$
$$+ F[\partial_{z_i} \tilde{\varphi}(t), \partial_{z_j} \tilde{\varphi}(t)] + F[\partial_{z_j} \tilde{\varphi}(t), \partial_{z_i} \tilde{\varphi}(t)]$$

Since $\tilde{\varphi} \in C^1(\mathbb{R}; \mathcal{H}_f) \cap C(\mathbb{R}; L^2_{\text{loc}}(\mathbb{R}^d; H^2(\mathbb{T}^d; \mathbb{C})))$ one has using Lemma B.3 that for $i, j \in \{1, \ldots, d\}$ and $t \in \mathbb{R}$

$$\partial_t \tilde{W}_\varphi, \partial_{z_j} \tilde{W}_\varphi, \partial_{z_i z_j} \tilde{W}_\varphi \in C(\mathbb{R}; L^\infty_{z,p,\eta,\kappa})$$

and furthemore

$$|F[V\tilde{\varphi}(t), \tilde{\varphi}(t)](z, p, \eta, \kappa)|$$

$$= \left| \int_{\mathbb{T}^d} dy \int_{\left[-\frac{1}{2}, \frac{1}{2}\right]^d} d\theta e^{-4\pi i \theta \cdot p} e^{-4\pi i \kappa \cdot y} V(z+y) \tilde{\varphi}(\eta+\theta, z+y) \tilde{\varphi}^*(\eta-\theta, z-y) \right|$$

$$\leq \|V\|_{L^\infty(\mathbb{R}^d)} \|\tilde{W}_\varphi\|_{L^\infty_{z,p,\eta,\kappa}} \leq \|V\|_{L^\infty} \|\tilde{\varphi}\|^2_{\mathcal{H}_\gamma}$$

and a similar bound holds for $|F[\tilde{\varphi}(t), V\tilde{\varphi}(t)](z, p, \eta, \kappa)|$. Hence also

$$F[V\tilde{\varphi}, \tilde{\varphi}], F[\tilde{\varphi}, V\tilde{\varphi}] \in C(\mathbb{R}; L^\infty_{z,p,\eta,\kappa})$$

Now, assume that $\varphi \in C(\mathbb{R}; H^2(\mathbb{R}^d; \mathbb{C})) \cap C^1(\mathbb{R}; L^2(\mathbb{R}^d; \mathbb{C}))$ satisfies the Schrödinger equation (1.1). We have that by applying the BFZ transform to both sides of the Schrödinger equation that

$$i\widetilde{\partial_t \varphi} = -\widetilde{\Delta \varphi} + \varepsilon^{1/2} \widetilde{V\varphi}$$

Multiplying by $-i$ on both sides

$$\widetilde{\partial_t \varphi} = i\widetilde{\Delta \varphi} - i\varepsilon^{1/2} \widetilde{V\varphi}$$

and using lemma A.5 and

$$\widetilde{V\varphi}(\theta, x) = \sum_{m \in \mathbb{Z}^d} e^{2\pi i \theta \cdot (x-m)} V(x-m) \varphi(t, x-m)$$

$$= V(x) \sum_{m \in \mathbb{Z}^d} e^{2\pi i \theta \cdot (x-m)} \varphi(t, x-m) = V(x) \tilde{\varphi}(\theta, x)$$

we get that
$$\partial_t \tilde{\varphi}(t, \theta, x) = i\widetilde{\Delta \varphi}(t, \theta, x) - i\varepsilon^{1/2} V(x) \tilde{\varphi}(t, \theta, x)$$

Plugging this into equation B.2 one has
$$\partial_t \tilde{W}_\varphi(t) = F[\partial_t \tilde{\varphi}(t), \tilde{\varphi}(t)] + F[\tilde{\varphi}(t), \partial_t \tilde{\varphi}(t)]$$
$$= F[i\widetilde{\Delta \varphi}(t), \tilde{\varphi}(t)] + F[-i\varepsilon^{1/2} V\tilde{\varphi}(t), \tilde{\varphi}(t)] + F[\tilde{\varphi}(t), i\widetilde{\Delta \varphi}(t)] + F[\tilde{\varphi}(t), -i\varepsilon^{1/2} V\tilde{\varphi}(t)]$$

since $F$ is a Hermitian form, this is
$$= i[F[\widetilde{\Delta \varphi}(t), \tilde{\varphi}(t)] - F[\tilde{\varphi}(t), \widetilde{\Delta \varphi}(t)]] + i\varepsilon^{1/2}[F[\tilde{\varphi}(t), V\tilde{\varphi}(t)] - F[V\tilde{\varphi}(t), \tilde{\varphi}(t)]]$$

The terms with the potential can be computed first
$$i\varepsilon^{1/2}[F[\tilde{\varphi}(t), V\tilde{\varphi}(t)] - F[V\tilde{\varphi}(t), \tilde{\varphi}(t)]](z, p, \eta, \kappa)$$
$$= i\varepsilon^{1/2} \int_{\mathbb{T}^d} dy \int_{[-\frac{1}{2},\frac{1}{2}]^d} d\theta e^{-4\pi i\theta \cdot p} e^{-4\pi i\kappa \cdot y} \tilde{\varphi}(\eta + \theta, z + y) \tilde{\varphi}^*(\eta - \theta, z - y)$$
$$[V(z - y) - V(z + y)]$$
$$= i\varepsilon^{1/2} \int_{\mathbb{T}^d} dy \int_{[-\frac{1}{2},\frac{1}{2}]^d} d\theta e^{-4\pi i\theta \cdot p} e^{-4\pi i\kappa \cdot y} \tilde{\varphi}(\eta + \theta, z + y) \tilde{\varphi}^*(\eta - \theta, z - y)$$
$$\left[ \sum_{n \in \mathbb{Z}^d} e^{2\pi i n \cdot (z-y)} \hat{V}(n) - e^{2\pi i n \cdot (z+y)} \hat{V}(n) \right]$$

by using Fubini's theorem to interchange the sum and integrals, this is
$$= i\varepsilon^{1/2} \sum_{n \in \mathbb{Z}^d} e^{2\pi i n \cdot z} \hat{V}(n) \int_{\mathbb{T}^d} dy \int_{[-\frac{1}{2},\frac{1}{2}]^d} d\theta e^{-4\pi i\theta \cdot p} \tilde{\varphi}(\eta + \theta, z + y) \tilde{\varphi}^*(\eta - \theta, z - y)$$
$$\left[ e^{-4\pi i \left(\kappa + \frac{n}{2}\right) \cdot y} - e^{-4\pi i \left(\kappa - \frac{n}{2}\right) \cdot y} \right]$$
$$= i\varepsilon^{1/2} \sum_{n \in \mathbb{Z}^d} e^{2\pi i n \cdot z} \hat{V}(n) \left[ \tilde{W}_\varphi\left(t, z, p, \eta, \kappa + \frac{n}{2}\right) - \tilde{W}_\varphi\left(t, z, p, \eta, \kappa - \frac{n}{2}\right) \right]$$

Now, consider the terms in the Laplacian,
$$i[F[\widetilde{\Delta \varphi}(t), \tilde{\varphi}(t)] - F[\tilde{\varphi}(t), \widetilde{\Delta \varphi}(t)]](z, p, \eta, \kappa)$$

we temporarily leave out $t$ out and shorten $\tilde{\varphi}(\eta + \theta, z + y)$ to $\tilde{\varphi}$ and $\tilde{\varphi}^*(\eta - \theta, z - y)$ to $\tilde{\varphi}^*$ to improve legibility in the computations below
$$= i \int_{\mathbb{T}^d} dy \int_{[-\frac{1}{2},\frac{1}{2}]^d} d\theta e^{-4\pi i\theta \cdot p} e^{-4\pi i\kappa \cdot y} \widetilde{\Delta \varphi}(\eta + \theta, z + y) \tilde{\varphi}^*$$
$$- i \int_{\mathbb{T}^d} dy \int_{[-\frac{1}{2},\frac{1}{2}]^d} d\theta e^{-4\pi i\theta \cdot p} e^{-4\pi i\kappa \cdot y} \tilde{\varphi} \widetilde{\Delta \varphi}^*(\eta - \theta, z - y)$$

one uses equations (A.4) and (A.5) to get

$$= i\int_{\mathbb{T}^d} dy \int_{[-\frac{1}{2},\frac{1}{2}]^d} d\theta e^{-4\pi i\theta \cdot p} e^{-4\pi i\kappa \cdot y} \tilde{\varphi}^*[(\Delta_z - 4\pi^2|\eta+\theta|^2 - 4\pi i(\eta+\theta)\cdot\nabla_z)\tilde{\varphi}]$$

$$- i\int_{\mathbb{T}^d} dy \int_{[-\frac{1}{2},\frac{1}{2}]^d} d\theta e^{-4\pi i\theta \cdot p} e^{-4\pi i\kappa \cdot y} \tilde{\varphi}[(\Delta_z - 4\pi^2|\eta-\theta|^2 + 4\pi i(\eta-\theta)\cdot\nabla_z)\tilde{\varphi}^*]$$

Splitting this into three parts, one has

$$i[F[\widetilde{\Delta\varphi},\tilde{\varphi}] - F[\tilde{\varphi},\widetilde{\Delta\varphi}]](z,p,\eta,\kappa) = A_1 + A_2 + A_3$$

where

$$A_1 = -4\pi^2 i \int_{\mathbb{T}^d} dy \int_{[-\frac{1}{2},\frac{1}{2}]^d} d\theta e^{-4\pi i\theta\cdot p} e^{-4\pi i\kappa\cdot y}[|\eta+\theta|^2 - |\eta-\theta|^2]\tilde{\varphi}\tilde{\varphi}^*$$

$$A_2 = i\int_{\mathbb{T}^d} dy \int_{[-\frac{1}{2},\frac{1}{2}]^d} d\theta e^{-4\pi i\theta\cdot p} e^{-4\pi i\kappa\cdot y}[\Delta_z\tilde{\varphi}\tilde{\varphi}^* - \tilde{\varphi}\Delta_z\tilde{\varphi}^*]$$

$$A_3 = 4\pi \int_{\mathbb{T}^d} dy \int_{[-\frac{1}{2},\frac{1}{2}]^d} d\theta e^{-4\pi i\theta\cdot p} e^{-4\pi i\kappa\cdot y}(((\eta+\theta)\cdot\nabla_z\tilde{\varphi})\tilde{\varphi}^* + ((\eta-\theta)\cdot\nabla_z\tilde{\varphi}^*)\tilde{\varphi})$$

One can compute that

$$A_1 = -4\pi^2 i \int_{\mathbb{T}^d} dy \int_{[-\frac{1}{2},\frac{1}{2}]^d} d\theta e^{-4\pi i\theta\cdot p} e^{-4\pi i\kappa\cdot y}[|\eta|^2 + |\theta|^2 + 2\eta\cdot\theta - |\eta|^2 - |\theta|^2 + 2\eta\cdot\theta]\tilde{\varphi}\tilde{\varphi}^*$$

$$= 4\pi \int_{\mathbb{T}^d} dy \int_{[-\frac{1}{2},\frac{1}{2}]^d} d\theta e^{-4\pi i\theta\cdot p} e^{-4\pi i\kappa\cdot y}(-4\pi i\eta\cdot\theta)\tilde{\varphi}\tilde{\varphi}^*$$

$$= 4\pi \int_{\mathbb{T}^d} dy \int_{[-\frac{1}{2},\frac{1}{2}]^d} d\theta \eta\cdot\nabla_p(e^{-4\pi i\theta\cdot p}) e^{-4\pi i\kappa\cdot y}\tilde{\varphi}\tilde{\varphi}^*$$

$$= 4\pi\eta\cdot\nabla_p\tilde{W}_\varphi(t,z,p,\eta,\kappa)$$

Next, for the term

$$A_2 = i\int_{\mathbb{T}^d} dy \int_{[-\frac{1}{2},\frac{1}{2}]^d} d\theta e^{-4\pi i\theta\cdot p} e^{-4\pi i\kappa\cdot y}[\Delta_z\tilde{\varphi}\tilde{\varphi}^* - \tilde{\varphi}\Delta_z\tilde{\varphi}^*]$$

we note that

$$\nabla_z\tilde{\varphi} = \nabla_z\tilde{\varphi}(\eta+\theta, z+y) = \nabla_y\tilde{\varphi}(\eta+\theta, z+y)$$

and similarly

$$\nabla_z\tilde{\varphi}^* = -\nabla_y\tilde{\varphi}^*, \qquad \Delta_z\tilde{\varphi} = \Delta_y\tilde{\varphi}, \qquad \Delta_z\tilde{\varphi}^* = \Delta_y\tilde{\varphi}^*$$

So

$$A_2 = i\int_{\mathbb{T}^d} dy \int_{[-\frac{1}{2},\frac{1}{2}]^d} d\theta e^{-4\pi i\theta\cdot p} e^{-4\pi i\kappa\cdot y}[\Delta_y\tilde{\varphi}\tilde{\varphi}^* - \tilde{\varphi}\Delta_y\tilde{\varphi}^*]$$

$$= i\int_{\mathbb{T}^d} dy \int_{[-\frac{1}{2},\frac{1}{2}]^d} d\theta e^{-4\pi i\theta\cdot p} e^{-4\pi i\kappa\cdot y}[(\operatorname{div}\nabla_y\tilde{\varphi})\tilde{\varphi}^* - \tilde{\varphi}(\operatorname{div}\nabla_y\tilde{\varphi}^*)]$$

since $a \operatorname{div}(v) = \operatorname{div}(av) - v \cdot \nabla a$, we have that

$$a^* \operatorname{div}(v) - a \operatorname{div}(v^*) = \operatorname{div}(a^* v - a v^*) - v \cdot \nabla a^* + v^* \cdot \nabla a$$

and since in our case $a = \tilde{\varphi}$, $v = \nabla_y \tilde{\varphi}$ the last two terms cancel since

$$-v \cdot \nabla a^* + v^* \cdot \nabla a = -\nabla_y \tilde{\varphi} \cdot \nabla_y \tilde{\varphi}^* + \nabla_y \tilde{\varphi}^* \cdot \nabla_y \tilde{\varphi} = 0$$

Hence

$$A_2 = i \int_{\mathbb{T}^d} \mathrm{d}y \int_{\left[-\frac{1}{2}, \frac{1}{2}\right]^d} \mathrm{d}\theta e^{-4\pi i \theta \cdot p} e^{-4\pi i \kappa \cdot y} \operatorname{div}[(\nabla_y \tilde{\varphi}) \tilde{\varphi}^* - \tilde{\varphi}(\nabla_y \tilde{\varphi}^*)]$$

integrating by parts in $y$ one has

$$= -i \int_{\mathbb{T}^d} \mathrm{d}y \int_{\left[-\frac{1}{2}, \frac{1}{2}\right]^d} \mathrm{d}\theta e^{-4\pi i \theta \cdot p} \nabla_y (e^{-4\pi i \kappa \cdot y})[(\nabla_y \tilde{\varphi}) \tilde{\varphi}^* - \tilde{\varphi}(\nabla_y \tilde{\varphi}^*)]$$

$$= -4\pi \int_{\mathbb{T}^d} \mathrm{d}y \int_{\left[-\frac{1}{2}, \frac{1}{2}\right]^d} \mathrm{d}\theta e^{-4\pi i \theta \cdot p} e^{-4\pi i \kappa \cdot y}[(\kappa \cdot \nabla_y \tilde{\varphi}) \tilde{\varphi}^* - \tilde{\varphi}(\kappa \cdot \nabla_y \tilde{\varphi}^*)]$$

$$= -4\pi \int_{\mathbb{T}^d} \mathrm{d}y \int_{\left[-\frac{1}{2}, \frac{1}{2}\right]^d} \mathrm{d}\theta e^{-4\pi i \theta \cdot p} e^{-4\pi i \kappa \cdot y}[(\kappa \cdot \nabla_z \tilde{\varphi}) \tilde{\varphi}^* + \tilde{\varphi}(\kappa \cdot \nabla_z \tilde{\varphi}^*)]$$

$$= -4\pi \kappa \cdot \nabla_z \tilde{W}_\varphi(t, z, p, \eta, \kappa)$$

Finally for the term

$$A_3 = 4\pi \int_{\mathbb{T}^d} \mathrm{d}y \int_{\left[-\frac{1}{2}, \frac{1}{2}\right]^d} \mathrm{d}\theta e^{-4\pi i \theta \cdot p} e^{-4\pi i \kappa \cdot y}[((\eta + \theta) \cdot \nabla_z \tilde{\varphi}) \tilde{\varphi}^* + ((\eta - \theta) \cdot \nabla_z \tilde{\varphi}^*) \tilde{\varphi}]$$

$$= 4\pi \int_{\mathbb{T}^d} \mathrm{d}y \int_{\left[-\frac{1}{2}, \frac{1}{2}\right]^d} \mathrm{d}\theta e^{-4\pi i \theta \cdot p} e^{-4\pi i \kappa \cdot y}[((\eta + \theta) \cdot \nabla_y \tilde{\varphi}) \tilde{\varphi}^* + ((\theta - \eta) \cdot \nabla_y \tilde{\varphi}^*) \tilde{\varphi}]$$

using integration by parts for the first term in the sum, this is

$$= 4\pi \int_{\mathbb{T}^d} \mathrm{d}y \int_{\left[-\frac{1}{2}, \frac{1}{2}\right]^d} \mathrm{d}\theta e^{-4\pi i \theta \cdot p} e^{-4\pi i \kappa \cdot y}[\tilde{\varphi}((-\eta - \theta) \cdot \nabla_y \tilde{\varphi}^*) + ((\theta - \eta) \cdot \nabla_y \tilde{\varphi}^*) \tilde{\varphi}]$$

$$- 4\pi \int_{\mathbb{T}^d} \mathrm{d}y \int_{\left[-\frac{1}{2}, \frac{1}{2}\right]^d} \mathrm{d}\theta e^{-4\pi i \theta \cdot p} (\eta + \theta) \cdot \nabla_y (e^{-4\pi i \kappa \cdot y}) \tilde{\varphi} \tilde{\varphi}^*$$

The terms with $\theta$ in the first expression cancel to give

$$= -4\pi \int_{\mathbb{T}^d} \mathrm{d}y \int_{\left[-\frac{1}{2}, \frac{1}{2}\right]^d} \mathrm{d}\theta e^{-4\pi i \theta \cdot p} e^{-4\pi i \kappa \cdot y}[\tilde{\varphi}(\eta \cdot \nabla_y \tilde{\varphi}^*)]$$

$$- 4\pi \int_{\mathbb{T}^d} \mathrm{d}y \int_{\left[-\frac{1}{2}, \frac{1}{2}\right]^d} \mathrm{d}\theta e^{-4\pi i \theta \cdot p} e^{-4\pi i \kappa \cdot y}[\tilde{\varphi}(\eta \cdot \nabla_y \tilde{\varphi}^*)]$$

$$+ 16\pi^2 i \int_{\mathbb{T}^d} \mathrm{d}y \int_{\left[-\frac{1}{2}, \frac{1}{2}\right]^d} \mathrm{d}\theta e^{-4\pi i \theta \cdot p} (\eta + \theta) \cdot \kappa e^{-4\pi i \kappa \cdot y} \tilde{\varphi} \tilde{\varphi}^*$$

using integration by parts once more, this is

$$=4\pi \int_{\mathbb{T}^d} dy \int_{[-\frac{1}{2},\frac{1}{2}]^d} d\theta e^{-4\pi i\theta\cdot p} e^{-4\pi i\kappa\cdot y}[(\eta\cdot\nabla_y\tilde\varphi)\tilde\varphi^* - \tilde\varphi(\eta\cdot\nabla_y\tilde\varphi^*)]$$

$$+16\pi^2 i \int_{\mathbb{T}^d} dy \int_{[-\frac{1}{2},\frac{1}{2}]^d} d\theta e^{-4\pi i\theta\cdot p}(\eta+\theta)\cdot\kappa e^{-4\pi i\kappa\cdot y}\tilde\varphi\tilde\varphi^*$$

$$+4\pi \int_{\mathbb{T}^d} dy \int_{[-\frac{1}{2},\frac{1}{2}]^d} d\theta e^{-4\pi i\theta\cdot p}\eta\cdot\nabla_y(e^{-4\pi i\kappa\cdot y})\tilde\varphi\tilde\varphi^*$$

$$=4\pi \int_{\mathbb{T}^d} dy \int_{[-\frac{1}{2},\frac{1}{2}]^d} d\theta e^{-4\pi i\theta\cdot p} e^{-4\pi i\kappa\cdot y}[(\eta\cdot\nabla_y\tilde\varphi)\tilde\varphi^* - \tilde\varphi(\eta\cdot\nabla_y\tilde\varphi^*)]$$

$$+16\pi^2 i \int_{\mathbb{T}^d} dy \int_{[-\frac{1}{2},\frac{1}{2}]^d} d\theta e^{-4\pi i\theta\cdot p}(\eta+\theta)\cdot\kappa e^{-4\pi i\kappa\cdot y}\tilde\varphi\tilde\varphi^*$$

$$-16\pi^2 i \int_{\mathbb{T}^d} dy \int_{[-\frac{1}{2},\frac{1}{2}]^d} d\theta e^{-4\pi i\theta\cdot p}\eta\cdot\kappa e^{-4\pi i\kappa\cdot y}\tilde\varphi\tilde\varphi^*$$

The $\eta$ terms in the last two expressions cancel to give

$$=4\pi \int_{\mathbb{T}^d} dy \int_{[-\frac{1}{2},\frac{1}{2}]^d} d\theta e^{-4\pi i\theta\cdot p} e^{-4\pi i\kappa\cdot y}[(\eta\cdot\nabla_z\tilde\varphi)\tilde\varphi^* + \tilde\varphi(\eta\cdot\nabla_z\tilde\varphi^*)]$$

$$+16\pi^2 i \int_{\mathbb{T}^d} dy \int_{[-\frac{1}{2},\frac{1}{2}]^d} d\theta e^{-4\pi i\theta\cdot p}\theta\cdot\kappa e^{-4\pi i\kappa\cdot y}\tilde\varphi\tilde\varphi^*$$

$$=4\pi\eta\cdot\nabla_z\tilde W_\varphi(t,z,p,\eta,\kappa) - 4\pi \int_{\mathbb{T}^d} dy \int_{[-\frac{1}{2},\frac{1}{2}]^d} d\theta e^{-4\pi i\theta\cdot p}(-4\pi i\theta\cdot\kappa)e^{-4\pi i\kappa\cdot y}\tilde\varphi\tilde\varphi^*$$

$$=4\pi\eta\cdot\nabla_z\tilde W_\varphi(t,z,p,\eta,\kappa) - 4\pi \int_{\mathbb{T}^d} dy \int_{[-\frac{1}{2},\frac{1}{2}]^d} d\theta\kappa\cdot\nabla_p(e^{-4\pi i\theta\cdot p})e^{-4\pi i\kappa\cdot y}\tilde\varphi\tilde\varphi^*$$

$$=4\pi\eta\cdot\nabla_z\tilde W_\varphi(t,z,p,\eta,\kappa) - 4\pi\kappa\cdot\nabla_p\tilde W_\varphi(t,z,p,\eta,\kappa)$$

Overall, one has

$$\partial_t\tilde W_\varphi(t,z,p,\eta,\kappa) = 4\pi\eta\cdot\nabla_p\tilde W_\varphi(t,z,p,\eta,\kappa) - 4\pi\kappa\cdot\nabla_z\tilde W_\varphi(t,z,p,\eta,\kappa)$$

$$+4\pi\eta\cdot\nabla_z\tilde W_\varphi(t,z,p,\eta,\kappa) - 4\pi\kappa\cdot\nabla_p\tilde W_\varphi(t,z,p,\eta,\kappa)$$

$$+i\varepsilon^{1/2}\sum_{n\in\mathbb{Z}^d} e^{2\pi i n\cdot z}\hat V(n)\left[\tilde W_\varphi\left(t,z,p,\eta,\kappa+\frac{n}{2}\right) - \tilde W_\varphi\left(t,z,p,\eta,\kappa-\frac{n}{2}\right)\right]$$

and this proves the claim. □

## B.3 Estimates on smoothing operators

LEMMA B.4. *For $\mathcal{J}_v$ defined in expression (3.19), we have that estimate (3.21) holds.*

PROOF. Let $(m,n) \in \{(1,1),(1,2),(2,2)\}$. Consider

$$\|\mathcal{J}_\nu \varphi\|_{E_n} = \sum_{|\beta| \leqslant n} \int_{\mathbb{R}^d} dp \sum_{\kappa \in \left(\frac{\mathbb{Z}}{2}\right)^d} \langle \kappa \rangle^n \|D_p^\beta \mathcal{J}_\nu \varphi(\cdot, p, \kappa)\|_{l_\xi^2}$$

We have that

$$\sum_{\kappa \in \left(\frac{\mathbb{Z}}{2}\right)^d} \langle \kappa \rangle^n \|D_p^\beta \mathcal{J}_\nu \varphi(\cdot, p, \kappa)\|_{l_\xi^2} = \sum_{\kappa \in \left(\frac{\mathbb{Z}}{2}\right)^d} \langle \kappa \rangle^n \left( \sum_{\xi \in \mathbb{Z}^d} \left| D_p^\beta \mathcal{J}_\nu \varphi(\xi, p, \kappa) \right|^2 \right)^{1/2}$$

$$= \sum_{\kappa \in \left(\frac{\mathbb{Z}}{2}\right)^d} \langle \kappa \rangle^n \left( \sum_{\xi \in \mathbb{Z}^d} \left| D_p^\beta \big( e^{-\nu^{1/2} \langle \kappa \rangle} (\varphi(\xi, \cdot, \kappa) *_p \phi_\nu(\cdot))(p) \big) \right|^2 \right)^{1/2}$$

$$= \sum_{\kappa \in \left(\frac{\mathbb{Z}}{2}\right)^d} e^{-\nu^{1/2} \langle \kappa \rangle} \langle \kappa \rangle^n \|D_p^\beta ((\varphi(\xi, \cdot, \kappa) *_p \phi_\nu(\cdot))(p))\|_{l_\xi^2}$$

Distributing the derivatives between the terms, we split $\beta = \alpha_1(\beta) + \alpha_2(\beta) = \alpha_1 + \alpha_2$, such that $|\alpha_1(\beta)| = \min(|\beta|, m)$

$$D_p^\beta((\varphi(\xi, \cdot, \kappa) *_p \phi_\nu(\cdot))(p)) = (D_p^{\alpha_1} \varphi(\xi, \cdot, \kappa) *_p D_p^{\alpha_2} \phi_\nu(\cdot))(p)$$

Hence by the triangle inequality for the $l^2$-norm

$$\|D_p^\beta((\varphi(\xi, \cdot, \kappa) *_p \phi_\nu(\cdot))(p))\|_{l_\xi^2} \leqslant \|(D_p^{\alpha_1} \varphi(\xi, \cdot, \kappa) *_p D_p^{\alpha_2} \phi_\nu(\cdot))(p)\|_{l_\xi^2}$$

By Minkowski's inequality, this is

$$\leqslant \int_{\mathbb{R}^d} dq \|D_p^{\alpha_1} \varphi(\cdot, q, \kappa)\|_{l_\xi^2} |D_p^{\alpha_2} \phi_\nu(p-q)|$$

Hence

$$\sum_{\kappa \in \left(\frac{\mathbb{Z}}{2}\right)^d} \langle \kappa \rangle^n \|D_p^\beta \mathcal{J}_\nu \varphi(\cdot, p, \kappa)\|_{l_\xi^2} = \sum_{\kappa \in \left(\frac{\mathbb{Z}}{2}\right)^d} \langle \kappa \rangle^m \langle \kappa \rangle^{n-m} \|D_p^\beta \mathcal{J}_\nu \varphi(\cdot, p, \kappa)\|_{l_\xi^2}$$

By Tonelli's theorem, this is

$$\leqslant \int_{\mathbb{R}^d} dq \sup_{\kappa \in \left(\frac{\mathbb{Z}}{2}\right)^d} e^{-\nu^{1/2} \langle \kappa \rangle} \langle \kappa \rangle^{n-m} \sum_{\kappa \in \left(\frac{\mathbb{Z}}{2}\right)^d} \langle \kappa \rangle^m \|D_p^{\alpha_1} \varphi(\cdot, q, \kappa)\|_{l_\xi^2} |D_p^{\alpha_2} \phi_\nu(p-q)|$$

Integrating in $p$, and using Tonelli's theorem once more, we have that this is

$$\int_{\mathbb{R}^d} dp \sum_{\kappa \in \left(\frac{\mathbb{Z}}{2}\right)^d} \langle \kappa \rangle^n \|D_p^\beta \mathcal{J}_\nu \varphi(\cdot, p, \kappa)\|_{l_\xi^2} \leqslant \sup_{\kappa \in \left(\frac{\mathbb{Z}}{2}\right)^d} e^{-\nu^{1/2} \langle \kappa \rangle} \langle \kappa \rangle^{n-m}$$

$$\int_{\mathbb{R}^d} dq \sum_{\kappa \in \left(\frac{\mathbb{Z}}{2}\right)^d} \langle \kappa \rangle^m \|D_p^{\alpha_1} \varphi(\cdot, q, \kappa)\|_{l_\xi^2} \int_{\mathbb{R}^d} dp |D_p^{\alpha_2} \phi_\nu(p-q)|$$

Next we estimate

$$\int_{\mathbb{R}^d} dp\, |D_p^{\alpha_2} \phi_\nu(p-q)| = \int_{\mathbb{R}^d} dr\, |D_p^{\alpha_2} \phi_\nu(r)| = \nu^{-\frac{d}{2}} \int_{\mathbb{R}^d} dr\, \left| D_p^{\alpha_2}\left(\phi\left(\frac{r}{\nu^{1/2}}\right)\right) \right|$$

$$= \nu^{-\frac{d+|\alpha_2|}{2}} \int_{\mathbb{R}^d} dr\, \left| D_p^{\alpha_2} \phi\left(\frac{r}{\nu^{1/2}}\right) \right| = \nu^{-\frac{|\alpha_2|}{2}} \int_{\mathbb{R}^d} dr'\, |D_p^{\alpha_2} \phi(r')| \lesssim \nu^{-\frac{|\alpha_2|}{2}} \lesssim \nu^{-\frac{n-m}{2}}$$

since $|\alpha_2| = |\beta| - |\alpha_1|$ due to how we choose to split the derivative and we also have that

$$\sup_{\kappa \in \left(\frac{\mathbb{Z}}{2}\right)^d} e^{-\nu^{1/2}\langle\kappa\rangle} \langle\kappa\rangle^{n-m} = \sup_x e^{-x} \langle x\nu^{-1/2}\rangle^{n-m} \lesssim \nu^{-\frac{n-m}{2}} \sup_x e^{-x} \langle x\rangle^{n-m} \lesssim \nu^{-\frac{n-m}{2}}$$

so putting this together

$$\int_{\mathbb{R}^d} dp \sum_{\kappa \in \left(\frac{\mathbb{Z}}{2}\right)^d} \langle\kappa\rangle^n \|D_p^\beta \mathcal{J}_\nu \varphi(\cdot, p, \kappa)\|_{l_\xi^2} \lesssim \nu^{-(n-m)} \int_{\mathbb{R}^d} dq \sum_{\kappa \in \left(\frac{\mathbb{Z}}{2}\right)^d} \langle\kappa\rangle \|D_p^{\alpha_1} \varphi(\cdot, q, \kappa)\|_{l_\xi^2}$$

Hence summing over $\beta$, one has that

$$\|\mathcal{J}_\nu \varphi\|_{E_n} \lesssim \nu^{-(n-m)} \sum_{|\beta| \leq n} \int_{\mathbb{R}^d} dq \sum_{\kappa \in \left(\frac{\mathbb{Z}}{2}\right)^d} \langle\kappa\rangle^m \|D_p^{\alpha_1(\beta)} \varphi(\cdot, q, \kappa)\|_{l_\xi^2}$$

since for every $\beta$ we chose $\alpha_1 : |\alpha_1| \leq 1$, at the price of worsening the constant, this is

$$\lesssim \nu^{-(n-m)} \sum_{|\alpha_1| \leq m} \int_{\mathbb{R}^d} dq \sum_{\kappa \in \left(\frac{\mathbb{Z}}{2}\right)^d} \langle\kappa\rangle^m \|D_p^{\alpha_1} \varphi(\cdot, q, \kappa)\|_{l_\xi^2}$$

$$\lesssim \nu^{-(n-m)} \|\varphi\|_{E_1}$$

$\square$

LEMMA B.5. *For $\mathcal{J}_\nu$ defined in expression (3.19), we have that estimate (3.22) holds.*

PROOF. Consider

$$\|(\mathcal{J}_\nu - \mathrm{Id})\varphi\|_{E_0} = \int_{\mathbb{R}^d} dp \sum_{\kappa \in \left(\frac{\mathbb{Z}}{2}\right)^d} \|(\mathcal{J}_\nu - \mathrm{Id})\varphi(\cdot, p, \kappa)\|_{l_\xi^2}$$

Now

$$(\mathcal{J}_\nu - \mathrm{Id})\varphi(\xi, p, \kappa) = e^{-\nu^{1/2}\langle\kappa\rangle}(\varphi(\xi, \cdot, \kappa) *_p \phi_\nu(\cdot))(p) - \varphi(\xi, p, \kappa)$$

$$= (e^{-\nu^{1/2}\langle\kappa\rangle} - 1)(\varphi(\xi, \cdot, \kappa) *_p \phi_\nu(\cdot))(p) - (\varphi(\xi, p, \kappa) - (\varphi(\xi, \cdot, \kappa) *_p \phi_\nu(\cdot))(p))$$

We handle the two terms separately, beginning with

$$\int_{\mathbb{R}^d} dp \sum_{\kappa \in \left(\frac{\mathbb{Z}}{2}\right)^d} \|(e^{-\nu^{1/2}\langle\kappa\rangle} - 1)(\varphi(\xi, \cdot, \kappa) *_p \phi_\nu(\cdot))(p)\|_{l_\xi^2}$$

$$= \int_{\mathbb{R}^d} dp \sum_{\kappa \in \left(\frac{\mathbb{Z}}{2}\right)^d} |e^{-\nu^{1/2}\langle\kappa\rangle} - 1| \|(\varphi(\xi, \cdot, \kappa) *_p \phi_\nu(\cdot))(p)\|_{l_\xi^2}$$

Now using that $|e^{-x} - 1| \leqslant x$ for $x > 0$, this is

$$\leqslant \nu^{1/2} \int_{\mathbb{R}^d} dp \sum_{\kappa \in \left(\frac{\mathbb{Z}}{2}\right)^d} \|(\varphi(\xi,\cdot,\kappa) *_p \phi_\nu(\cdot))(p)\|_{l^2_\xi}$$

by Minkowski's inequality, this is

$$\leqslant \nu^{1/2} \int_{\mathbb{R}^d} dp \sum_{\kappa \in \left(\frac{\mathbb{Z}}{2}\right)^d} \int_{\mathbb{R}^d} dq \|\varphi(\cdot,q,\kappa)\|_{l^2_\xi} \phi_\nu(p-q)$$

By using Tonelli's theorem, this is

$$\leqslant \nu^{1/2} \sum_{\kappa \in \left(\frac{\mathbb{Z}}{2}\right)^d} \int_{\mathbb{R}^d} dq \|\varphi(\cdot,q,\kappa)\|_{l^2_\xi} \int_{\mathbb{R}^d} dp \phi_\nu(p-q)$$

$$\leqslant \nu^{1/2} \int_{\mathbb{R}^d} dq \sum_{\kappa \in \left(\frac{\mathbb{Z}}{2}\right)^d} \|\varphi(\cdot,q,\kappa)\|_{l^2_\xi} = \nu^{1/2} \|\varphi\|_{E_0}$$

$$\leqslant \nu^{1/2} \|\varphi\|_{E_1}$$

For the second term, we consider

$$\int_{\mathbb{R}^d} dp \sum_{\kappa \in \left(\frac{\mathbb{Z}}{2}\right)^d} \|\varphi(\cdot,p,\kappa) - (\varphi(\cdot,\cdot,\kappa) *_p \phi_\nu(\cdot))(p)\|_{l^2_\xi}$$

Now

$$\varphi(\xi,p,\kappa) - (\varphi(\xi,\cdot,\kappa) *_p \phi_\nu(\cdot))(p) = \int_{\mathbb{R}^d} dq (\varphi(\xi,p-q,\kappa) - \varphi(\xi,p,\kappa)) \phi_\nu(q)$$

$$= \int_{\mathbb{R}^d} dr (\varphi(\xi,p-\nu^{1/2}r,\kappa) - \varphi(\xi,p,\kappa)) \phi(r)$$

Hence

$$\|\varphi(\cdot,p,\kappa) - (\varphi(\cdot,\cdot,\kappa) *_p \phi_\nu(\cdot))(p)\|_{l^2_\xi} \leqslant \int_{\mathbb{R}^d} dr \|\varphi(\cdot,p-\nu^{1/2}r,\kappa) - \varphi(\cdot,p,\kappa)\|_{l^2_\xi} \phi(r)$$

$$\leqslant \int_{\mathbb{R}^d} dr \left\| \int_0^{\nu^{1/2}} dt\, r\cdot \nabla \varphi(\cdot,p-tr,\kappa) \right\|_{l^2_\xi} \phi(r) \leqslant \int_{\mathbb{R}^d} dr \int_0^{\nu^{1/2}} dt \|r\cdot \nabla \varphi(\cdot,p-tr,\kappa)\|_{l^2_\xi}$$

$$\leqslant \int_{\mathbb{R}^d} dr |r| \phi(r) \int_0^{\nu^{1/2}} dt \|\nabla \varphi(\cdot,p-tr,\kappa)\|_{l^2_\xi}$$

Hence by Tonelli's theorem

$$\int_{\mathbb{R}^d} dp \sum_{\kappa \in \left(\frac{\mathbb{Z}}{2}\right)^d} \|\varphi(\cdot,p,\kappa) - (\varphi(\cdot,\cdot,\kappa) *_p \phi_\nu(\cdot))(p)\|_{l^2_\xi}$$

$$\leqslant \int_{\mathbb{R}^d} dr |r| \phi(r) \int_0^{\nu^{1/2}} dt \int_{\mathbb{R}^d} dp \sum_{\kappa \in \left(\frac{\mathbb{Z}}{2}\right)^d} \|\nabla \varphi(\cdot,p-tr,\kappa)\|_{l^2_\xi}$$

$$\leqslant \int_{\mathbb{R}^d} dr |r| \phi(r) \int_0^{\nu^{1/2}} dt \|\varphi\|_{E_1} \leqslant \nu^{1/2} \|\varphi\|_{E_1}$$

This concludes the proof. □

## C  Miscellaneous

LEMMA C.1. *The Fourier transform of a Gaussian is a Gaussian : For $a > 0$ and $f(x) := e^{-a\pi |x|^2}$, $x \in \mathbb{R}^d$, we have that*

$$\hat{f}(\xi) = \frac{1}{a^{\frac{d}{2}}} e^{-\frac{\pi |\xi|^2}{a}}$$

This is a standard result in Fourier analysis.

LEMMA C.2. *(Poisson summation formula) Let $f \in \mathscr{S}(\mathbb{R}^d; \mathbb{C})$. Then*

$$\sum_{m \in \mathbb{Z}^d} f(m) = \sum_{m = \mathbb{Z}^d} \hat{f}(m)$$

This is a standard result in harmonic analysis. For a proof, see for instance [19].